\documentclass[10pt,letterpaper,twocolumn,nofootinbib,prx,english,superscriptaddress]{revtex4-1} 
\usepackage{graphicx}
\usepackage{amssymb}
\usepackage{epstopdf}
\usepackage{amsmath}
\usepackage{amsfonts}
\usepackage[english]{babel}
\usepackage{bbold}
\usepackage{color}
\usepackage{amsthm,amsmath,amssymb,graphicx,comment} 
\usepackage{mathtools}
\usepackage{xfrac}
\usepackage{hyperref}
\usepackage[T1]{fontenc}
\usepackage[utf8]{inputenc}
\usepackage[font=small,labelfont=bf]{caption}
\usepackage{soul}
\newcommand{\ket}[1]{\left| #1 \right\rangle}
\newcommand{\bra}[1]{\left\langle #1 \right|}

\newcommand{\beq}{\begin{equation}}
\newcommand{\eeq}{\end{equation}}
\newcommand{\bea}{\begin{align}}
\newcommand{\eea}{\end{align}}

\usepackage{hyperref}

\begin{document}
\title{Why initial system-environment correlations do not imply the failure of complete positivity: a causal perspective}
\author{David Schmid}
\email{dschmid@perimeterinstitute.ca}
\affiliation{Perimeter Institute for Theoretical Physics, 31 Caroline Street North, Waterloo, Ontario Canada N2L 2Y5}
\affiliation{Institute for Quantum Computing and Department of Physics and Astronomy, University of Waterloo, Waterloo, Ontario N2L 3G1, Canada}
\author{Katja Ried}
\affiliation{Institut f\"{u}r Theoretische Physik, Universit\"{a}t Innsbruck, Technikerstra{\ss}e 21a, 6020 Innsbruck, Austria}
\author{Robert W. Spekkens}
\affiliation{Perimeter Institute for Theoretical Physics, 31 Caroline Street North, Waterloo, Ontario Canada N2L 2Y5}
\affiliation{Institute for Quantum Computing and Department of Physics and Astronomy, University of Waterloo, Waterloo, Ontario N2L 3G1, Canada}

\begin{abstract}
The common wisdom in the field of quantum information theory is that when a system is initially correlated with its environment, the map describing its evolution may fail to be completely positive.  If true, this would have practical and foundational significance. 
We here demonstrate, however, that the common wisdom is mistaken.
We trace the error to the standard proposal for how the evolution map ought to be defined. We summarize this standard proposal and then show that it sometimes fails to define a linear map or any map at all. Further, we show that these pathologies persist even in completely classical examples. 
Drawing inspiration from the framework of classical causal models, we argue that the correct definition of the evolution map is obtained by considering a counterfactual scenario wherein the system is reprepared independently of any systems in its causal past while the rest of the circuit remains the same, yielding a map that is always completely positive. In a post-mortem on the standard proposal, we highlight two distinct mistakes that retrospectively become evident (in its application to completely classical examples): (i) the types of constraints to which it appealed are constraints on what one can {\em infer} about the final state of a system based on its initial state, where such inferences are based not just on the cause-effect relation between them---which defines the correct evolution map---but also on the common cause of the two; (ii) in a (retrospectively unnecessary) attempt to introduce variability in the input state, it inadvertently introduced variability in the inference map itself, then tried to fit the input-output pairs associated to these different maps with a single map. 
\end{abstract}
\maketitle

\section{Introduction}

Quantum state evolution is typically represented by a completely positive linear map. 
%A map $\mathcal{E}_{A}: \mathcal{L}(\mathcal{H}_A) \to \mathcal{L}(\mathcal{H}_A)$, where $\mathcal{L}(\mathcal{H}_A)$ denotes the linear operators on $\mathcal{H}_A$, is said to be completely positive if the map $\mathcal{E}_{A} \otimes {\rm id}_{B}$ is positivity-preserving on the composite $AB$, where $B$ is any ancillary system and ${\rm id}_B$ is the identity map on $B$. That is, it is completely positive if for all $\rho_{AB} \ge 0$, one has $\mathcal{E}_{A} \otimes {\rm id}_{B} (\rho_{AB}) \ge 0$.  
There are two justifications that are usually given for this. %doing so. 
 %typically given for assuming complete positivity for quantum evolution maps. 
 The first (the axiomatic justification) is that one always has the freedom to implement a quantum evolution map on a subsystem of some composite system, and complete positivity is then required in order for the state of the composite to remain positive, which is in turn required in order for the Born rule to return positive numbers  as the probabilities for the outcomes of future measurements on the system.
 %\footnote{It is important to preserve the positivity of states, since otherwise there would be measurements for which the probabilities assigned to their outcomes (by the Born rule) would not be positive numbers. }  
 The second justification notes that the evolution of an isolated system is always described by a unitary map.  This implies that the most general sort of evolution of an open system arises by unitarily coupling the system to an environment (in some fixed state) 
 %, initially in some fixed quantum state, 
 and then tracing over the environment,
%Denoting the system by $S$ and the environment by $E$, the evolution map is $\mathcal{E}_S (\cdot_S) = {\rm Tr}_E \big( \mathcal{U} (\cdot_S \otimes \rho_E) \big)$. 
%Finally, one appeals to the fact that 
and such evolution is always represented by a completely positive map.
%the effective evolution of the system in this case is described by a completely positive map
% (in fact, {\em every} completely positive map can be realized in this fashion, a result known as the Stinespring dilation theorem \{[cite Stinespring]} ).
%\footnote{The justification based on dilation is appropriate for the `Church of the larger Hilbert space' view of quantum theory, while those preferring the `Church of the smaller Hilbert space' presumably prefer the axiomatic justification.}.

%This representation can be justified in several ways. First, there is the justification by dilation (church of the larger Hilbert space). If one believes that unitary dynamics is fundamental, then complete positivity arises for the marginal map on the system when the environment is presumed to be prepared in some fixed state, be coupled unitarily to the system, and then to be traced out. Second, there is the axiomatic derivation (church of the smaller Hilbert space): physically, if one demands that the joint state resulting from acting a transformation on half of an entangled state is a valid quantum state, then one finds that the map must be completely positive.

However, the common wisdom in the field of quantum information is that there is an exception to the rule that the evolution of quantum states is represented by completely positive maps, namely, when the initial state of the system-environment composite does not factorize. 
%\{[Provide reference to figure at this point?]} 
For instance, Nielsen and Chuang~[\cite{NC}, Sec. 8.5] state that 
\begin{quote}
``a quantum system which interacts with the degrees of freedom used to prepare that system after the preparation is complete will in general suffer a dynamics which is not adequately described within the quantum operations formalism'',
\end{quote}
(here, ``quantum operations'' refer to completely positive trace-preserving linear maps), and that 
\begin{quote}
``It is an interesting problem for further research to study quantum information processing beyond the quantum operations formalism.''
\end{quote}

A large body of literature has arisen to address this problem~\cite{Pechukas,Alicki,Preplies,stelm,Kraus,linear,sud,afraid,standard2,beyond,standard1,lidar,erratum,assign1,Cuffaro,brodutch,Dominy2015,Byrd,Dominy2016}.
This literature includes a great diversity of examples to support the claim that quantum state evolution cannot always be described by a completely positive map from initial to final states of the system.
%{\color{red} [Need to think about how to do citations here...]}

Some recent work in the literature, e.g. Ref.~\cite{Modi1,Modi4}, has criticized non-completely positive maps on the grounds that they lack operational significance. These authors then advocate for a more operational approach (building on Refs.~\cite{Mrole,Mprep}), introducing an alternative framework that avoids the issue of non-complete positivity by proposing a different type of map~\cite{ModiChar,Modi2,Modi1,Modi4,Modi3,ModiChir} as representing the evolution of the system.
%to `describe the evolution'
%~\footnote{Our work provides an independent motivation for this new type of map. However, in our view the map addresses an inference problem, and is not a direct description of evolution. See Section~\ref{causalmaps} for more discussion.}. 
  The relation between this prior work and our own proposal is discussed in Sec.~\ref{causalmaps}.

Here, we take a more radical stance against non-completely positive maps, arguing that they are explicitly at odds with the standard notion of evolution from elsewhere in physics, a fact that we highlight by taking the perspective of causal modeling. In particular, we do {\em not} assume that the evolution of a quantum state is constrained by the marginal states of the system before and after the evolution, as has previously been assumed; indeed, we show that this constraint does not hold.
%no more captured by a time series of states (as assumed in all prior work) than causal influences are captured by statistical correlations. 

In the framework of classical causal models, the evolution of a system in any setting is described by `do-conditionals'~\cite{Pearl}. These were introduced to describe the inferences that can be made from one variable to another based solely on their {\em cause-effect relation}, even when these variables also exhibit correlations due to a common cause. In particular, the correlations that arise between the initial and final versions of a system undergoing Markovian evolution is an example of correlations based solely on a cause-effect relation, whereas correlations between the initial and final versions of a system when the system is initially correlated with an environment with which it subsequently interacts are an example of correlations that are due to both a cause-effect and a common-cause relation. 
%a system and its environment, considered at the same time, are an example of correlations based solely on a common-cause relation. 
%This common cause relation obviously changes one's {\em inferences} about the state of the system over time,  but it plays no role in the description of {\em evolution}.  
 The presence of a nontrivial common cause relation changes the sorts of {\em inferences} one can make about the final version of the system based on knowledge of the initial version of the system, but the map describing such inferences does not describe the system's evolution.

 %\color{cyan} [Last sentence really needs an example of correlations between a system at one time and another time that are connected by a common cause.] \color{black}

Our criticism of the standard proposal for how to define an evolution map rests on the fact that for purely classical examples, it contradicts the answer that one obtains by defining an evolution map in the presence of initial system-environment correlations using a do-conditional, as the framework of causal inference would have us do. The thrust of our paper is to argue that once one separates out inference and influence, one sees that the mathematical relation that the standard proposal focuses on has no relevance for the map which describes the dynamics of the system; it need not constrain the {\em evolution} map in any way. This is in explicit contradiction with the common wisdom.

Consequently, we argue that the common wisdom is mistaken, and that the evolution of a quantum system in time is always represented by a linear and completely positive map.
% from initial to final states of the system.
Our definition of the evolution map, therefore, does not lead to any of the pathological conclusions of the standard proposal, such as the failure of complete positivity, the failure of linearity or the failure to define a map.  

Pearl has expressed a dictum regarding what counts as a satisfactory resolution of an apparent paradox~\cite{Why}, namely, that the resolution should allow one to retrospectively identify why there was an appearance of paradox.  The framework of causal modelling allows us to identify two distinct mistaken assumptions in the standard proposal and to understand precisely how the pathological conclusions arise from these mistakes.

We show that the correct evolution map depends on the marginal state of the environment, but does {\em not} depend on the correlations between the system and the environment. This fact is {\em derived} from our conception of evolution, and is not assumed {\em a priori}. Further, it holds true regardless of the nature of the correlations or the operational procedure by which the initial state was prepared. 

%As we noted above, some prior work has also suggested that the dynamics of the system ought to be represented by a map that is distinct from the one advocated by the standard proposal, in particular, a map that is completely positive.  The relation between this prior work and our own proposal is discussed in Sec.~\ref{causalmaps}.

Pursuing a quantum generalization of the classical framework of causal modelling has already had many interesting applications in quantum foundations, including:  revealing a quantum advantage for causal inference~\cite{Katja2}, uncovering new experimental scenarios wherein there is a gap between quantum and classical correlations \cite{Fritz2012,Pusey2014,Chaves2014,Chaves2015,Colbeck,Pienaar2017,Chaves2018}, uncovering a promising approach to achieving a causal explanation of Bell inequality violations without fine-tuning~\cite{Wood, bayes, EricLal,QCM1,Cavalcanti}, expanding the set of experimental configurations wherein one can achieve quantum state pooling \cite{Leifer2014}, and exploring the possibility of quantum uncertainty about the causal structure~\cite{Katja1,Giulio,Brukner2017}.

This article continues this trend by showing that the correct definition of a quantum evolution map in the presence of initial system-environment correlations requires a quantum generalization of a key notion from classical causal modelling, that of the do-conditional, thereby demonstrating the conceptual significance of the quantum version of this notion.

\subsection{Outline of the paper}

%Section~\ref{defns} lays out some basic definitions. 
%Section~\ref{sec:standard} lays out the standard argument that is found in the literature for how to compute an evolution map and for why the evolution map is sometimes not given by a completely positive map. 
Section~\ref{sec:standard} lays out the standard argument for the inadequacy of completely positive maps for describing evolution,
% in the presence of initial system-environment correlations, 
abstracted from the various perspectives on the subject found in the literature.
% , for how to compute an evolution map and for why the evolution map is sometimes not given by a completely positive map. 
The argument is based on a proposal, which we term the {\em standard proposal}, for how to define the evolution map in the scenarios of interest.
We give several examples in which the standard proposal leads to problematic conclusions, and we explain why these conclusions cast doubt on the validity of the proposal. 
% and we ultimately show that it does not provide a valid definition of the evolution map.

In Section~\ref{sec:classical}, we show that the pathological consequences of 
%inconsistencies in
 the standard proposal arise even in purely classical scenarios. In Section~\ref{causalpersp}, we describe how to define a classical evolution map in the presence of initial system-environment correlations, using the notion of do-conditionals from the 
 %well-understood 
 framework of classical causal models. In Section~\ref{sec:prob}, we show that the prescription of the standard proposal generally fails to reproduce it,
 %the correct classical evolution map, 
 and we leverage the framework of classical causal models to elucidate the underlying mistaken assumptions 
 %of the standard argument 
 within the scope of classical scenarios. 

In Section~\ref{sec:quantumdomaps}, we show that the definition of the classical evolution map in terms of do-conditionals generalizes naturally to a definition of the quantum evolution map, which is seen to be always completely positive and to avoid the pathologies of the standard proposal. We then assess the mistakes of the standard proposal in the quantum sphere.%Furthermore, we see that the mistaken assumptions of the standard argument in quantum scenarios are precisely the same mistaken assumptions that are made by the standard argument in classical scenarios. These mistakes lay in the standard argument's reliance on observational data only, rather than incorporating facts about causal structure.

Finally, in Section~\ref{disc} we discuss some of the implications of our definition of the quantum evolution map, we explain how it can be extracted from experimental data, and we advocate for the study of open-system dynamics from the causal modelling perspective.
%a shift in research focus for open-system quantum dynamics, towards an approach based in causal modeling.

\section{Problems with the standard proposal}
%Pathological implications of the standard argument
 %for the inadequacy of completely positive maps
\label{sec:standard}

\subsection{Preliminaries} \label{defns}

First we recall some mathematical facts. Let $\mathcal{L}(\mathcal{H_S})$ denote the space of linear operators on the Hilbert space $\mathcal{H_S}$ describing system $S$. We denote a map from $\mathcal{L}(\mathcal{H}_{S_1})$ to $\mathcal{L}(\mathcal{H}_{S_2})$ by $\mathcal{E}_{S_2|S_1}$. Such a map is said to be {\em trace-preserving} if $\forall \rho_{S_1} \in \mathcal{L}(\mathcal{H}_{S_1}): {\rm Tr}_{S_2}[ \mathcal{E}_{S_2|S_1}(\rho_{S_1})] = {\rm Tr}_{S_1} (\rho_{S_1})$. It is said to be {\em positivity-preserving}, or simply {\em positive}, if it takes all positive operators to positive operators, $\forall \rho_{S_1} \ge 0: \mathcal{E}_{S_2|S_1}(\rho_{S_1}) \ge 0$. It is said to be {\em completely positivity-preserving}, or simply {\em completely positive}, if its action on a composite system is also positive; that is, for any ancillary system, denoted $E_1$ at the initial time and $E_2$ at the final time, and evolving by the identity map, ${\rm id}_{E_2|E_1}$, we have $\forall \rho_{S_1 E_1} \ge 0,\; (\mathcal{E}_{S_2|S_1} \otimes {\rm id}_{E_2|E_1} ) (\rho_{S_1 E_1}) \ge 0$.

Suppose that the principal system $S$ is coupled to an ancillary system $E$ by a unitary $U_{SE}$, that $E$ is prepared at the initial time in the state $\rho_{E}$ and that one traces over the ancillary system to obtain the final state of the principal system. The evolution of the principal system is then represented by the map $\mathcal{E}_{S_2|S_1}(\framebox(5,5){}_{S_1}\!):\mathcal{L}(\mathcal{H}_{S_1}) \rightarrow \mathcal{L}(\mathcal{H}_{S_2})$ defined by
\beq \label{product}
\mathcal{E}_{S_2|S_1}(\framebox(5,5){}_{S_1}\!) = {\rm Tr}_E[U_{S_1E}(\framebox(5,5){}_{S_1}\! \otimes \rho_E )U_{S_1E}^{\dagger}].
\eeq
Note that throughout this article we denote the argument of a map by $\framebox(5,5){}_{A}$, where the subscript specifies the type of system at the input.
%$A$ labels the relevant system. 
Clearly, such a map is completely positive and trace-preserving. It turns out, furthermore, that {\em any} completely positive trace-preserving map can be realized in this fashion, a result known as the {\em Stinespring dilation} theorem~\cite{Stinespring}.

\subsection{The standard argument for the inadequacy of completely positive maps}\label{standardargument4inadequacyCP}

We now review the standard argument for the inadequacy of completely positive maps in describing the evolution of the quantum state of the principal system (which we henceforth simply call `the system') when it is initially correlated with the environment~\cite{NC,Pechukas,Alicki,afraid,sud,standard1,beyond,assign1,stelm,Mrole,lidar,brodutch,Kraus,linear}. 

\begin{figure}[htb!]
\centering
\includegraphics[width=0.22\textwidth]{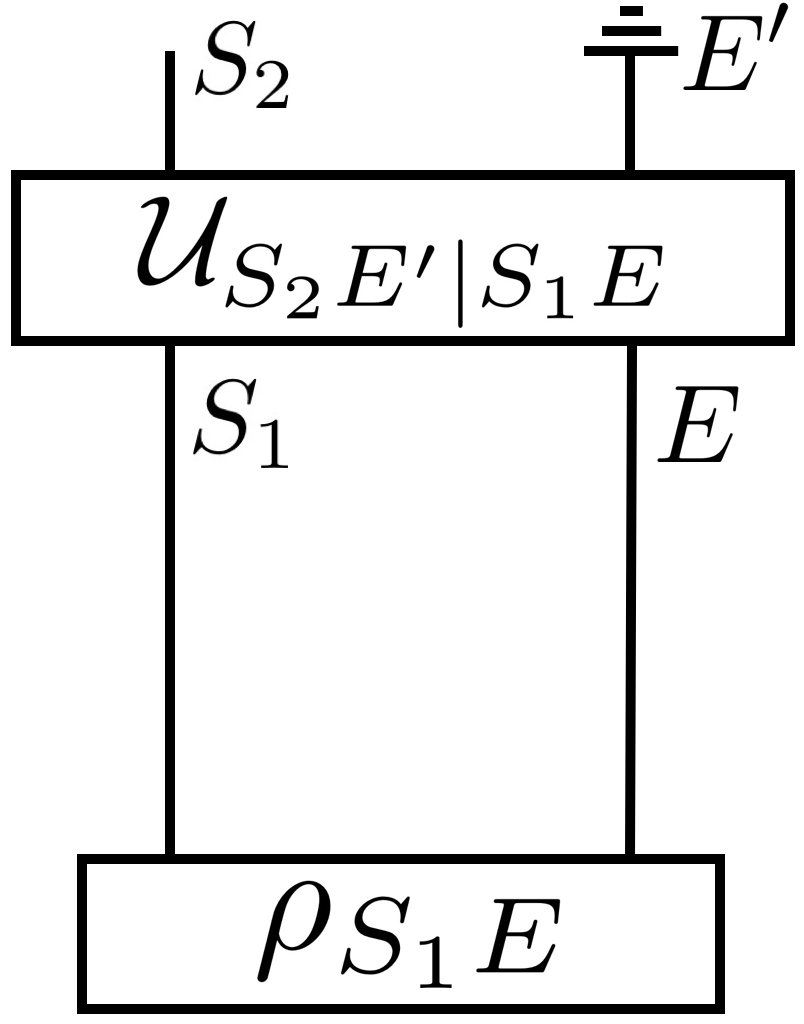}
\caption{The scenario of interest.  An initially correlated state of system and environment, $\rho_{S_1E}$, evolves according to a quantum channel $\mathcal{E}_{S_2|S_1 E}$.}
\label{onestate}
\end{figure}

To assume that the system and environment are initially correlated is to assume that their joint state does not factorize, that is, $\rho_{S_1 E}\ne \rho_{S_1} \otimes \rho_E$.
The system and environment are imagined to subsequently interact according to the map
%some unitary superoperator 
\begin{equation} \label{supop}
\mathcal{U}_{S_2E'|S_1 E}(\framebox(5,5){}_{S_1 E}\!):=U \ \framebox(5,5){}_{S_1 E}U^{\dagger},
\end{equation}
where $U: \mathcal{H}_{S_1}\otimes\mathcal{H}_{E}\rightarrow \mathcal{H}_{S_2}\otimes\mathcal{H}_{E'}$ is a unitary operator.
%\mathcal{L}(\mathcal{H}_{S_1}\otimes\mathcal{H}_{E}) \rightarrow \mathcal{L}(\mathcal{H}_{S_2})$. 
This scenario is depicted in Fig.~\ref{onestate}.

The standard argument is predicated on a particular proposal for how to define the evolution map from $S_1$ to $S_2$, which we call the {\em standard proposal}. (We will argue in the following that the standard argument is mistaken precisely because the standard proposal is conceptually misguided.) The standard proposal begins by assuming that the evolution map,
%The standard argument begins by presuming that the map representing the system's evolution, 
which we denote here by $\mathcal{E}_{S_2|S_1}$, is constrained by the following equation:
\beq \label{corr}
\rho_{S_2} = \mathcal{E}_{S_2|S_1}(\rho_{S_1}),
\eeq
where $\rho_{S_1}$ denotes the marginal on $S_1$,
\beq
\rho_{S_1} := {\rm Tr}_E (\rho_{S_1 E}),
\eeq
and $\rho_{S_2}$ denotes the marginal on $S_2$,
%\beq
%\rho_{S_2} := {\rm Tr}_E (U_{S_1 E}\rho_{S_1 E}U_{S_1 E}^{\dagger}).
%\eeq
\beq
\rho_{S_2} := {\rm Tr}_{E'} (\mathcal{U}_{S_2 E'|S_1 E}(\rho_{S_1 E})).
\eeq

However, Eq.~\eqref{corr} only specifies how a single state of the system, namely $\rho_{S_1}$, is transformed. This is clearly not sufficient to determine how an arbitrary state on $S_1$ is transformed, and therefore Eq.~\eqref{corr} does not serve to define a map uniquely.\footnote{In particular, any pair of states $\rho_{S_1}$ and $\rho_{S_2}$ is consistent with the map that ignores the state of $S_1$ and simply prepares $\rho_{S_2}$, $\mathcal{E}_{S_2|S_1} (\framebox(5,5){}_{S_1}\!)= \rho_{S_2} {\rm Tr}_{S_1}(\framebox(5,5){}_{S_1}\!)$.}

In order to define a map uniquely, it is critical that its action be specified on many different input states. Towards this end, most articles on the topic do not consider a single joint state on system and environment, but rather a set 
%$\{\rho_{S_1E}^{(j,k)}\}_{j,k}$
 of such states with differing marginal states for the system. The specific means by which this variation is generated differs among proposals: one might apply a transformation on the system, or one might apply a joint transformation on the system and the environment, or one might imagine performing some non-destructive measurement on the system. Physically, it must be that some sort of laboratory operation induces variation on the initial system-environment state. We denote the random variable which encodes the setting of this operation by $J$ and the random variable which encodes its outcome by $K$. 
%In special cases, either $J$ or $K$ (but not both) may be trivial. 
For each pair of values $(j,k)$ for these variables, the system-environment composite is prepared in a corresponding state $\rho^{(j,k)}_{S_1E}$. 

\begin{figure}[htb!]
\centering
\includegraphics[width=0.22\textwidth]{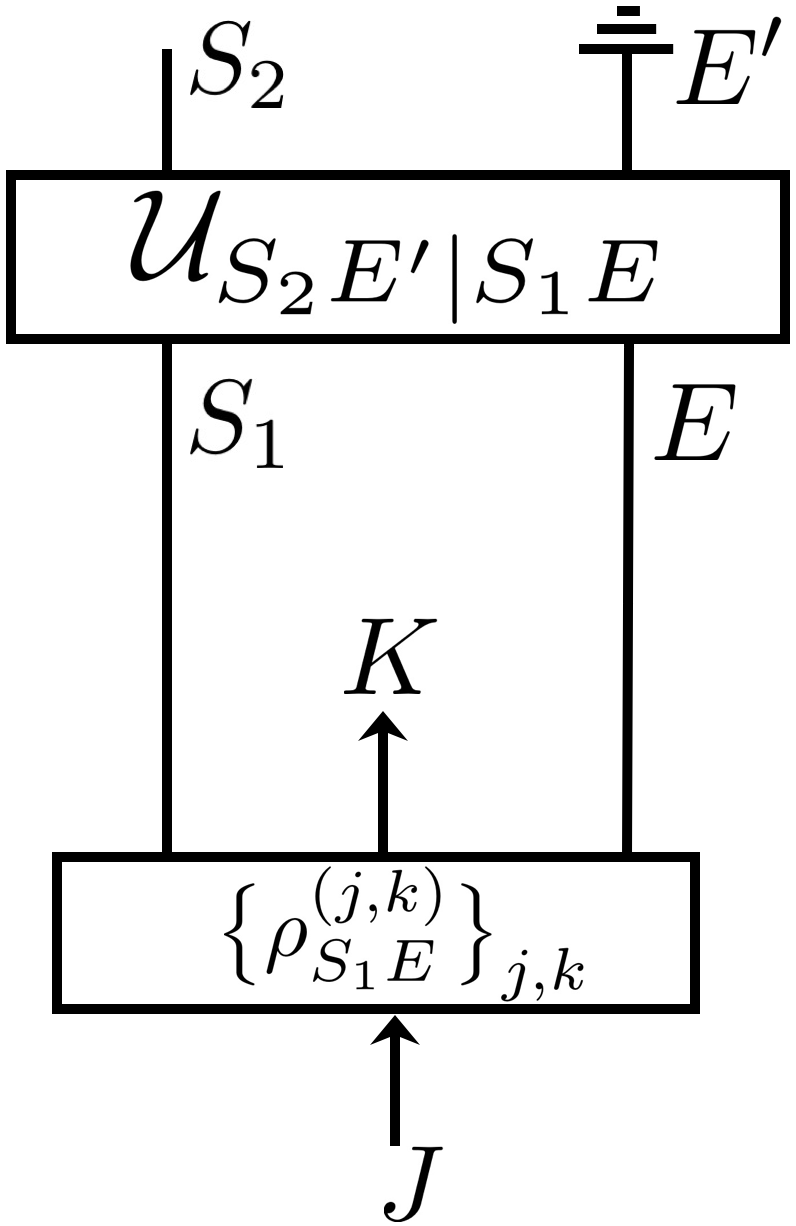}
\caption{ The most general circuit considered in the standard proposal. 
%(a) The most general circuit of interest. The state preparation protocol---denoted by a dashed box---has input $J$ and outcome $K$. The second state in the $(j,k)$-th ordered pair of $\mathfrak{R}$ is the state of $S_2$ defined by this circuit, for $J=j$ and $K=k$. The first state in the $(j,k)$-th ordered pair of $\mathfrak{R}$ is the state of $S_1$ defined by a fragment of the full circuit, shown in (b), for $J=j$ and $K=k$.
}
\label{fig:standard}
\end{figure}

The general circuit diagram representing the class of scenarios studied in the literature, then, is that of Fig.~\ref{fig:standard}. 
%In all of our figures, we highlight the part of the circuit which represents the preparation of the various initial joint states by a dashed grey box.

The standard proposal asks us to consider the mathematical relation
%Consider now the mathematical relation
\beq
\mathfrak{R} := \{ (\rho^{(j,k)}_{S_1},\rho^{(j,k)}_{S_2})\}_{j,k}
\eeq
where 
\beq
\rho^{(j,k)}_{S_1}:= {\rm Tr}_E (\rho^{(j,k)}_{S_1E})\label{rhoS1}
\eeq
and 
\beq
\rho^{(j,k)}_{S_2}:= {\rm Tr}_{E'}[ \mathcal{U}_{S_2 E'| S_1E}(\rho^{(j,k)}_{S_1E} )]\label{rhoS2},
\eeq
which we term the {\em input-output relation}.
%We refer to $\mathfrak{R}$ as the {\em input-output relation}.
It asserts that the evolution map $\mathcal{E}_{S_2|S_1}: \mathcal{L}(\mathcal{H}_{S_1}) \to \mathcal{L}(\mathcal{H}_{S_2})$ should satisfy the constraints
\beq
 \forall j,k: \quad \mathcal{E}_{S_2|S_1} [\rho^{(j,k)}_{S_1}] = \rho^{(j,k)}_{S_2}.\label{mapfromrelation}
\eeq
%i.e., that if the first state in any ordered pair is its input, then the second state in that ordered pair is its output.

Note that one could only hope to uniquely define a map on all input states in this manner if the map is assumed to be linear and the domain of the input-output relation includes an informationally complete set of states (i.e., the set $\{ \rho^{(j,k)}_{S_1}\}_{j,k}$ forms a basis of the operator space $\mathcal{L}(\mathcal{H}_{S_1})$).

The standard argument for the inadequacy of completely positive maps for describing the evolution of the quantum state of the system is concluded by noting that in many scenarios, the map which is defined by Eq.~\eqref{mapfromrelation} fails to be completely positive. We provide a simple example of this type in the next section.
%, motivated by examples from Ref.~\cite{Katjathesis} and Ref.~\cite{beyond} in the next section. 

In fact, one can find examples wherein the prescription that is endorsed by the standard argument defines a map that is not linear, and other examples where it does not define any map at all. We provide such examples in the two subsequent sections. For each type of failure---the failure of complete positivity, the failure of linearity, and the failure to define a map---we explain why it casts doubt on the standard argument. 

Before moving to these examples, we pause to note a problematic feature that lies at the very base of the standard proposal.  

The question that we believe ought to be answered is the following one:
 %is the important one to answer, namely, 
 \begin{quote}
$Q$: What is the evolution map from $S_1$ to $S_2$ in the experimental scenario of Fig.~\ref{onestate}? 
\end{quote}
% (and which we will ultimately also remedy). 
However, the standard proposal immediately substitutes this question for a new one:
\begin{quote}
$Q'$: What is the evolution map from $S_1$ to $S_2$ in the experimental scenario of Fig.~\ref{fig:standard}? 
\end{quote}
As noted earlier, the motivation for the substitution is that there seems to be {\em no way} to answer question $Q$ if one is committed to the standard proposal, that is, if one is committed to defining the evolution map using an input-output relation for the input states that are realized in the scenario at hand.  We will ultimately argue that this commitment is mistaken and that there is consequently no need to retreat from $Q$ to $Q'$.   For the moment, however, we wish simply to note a problem with any such retreat.  Namely, it implies a violation of 
% we note will critize this retreat on the grounds that it violates
  a {\em criterion of universality} that we believe ought to be upheld in such investigations: 
  \begin{quote}
 Any proposal for the evolution map should be applicable to any scenario. 
  \end{quote}
 In particular, it should be applicable to the natural scenario depicted in Fig.~\ref{onestate}.  
  %we believe that it is meaningful to ask about the evolution 

%point out that to retreat in this fashion is to compromise on a principle that 
%[Connect to the business about reasoning about counterfactuals]

\color{black}

\subsection{An example where the standard proposal implies a map that is not completely positive} \label{transpose}

This example is motivated by related examples in Refs.~\cite{beyond} and ~\cite{Katjathesis}.

Imagine that one achieves a variation over the initial state of the system-environment composite as follows: one first prepares it in the maximally entangled state with the environment, $\ket{\phi^+}_{S E}:= \frac{1}{\sqrt{2}} (\ket{0}_{S}\ket{0}_E +\ket{1}_{S}\ket{1}_E)$, and then for each value of a setting variable $J$, one implements upon it the binary-outcome measurement associated to the orthogonal basis $\{ \ket{\psi_{j,1}}, \ket{\psi_{j,2}}\}$ with von Neumann-L\"{u}ders state update rule and one post-selects on obtaining the first outcome, $K=1$. Here, $\ket{\psi_{j,1}}$ and $\ket{\psi_{j,2}}$ form an orthogonal basis for the qubit Hilbert space. 
% $\mathcal{H}_{S}$.
%$\langle \psi_{j,1}|\psi_{j,2} \rangle =0$, so 

\begin{figure}[htb!]
\centering
\includegraphics[width=0.28\textwidth]{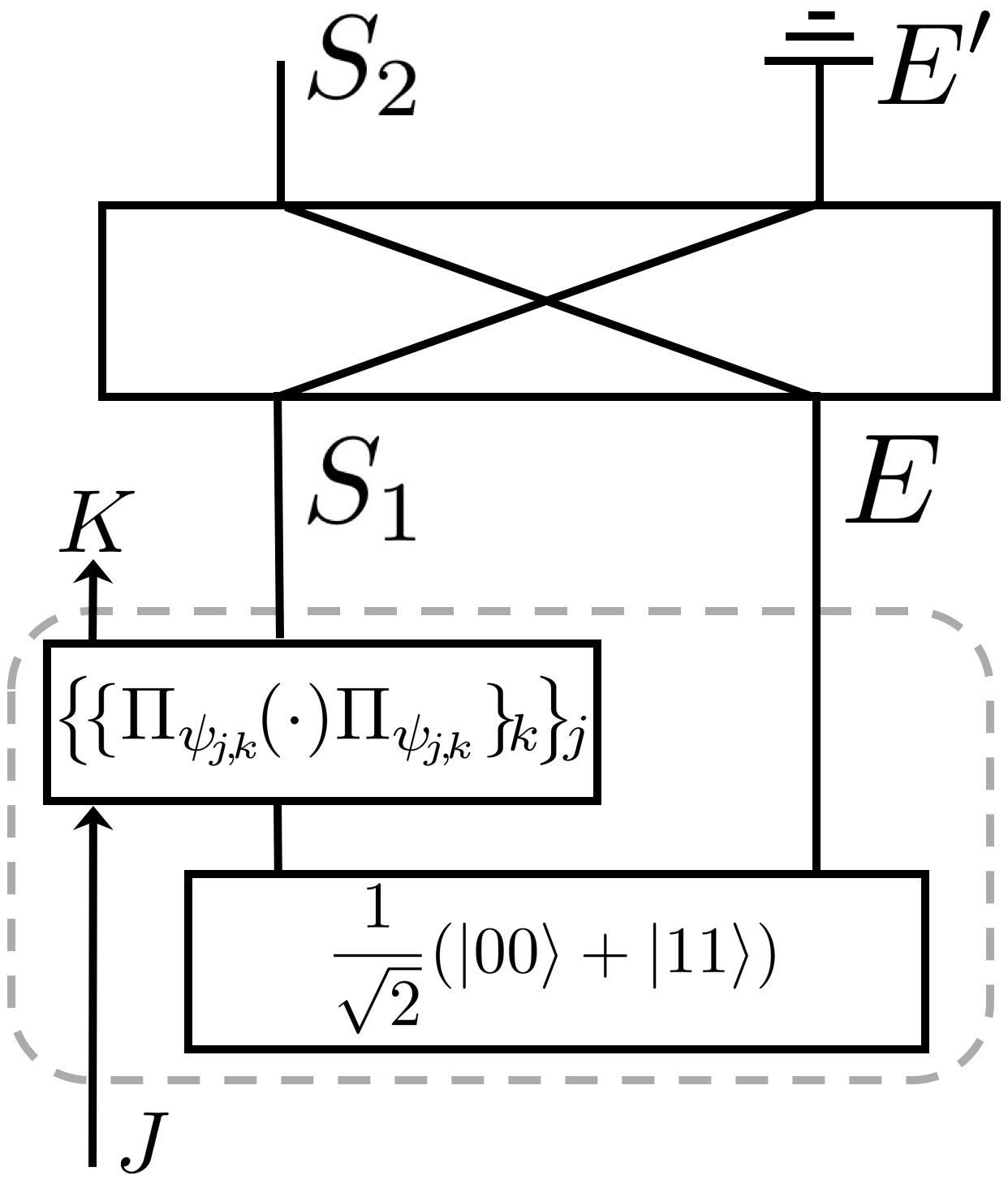}
\caption{An example where the input-output relation corresponds to a map that is linear but not completely positive.
%'s relation corresponds to a non-completely positive map. 
%The gate which prepares the initial state of the composite, conditioned on $J$ and $K$, is shown in a dashed box. } 
%The state preparation protocol---denoted by a dashed box---involves post-selecting on the first outcome $K=1$ of a measurement in the basis $J=j$ defined by $\{ \ket{\psi_{j,1}}, \ket{\psi_{j,2}}\}$ (on a maximally entangled state of two qubits). The system-environment interaction is a swap gate.
}
\label{Transpose}
\end{figure}

This scenario is depicted in Fig.~\ref{Transpose}. The part of the circuit that is conceptualized as the preparation of the joint state of system and environment (that is, the part which corresponds to the first gate in Fig.~\ref{fig:standard}) is highlighted by a dashed box.  This convention is followed in all of the examples we consider. 

Because the environment is initially correlated with the system, 
%If the system state prior to the projective measurement has correlations with the environment state, then 
the state that one infers for it from the post-selection depends on the basis choice and the outcome, a phenomenon that is often termed {\em quantum steering}~\cite{EPR,schrod}. Specifically, when one learns from the measurement that the quantum state of the system is $\ket{\psi_{j,1}}$, one infers that the state of the environment is $\ket{\psi_{j,1}^T}$, where $T$ denotes transposition in the computational basis 
(since $\langle \psi | \big( \ket{\phi^+}\bra{\phi^+}\big)| \psi \rangle = \ket{\psi^T}\bra{\psi^T}$ for any $\psi$).
%(since $_S\langle \psi\ket{\phi^+}_{SE}\bra{\phi^+} \psi \rangle_S = \ket{\psi^T}\bra{\psi^T}_E$ for any $\psi$).
 
Thus by varying the parameter $J$ and conditioning on obtaining the first outcome of the measurement, $K=1$, one obtains the set $\{\rho_{S_1 E}^{(j,1)} \}_j$ where
\beq
 \rho_{S_1 E}^{(j,1)} = \ket{\psi_{j,1}} \bra{\psi_{j,1}}_{S_1} \otimes \ket{\psi_{j,1}^T}\bra{\psi_{j,1}^T}_{E}.
\eeq

The subsequent system-environment interaction is presumed to implement the swap operation on their states. That is, $\mathcal{U}_{S_2 E'|S_1 E}$ is defined, via Eq.~\eqref{supop}, by the operator
\beq
U = {\rm SWAP},
\eeq
where 
${\rm SWAP} (|\phi\rangle \otimes |\chi\rangle) := |\chi\rangle \otimes |\phi\rangle$ for all $\chi$ and $\phi$.
%\beq
%U_{S_1 E} = {\rm SWAP}_{S_1 E},
%\eeq
%where 
%${\rm SWAP}_{S_1 E} (|\phi\rangle_{S_1} \otimes |\chi\rangle_{E}) := |\chi\rangle_{S_1} \otimes |\phi\rangle_{E}$ for all $\chi$ and $\phi$.

By Eq.~\eqref{rhoS2}, we infer that the marginal states of the system (for each value of $J$) after the system-environment interaction are given by 
\beq
\rho^{(j,1)}_{S_2} = \ket{\psi_{j,1}^T}\bra{\psi_{j,1}^T}_{S_2}.
\eeq
%It follows that the final state of the system is $ \ket{\psi_j^T}\bra{\psi_j^T}_{S_2}$.

The input-output relation in this case, therefore, is 
\beq
\mathfrak{R} = \big\{ \big(\ket{\psi_{j,1}}\bra{\psi_{j,1}}_{S_1}, \ket{\psi_{j,1}^T}\bra{\psi_{j,1}^T}_{S_2}\big) \big\}_j.
\eeq
 
 If the projectors $\{ \ket{\psi_{j,1}}\bra{\psi_{j,1}}_{S_1} \}_j$ form a basis for the operator space $\mathcal{L}(\mathcal{H}_{S_1})$ (i.e., an informationally complete set), one can conclude that there is a unique linear map defined (via Eq.~\eqref{mapfromrelation}) by this relation, namely, the transpose map. This is the canonical example of a map that is positive but not completely positive~\cite{NC}. 

\subsubsection{Problematic implications of the failure of complete positivity} \label{problemimp}

Even though the failure of completely positivity is obviously acknowledged by proponents of the standard argument (it is the reason this field of research even exists), we believe that it already provides good reasons for being suspicious of the argument.

%The example we have just presented leads to a map that `merely' fails to be completely positive. The examples we will present next involve more radical pathologies that will make it clear that something is wrong with the standard argument. Already, however, the fact that one obtains a map that is not completely positive is reason to be suspicious of the claim that the map associated to the system's evolution is obtained in the manner described above. 

An immediate worry is that such maps could lead to output states on system-ancilla composites that fail to be positive. 
%(Recall that such output states must be avoided because they do not predict positive probabilities for the outcomes of all measurements.) 

The standard response to this worry (found, for instance, in Ref.~\cite{afraid,sud}) is that 
the map describing the evolution of the system in this circumstance is only applicable on a limited domain of input states, and that this domain of input states does not include the marginals of the set of entangled states which manifest the failure of the map to be completely positive. 

Note, first of all, that this response is a denial of the axiomatic justification of complete positivity which we discussed in the introduction.
The problem we see with this denial is that it forces one to give up on the notion that the map describing the evolution can support inferences about counterfactuals. This notion is central to the notion of evolution in physics:
%Such a conclusion is mathematically consistent, but gives up a basic desideratum for defining evolution in physics: an essential feature of maps that describe the evolution of a state is that they {\em support inferences about counterfactuals}.
laws of motion are not just {\em descriptions} of historically actual motions, but {\em prescriptions} for determining what motion {\em would} occur for {\em any} initial condition. 
The possibility of making inferences about counterfactual scenarios is precisely what makes laws of motion so useful in practice. 
%\begin{comment}
%\footnote{For instance, Goodman summarizes the point as follows {[reference N. Goodman's `The problem of counterfactual conditionals' ]}:
%\begin{quote}
%As a first approximation then, we might say that a law is a true sentence used for making predictions. That laws are used predictively is of course a simple truism, and I am not proposing it as a novelty. I want only to emphasize the idea that rather than a sentence being used for prediction because it is a law, it is called a law because it is used for prediction; and that rather than the law being used for prediction because it describes a causal connection, the meaning of the causal connection is to be interpreted in terms of predictively used laws.
%\end{quote}
%}
%\end{comment}
%If one endorses the idea that 
To entertain the idea that a map {\em only} describes the evolution of a system {\em when that system is assured to be in one of a restricted set of states} is to retreat from the usual conception of an evolution map.
%a law of motion.

%\{Perhaps the restricted domain argument only works for circuits in which the preparations are generated by transformations on the system alone? Looking back at `Dynamics of initially entangled open quantum systems': what they do is confusing, but seems to not be motivated by any sort of circuit at all. Rather, I think they imagine fixing correlation terms in the state (when decomposed into tensor products of Pauli matrices) and then varying the other parameters in the state. }

%But there are further reasons to doubt the validity of the standard argument. As we now show, the standard argument is generically inconsistent with the linearity of quantum transformations.

%{\color{red} Ref.~\cite{Cuffaro} also argued that non-completely positive maps do not represent evolution, but only the mathematical extension of a map with a limited domain. [Not sure we should/need to include this, since there remain many problems with this point of view.]}

\subsection{An example where the standard proposal implies a map that is not linear} \label{NC}

The following example is a simplified version of the one presented in Nielsen and Chuang~\cite{NC}. 

Here, one begins with the product state $|0\rangle_{S_1} |0\rangle_E$ and, conditioned on the classical control $J \in \{0,1\}$, one applies
%Here, one initiates the system and environment in the state $|0\rangle_{S_1} |0\rangle_E$. $K$ is trivial, and  $J \in \{0,1\}$ acts as the control for 
%for each value of $J \in \{0,1\}$ one prepares a distinct initial joint state by performing 
a controlled-Hadamard gate on the system and a controlled-NOT gate on the environment. This procedure does not involve any measurement, and hence $K$ is trivial. The system-environment composite is therefore prepared in one of the states $\{ \rho_{S_1 E}^{(j)}\}_j$ where 
\begin{align}
\rho_{S_1 E}^{(0)} &= \ket{0}\bra{0}_{S_1} \otimes \ket{0}\bra{0}_E\\
\rho_{S_1 E}^{(1)} &= \ket{+}\bra{+}_{S_1} \otimes \ket{1}\bra{1}_E,
\end{align}
with $|+\rangle:=\frac{1}{\sqrt{2}}(\ket{0}+\ket{1})$.

 Next, there is a controlled-$XH$ gate with the environment qubit as control and the system qubit as target, 
 %with unitary description
 \beq
 U = \mathbb{1} \otimes \ket{0}\bra{0} + XH \otimes \ket{1}\bra{1},
\eeq
where $XH$ denotes the unitary gate obtained by performing the Hadamard gate $H$ followed by the Pauli gate $X$.

\begin{figure}[htb!]
\centering
\includegraphics[width=0.23\textwidth]{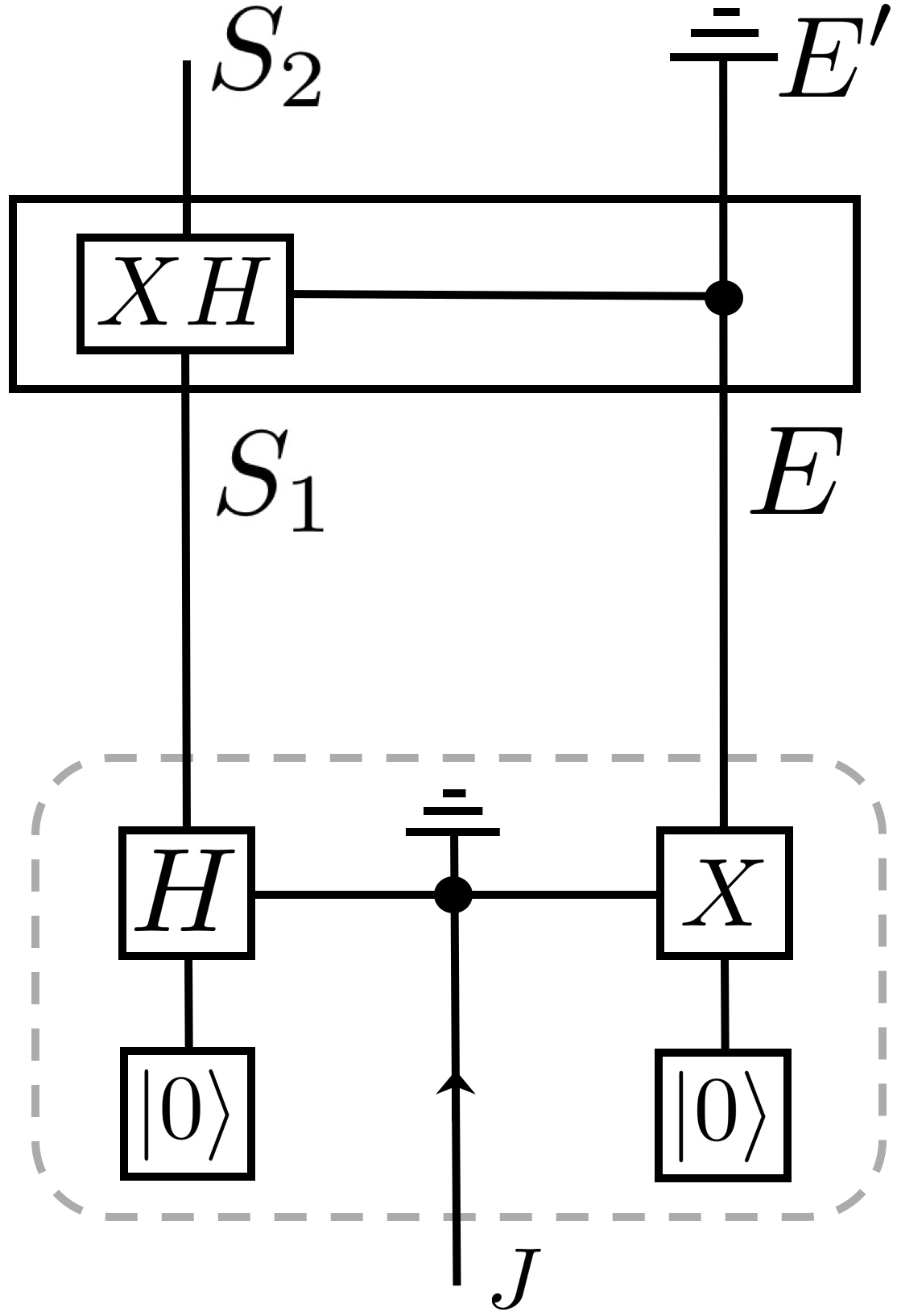}
\caption{An example where the input-output 
%standard argument's 
relation corresponds to a map that is nonlinear. 
%The gate which prepares the initial state of the composite, conditioned on $J$ and $K$, is shown in a dashed box. 
}%The state preparation protocol---denoted by a dashed box---involves a controlled Hadamard on the system and a controlled-NOT on the environment, both controlled on the classical ancilla $J$. The system-environment interaction is a controlled $XH$ gate with the system as the target.}
\label{NonlinearExample}
\end{figure}

This circuit is shown in Fig.~\ref{NonlinearExample}.

By Eq.~\eqref{rhoS2}, we infer that the marginal states of the system after the system-environment interaction (for each value of $J$)  are given by 
\begin{align}
 &\rho^{(0)}_{S_2} = \ket{0}\bra{0}_{S_2}\\
 &\rho^{(1)}_{S_2} = \ket{1}\bra{1}_{S_2}.
\end{align}

The input-output relation in this case, therefore, is 
\begin{align}
\mathfrak{R} = \big\{ &(\ket{0}\bra{0}_{S_1}, \ket{0}\bra{0}_{S_2}),\\
&(\ket{+}\bra{+}_{S_1}, \ket{1}\bra{1}_{S_2}) \big\}.
\end{align}

Note that the map implied by the standard proposal is not completely specified by these constraints.
% because $\ket{0}\bra{0}_{S_1}$ and $\ket{+}\bra{+}_{S_1}$ do not span $\mathcal{L}(\mathcal{H}_{S_1})$. 
However, any map consistent with this relation must take nonorthogonal states to orthogonal states, and every such map is nonlinear. (Note that once one allows for the failure of linearity, even having a set of input states that span $\mathcal{L}(\mathcal{H}_{S_1})$ becomes insufficient to determine how the map acts on {\em all} states.)
%implies $|0\rangle \to |0\rangle$, and $|+\rangle \to |1\rangle$, so it 

\subsubsection{Problematic implications of the failure of linearity}

The fact that the standard proposal does not always define a linear map is troubling, because the linearity of transformations can be justified on numerous physical grounds. 
 For example, representing a process by a nonlinear map violates the principle that 
processing of a system cannot increase the amount of information it contains about another system. Specifically, every such map violates the data processing inequality~\cite{NC}. Such violations have physically problematic implications, such as the possibility of superluminal signalling~\cite{nonlinear1,nonlinear2,nonlinear3}.

\subsection{An example where the standard proposal implies a relation that is not a map}\label{CNOT}

\begin{figure}[htb!]
\centering
\includegraphics[width=0.26\textwidth]{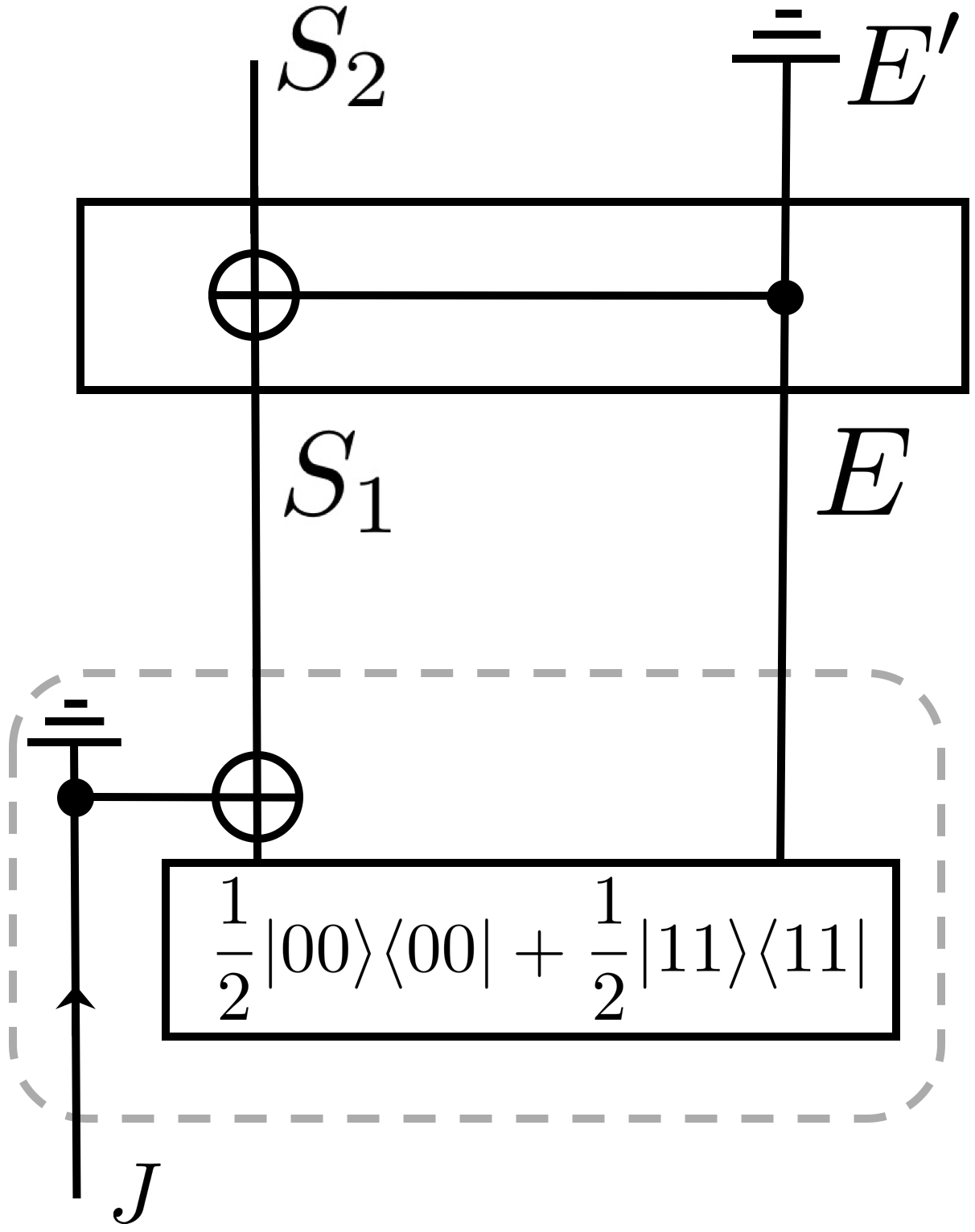}
\caption{An example where the input-output relation does not correspond to any map. 
%The circuit which prepares the initial state of the composite, conditioned on $J$ and $K$, is shown in a dashed box.
 }%The state preparation protocol---denoted by a dashed box---involves a controlled-NOT gate with the classical ancilla $I$. The system-environment interaction is a CNOT gate with the system as the target.}
\label{OneToMany}
\end{figure}

Most pathologically, the standard proposal can yield a relation which is inconsistent with {\em any} map whatsoever. This is illustrated by a simple example, motivated by one from Ref.~\cite{stelm} 
%\color{cyan}[Should we also mention the example surrounding Fig. 4 in Modi's 1708.00769? David: I don't think so.] \color{black} 
and pictured in Fig.~\ref{OneToMany}. 

%{ Here the preparation begins with the correlated state $\frac{1}{2} \ket{00}\bra{00}+\frac{1}{2}\ket{11}\bra{11}$, followed by a controlled-NOT gate on the system with $J \in \{0,1\}$ as the control. As before, $K$ is trivial.}
Here, $K$ is again trivial, and for each value of $J \in \{0,1\}$, one prepares a distinct initial joint state by performing a controlled-NOT gate on the system with $J$ as the control. 
In this way, the system-environment composite is prepared in one of the states $\{ \rho_{S_1 E}^{(j)}\}_j$, namely
\begin{align} 
 \!\rho_{S_1 E}^{(0)}=&\frac{1}{2} \ket{0}\!\bra{0}_{S_1}\! \otimes \ket{0}\!\bra{0}_E + \frac{1}{2} \ket{1}\!\bra{1}_{S_1} \!\otimes \ket{1}\!\bra{1}_E
\\
\!\rho_{S_1 E}^{(0)}=&\frac{1}{2} \ket{1}\!\bra{1}_{S_1}\! \otimes \ket{0}\!\bra{0}_E + \frac{1}{2} \ket{0}\!\bra{0}_{S_1} \!\otimes \ket{1}\!\bra{1}_E
\end{align}

Next, there is a controlled-NOT gate with the environment qubit as control and the system qubit as target, with unitary description
 \beq
 U = \mathbb{1} \otimes \ket{0}\bra{0} + X \otimes \ket{1}\bra{1}.
\eeq

By Eq.~\eqref{rhoS2}, we infer that the marginal states of the system (for each value of $J$) after the system-environment interaction are given by 
\begin{align}
 &\rho^{(0)}_{S_2} = \ket{0}\bra{0}_{S_2}\\
 &\rho^{(1)}_{S_2} = \ket{1}\bra{1}_{S_2}.
\end{align}

The input-output relation defined by this scenario is 
\begin{align}
\mathfrak{R} = \big\{ &\big(\frac{1}{2} \mathbb{1}_{S_1},\ket{0}\bra{0}_{S_2}\big),\nonumber\\
&\big(\frac{1}{2} \mathbb{1}_{S_1},\ket{1}\bra{1}_{S_2}\big) \big\}.
\end{align}
 Because this relation is one-to-many, it does not define a map from $\mathcal{L}(\mathcal{H}_{S_1})$ to $\mathcal{L}(\mathcal{H}_{S_2})$.

%[This example is very close to the example in Stelmachovic and Buzek. Some of the other examples might have been explicitly considered elsewhere, e.g. Katja's thesis, Joe's course notes, and elsewhere in the literature.] 
%[Giving names to the examples: we could name them according to various organizational principles, as follows.\\
%-system-post-selected, system-controlled, system/environment-controlled\\
%-Nielson-Chuang, Reid-Spekkens, Stelmachovick-Buzek\\
%-CNOT, SWAP, controlled-Y]

\subsubsection{Problematic implications of the failure of the standard proposal to define a map}

The standard proposal purports to define the evolution map. However, we have just seen that it sometimes fails to define {\em any} map whatsoever. This failure strikes us as a decisive criticism. Proponents of the standard proposal have not offered any satisfactory account of how their scheme can be salvaged in the face of this failure.

\subsection{Every pathology can be obtained from every circuit type} \label{genericrep}

We have provided three examples which illustrate distinct pathologies of the standard proposal. Much previous work~\cite{Pechukas,Alicki,Preplies,afraid,sud,assign1,standard1,beyond,Mrole,lidar,brodutch,Kraus,linear} has focused on these distinctions, e.g., by seeking necessary or sufficient conditions for the relation to define a completely positive map, a linear map, and so on. The coming analysis shows that these questions are misguided: evolution maps are {\em always} completely positive, and all of the apparent counterexamples are really just indicative of the fact that the standard proposal is {\em not} the correct way to define the evolution map. 
%Maybe also these: which attempt to classify what types of maps best fit the input-output relation.

Our examples differed also in the way in which they 
%laboratory operations by which one 
introduced variability in the initial state of the system-environment composite. Specifically, the variability was introduced (respectively) by (a) the choice of transformation on the system-environment composite,
(b) the choice of measurement on the system and the choice of post-selection on its outcomes, and (c) the choice of transformation on the system alone.
Roughly, all examples in the literature fit into one of these three special cases of the circuit of Fig.~\ref{fig:standard}. Some articles, however, left the operational (circuit) description of the problem unspecified~\cite{assign1,sud,afraid,standard1,stelm,beyond,lidar,Kraus}, and took the problem description to be a (possibly continuous) set of initial states on $S_1 E$ 
%$\{ \rho_{S_1 E}^{(j,k)} \}_{j,k}$
 and a unitary system-environment interaction, $\mathcal{U}_{S_2 E'| S_1 E}$. One consequence of our work is to show that this version of the problem is not well-posed; 
%this is not a physically well-posed question. 
knowing the causal structure of the circuit is in fact critical to defining the evolution map.\footnote{There is, however, some prior acknowledgement of the importance of the form of the circuit in Refs.~\cite{Mprep,Mrole}.} 

Apart from their simplicity, there is nothing special about the examples we have chosen. In Appendix~\ref{qex}, we show that each of the three circuit types can generate each of the pathologies of the standard proposal. In particular, this means that one cannot evade such pathologies by restricting attention to one of the three special classes of circuits.

As we argued at the end of Sec.~\ref{standardargument4inadequacyCP}, any sensible definition of the evolution map should satisfy a criterion of {\em universality},
%[Harmonize this bit with the new text!] 
%Furthermore, any sensible definition of the evolution map should satisfy a criterion of {\em universality},
 namely, that it should be applicable regardless of the scenario.  
%Many previous proposals have failed to satisfy this criterion.
% Most notably, we saw that the prescription of the standard argument did not define an evolution map in the simple scenario of Fig.~\ref{onestate}. 
As we already noted in that section, the standard proposal fails to satisfy this criterion because it does 
%did 
not define an evolution map in the simple scenario of Fig.~\ref{onestate}. 
%In other cases, 
We are also not satisfied with the standard responses to the problematic implications of the standard proposal just outlined, because they too
%that we have just outlined often 
compromise on the criterion of universality.  Indeed, they do so in a particularly unsatisfying way.  
Namely, they assert that their proposed definition of the evolution map is only
%Specifically, insofar as they proposed a definition of the evolution map that was only
%the proposed definition of the evolution map was 
applicable to a restricted class of circuits, 
%where
but the restriction is {\em ad hoc} insofar as no justification is given other than to avoid the pathologies that would otherwise result.
%For example, Ref~\cite{tradscheme} acknowledged the possibility that the relation of the standard argument might be one-to-many, and so demanded that one should restrict the class of scenarios in which the standard argument can be applied, in order to exclude these cases. 
For example, Refs.~\cite{tradscheme,Dominy2015} recognized that certain scenarios yielded the pathology discussed in Section~\ref{CNOT},  and sought to avoid it by excluding such scenarios {\em by fiat}. 
\color{black}

\color{black}

%\section{Persistence of the pathologies of the standard argument in the classical sphere}
\section{Persistence of the problems of the standard proposal in the classical sphere}
\label{sec:classical}

In a classical setting, one can also consider the evolution map for a system when there are initial correlations between the system and environment. In this section, we demonstrate that if one tries to define this classical evolution map using the prescription endorsed by the standard proposal, then one obtains all of the same problematic implications that one saw quantumly. (This should already be evident given that for many of the quantum examples we presented, both above and in the appendix, all the states and maps could be dephased in the computation basis without affecting our conclusions.) Therefore, although it is a widely held belief that the surprising form of quantum evolution maps in cases of initial system-environment correlations is just another example of a counterintuitive feature of quantum theory (so-called `quantum weirdness'), we will demonstrate that it should be taken instead as evidence of the fallacy of the standard proposal for how to define the evolution map. 

\subsection{Classical Preliminaries}

A classical system is described by a set of physical states, which can be encoded as values of a random variable $S$. 
%One's description of the 
The statistical state of the classical system is given by a probability distribution $P_S$ over the possible physical states of the system.
More precisely, if $\Lambda_S$ denotes the set of possible values of $S$ (here assumed discrete), then $P_S: \Lambda_S \to [0,1]$ where
%denotes the function over $\Lambda_S$, and 
%$P_S$ is a function from the set of physical states (here assumed discrete) to the interval $[0,1]$, so that 
$P_S(s)$ denotes the probability that $S=s$. $P_S$ satisfies the normalization condition $\sum_{S} P_{S}=1$, where $\sum_{S} P_{S}$ denotes the function on $S$ defined by $\forall s\in \Lambda_S: (\sum_S P_S)(s):=\sum_{s} P_{S}(s)$ and where the right-hand side of the condition denotes the function that takes value $1$ for all $s$.
%A classical system can be described by a classical random variable $S$. If $\mathcal{S}$ denotes the range of values of $S$, then $P_S$ is a function from $\mathcal{S}$ to the interval $[0,1]$. The classical analogue of a quantum state on system is a probability distribution $P_S$ over $S$. 

We are interested in the evolution of these probability distributions for a system interacting with an environment. We will denote the space of probability distributions on $S$ by $\mathfrak{P}(S)$, and for later convenience we introduce the following notation:
\begin{align}
&P_A = [a] \ \text{ means } \ P_A(a) = 1,\\ 
&P_{AB} = [a][b] \equiv [ab] \ \text{ means } \ P_{AB}(a,b)=1.
\end{align}

%The classical analogue of a completely-positive trace-preserving linear map is a stochastic map. 
%If the space of physical states on system is denoted $S$ and 
%If the space of probability distributions over $S$ is denoted $\mathcal{P}(\Lambda_S)$, then 
A map $\Gamma_{S_2|S_1}: \mathfrak{P}(S_1) \to \mathfrak{P}(S_2)$ %:: P_{S_1} \mapsto P_{S_2}$,
is called {\em stochastic} if there exists a conditional probability distribution $P_{S_2|S_1}$ such that 
\begin{align} \label{stochdefn}
\Gamma_{S_2|S_1} (P_{S_1}) = \sum_{S_1} P_{S_2|S_1} P_{S_1}.
\end{align}
%where we have introduced the shorthand notation $\sum_{S_1} P_{S_1} := \sum_{\lambda \in \Lambda_{S_1}} P_{S_1}(\lambda)$. %(Arguably, a better classical analogue of completely positive maps are stochastic maps that arise from reduction of a deterministic Hamiltonian evolution.)
Note that this is a linear map.

The most general manner in which a probability distribution on a classical system can evolve is by a stochastic map. 
%As with one of the justifications of the use of completely positive maps in quantum theory, one can justify the use of stochastic maps in classical theories as follows: 
This can be justified, analogously to complete positivity in the quantum case, as follows:
if one imagines that an ancillary system is prepared in some fixed distribution, and the system-ancilla composite is subjected to a deterministic dynamics, and then the ancilla is ignored, then the resulting map on the probability distribution over the system is always stochastic. Stochastic maps are also the most general type of map that preserve positivity for all input distributions.

%Just as there are two ways to justify the use of completely-positive linear maps to describe evolution in quantum theory, one can justify the use of stochastic maps in classical theories in two ways. First, if one imagines that an ancillary system is prepared in some fixed distribution, and the system-ancilla composite is subjected to a deterministic dynamics, followed by marginalizing over the ancilla, then the resulting map on the probability distribution over the system is always stochastic. Second, a stochastic map is the most general map that always preserves normalization and positivity when acting on a subsystem of a composite system prepared in an arbitrary joint distribution. 

%\st{Note that stochastic maps are linear and therefore as models of a physical process, they respect the principle that any processing of a system cannot increase the amount of information it has about another system. In particular, they respect the classical data processing inequality.}

\subsection{What the standard proposal stipulates in a classical scenario}

The standard proposal applies just as well to classical scenarios as to quantum scenarios, since the former are a strict subset of the latter, where all operators are diagonal in some fixed basis. A classical scenario can be described by the same sort of circuit as a quantum scenario: one simply replaces quantum states by probability distributions and unitary operations by deterministic functions. 
Hence we can consider a direct analogy for each of the quantum circuits considered previously.

\begin{figure}[htb!]
\centering
\includegraphics[width=0.47\textwidth]{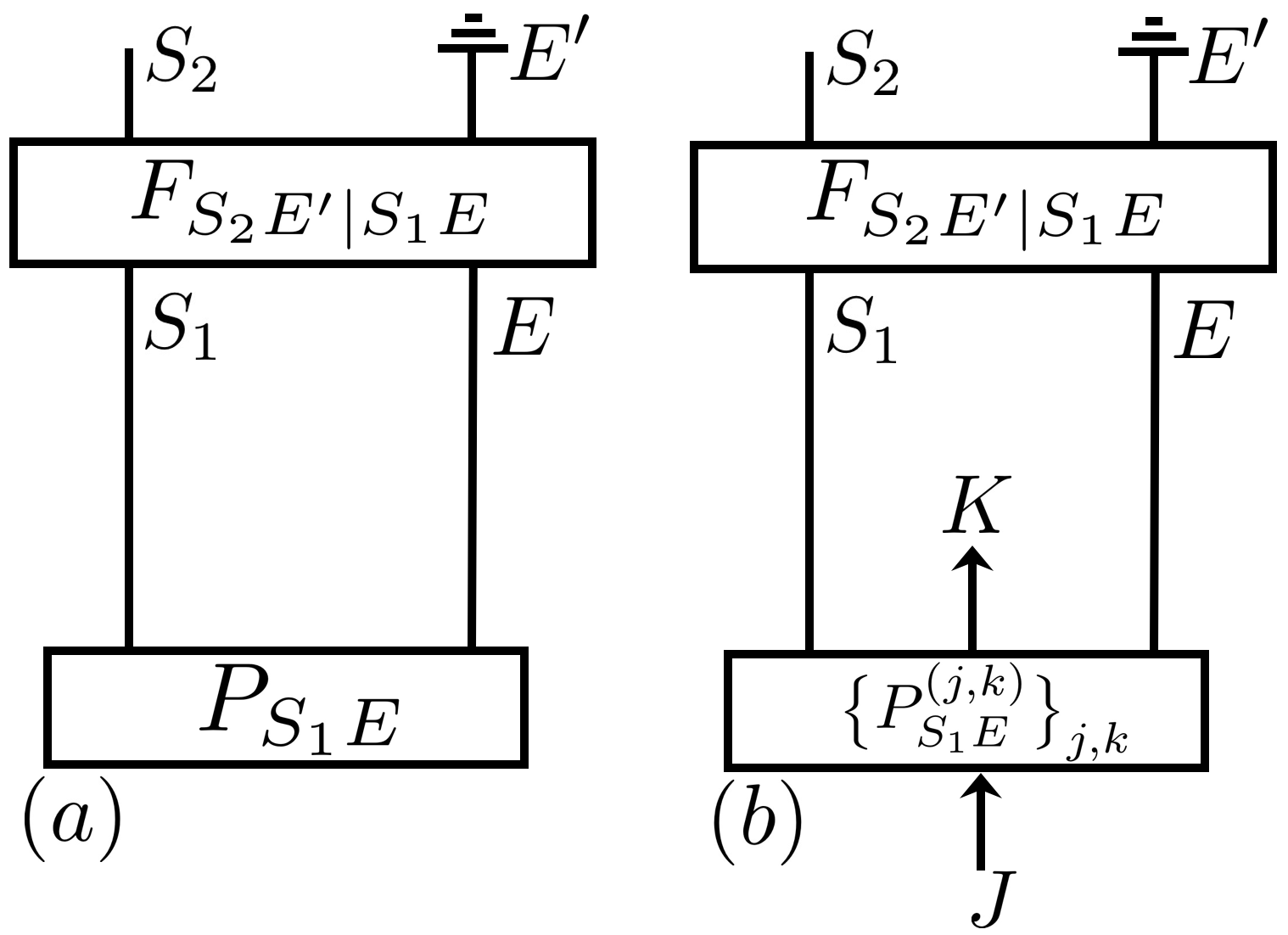}
\caption{(a) The classical analogue of the circuit in Fig.~\ref{onestate}. (b) The classical analogue of the circuit in Fig.~\ref{fig:standard}}.% The state preparation protocol---denoted by a dashed box---has input $J$ and outcome $K$. The labels of wires now refer to classical systems and the circuit elements are classical gates.}
\label{cgeneral}
\end{figure}

The classical analogue for the general circuit from Fig.~\ref{onestate} is shown in Fig.~\ref{cgeneral}(a). If one followed the standard proposal, one would demand that a constraint on the evolution map,
which we denote by $\Gamma_{S_2|S_1}$, is:
% constrained by the following equation:
\beq
P_{S_2} = \Gamma_{S_2|S_1}(P_{S_1}),
\eeq
with $P_{S_1} := \sum_E P_{S_1 E}$ and $P_{S_2} := \sum_{E'} F_{S_2 E'|S_1 E}(P_{S_1 E})$.
One would again be compelled to introduce variability in the input state on $S_1$ in a manner analogous to what was done in the quantum sphere in Fig.~\ref{fig:standard}.

The resulting scenario, a classical analogue for the general circuit from Fig.~\ref{fig:standard}, is shown in Fig.~\ref{cgeneral}(b), where $F_{S_2 E'|S_1 E}$ represents a stochastic map $\mathfrak{P}(S_1) \otimes \mathfrak{P}(E) \to \mathfrak{P}(S_2)\otimes \mathfrak{P}({E'})$ induced by a deterministic dynamics, so that 
\begin{equation} \label{defnstoch}
F_{S_2 E'|S_1 E}(\framebox(5,5){}_{S_1E})=\sum_{S_1 E} \delta_{S_2,f(S_1,E)}\delta_{E',g(S_1,E)}  \framebox(5,5){}_{S_1E}
\end{equation}
for some functions $f$ and $g$.

%As in the quantum setting, we imagine starting with a complete description of the circuit, which for the classical case can be viewed as a set of variables and their probabilistic dependences. (This notion is formalized as a causal model in Section~\ref{classicaldo}.) Given this information, the standard argument of Section~\ref{sec:standard} instructs an agent to compute the probability distribution over $S_2$ given an input probability distribution over $S_1$, for each value of $JK$. 

Applying the standard proposal to these classical scenarios, one simply computes 
\beq
P^{(j,k)}_{S_1} = \sum_{E} P^{(j,k)}_{S_1 E}
\eeq 
and 
\beq
P^{(j,k)}_{S_2} = \sum_{E'} F_{S_2 E'|S_1 E} (P^{(j,k)}_{S_1 E}),
\eeq 
generating an input-output relation 
\beq
\mathfrak{R} = \{(P^{(j,k)}_{S_1},P^{(j,k)}_{S_2})\}_{j,k},
\eeq
as in the quantum case. 
%
%An analogy to the proposed prescription for inferring the evolution map in the quantum case would suggest that the map describing the evolution should be taken to be one which satisfies the constraints encoded in this relation: that is, a map $_{S_2|S_1}: \mathcal{P}(\Lambda_{S_1}) \to \mathcal{P}(\Lambda_{S_2})$ for which
The standard proposal dictates that the map describing the evolution should be taken to be one which satisfies the constraints encoded in this relation; that is, a map $\Gamma_{S_2|S_1}: \mathfrak{P}(S_1) \to \mathfrak{P}(S_2)$ for which
\beq
 \forall j,k: \quad \Gamma_{S_1|S_1}[P^{(j,k)}_{S_1}] = P^{(j,k)}_{S_2}.
\eeq

%\subsection{Persistence of the problematic implications for classical examples}

%\subsection{The same problematic implications of the standard argument arise in the classical sphere} \label{classex}

As we now show, this prescription for how to define the evolution map leads to the same sorts of pathologies we saw in the quantum case. 
 %has problematic implications even in classical cases. 
 There exist simple physical scenarios that generate relations which imply maps that are not stochastic, others which imply maps that are not linear, and still others which do not define any map at all. 

Our examples of each of the three failures are chosen to be analogous to the corresponding quantum examples. Note that $S_1$, $S_2$ and $E$ are now taken to denote classical random variables rather than being mere labels of systems.

\subsection{A classical example where the standard proposal implies a map that is not stochastic} \label{c:transpose}

Our example of the failure of stochasticity in a classical system is analogous to our example of the failure of complete positivity in a quantum system (described in Section~\ref{transpose}) and is realized with a circuit of exactly the same form (shown in Fig.~\ref{Transpose}).

Consider a system and an environment that each have four possible physical states, so that $S_1,E \in \{0,1,2,3\}$. The pair are first prepared in the joint distribution $P_{S_1 E} := \frac{1}{4}([00]+[11]+[22]+[33])_{SE}$. System $S_1$ is then subjected to one of three measurements, determined by the value of a ternary variable $J$, and each measurement having a binary outcome $K$. The $J=1$ measurement determines whether $S_1\in \{2,3\}$ or not, $J=2$ whether $S_1\in \{1,3\}$ or not, and $J=3$ whether $S_1\in \{1,2\}$ or not. $K=1$ labels the outcome wherein $S_1$ is found to be in the given set, and $K=0$ the complementary set. 
We further imagine that these measurements are not passive, but disturb the value of the system variable. Specifically, the update rule is such that the final state on $S_1$ depends only on $J$ and $K$: $S_1$ is prepared in a uniform distribution over the values of $S_1$ in the complementary set to the one that $S_1$ was found in. For instance, if the $J=1$ measurement is done and the outcome $K=1$ occurs, verifying that $S_1 \in \{2,3\}$, then $S_1$ is reprepared in the distribution $\frac{1}{2} [0]_{S_1} + \frac{1}{2} [1]_{S_1}$.

Finally, $S_1$ undergoes a swap operation with $E$. 

In our example, we consider only the cases where the outcome is found to be $K=1$. For each possible value $j$ of $J$, the joint distribution on $S_1$ and $E$, $P_{S_1 E}^{(j,1)}$, is as follows:
% Because of the perfect correlations in their initial joint probability distribution, both $S_1$ and $E$ are left in the same marginal probability distribution upon post-selection. Further, by the measurement disturbance on the system's physical state, $S_1$ and $E$ are left completely uncorrelated. In this way, the system-environment composite is prepared in one of the states $\{ P_{S_1 E}^{(j,1)}\}_j$, namely
\begin{align} \label{prep123}
 P_{S_1 E}^{(1,1)}=&\frac{1}{2}([0]+[1])_{S_1} \frac{1}{2}([2]+[3])_E \\
 P_{S_1 E}^{(2,1)}=&\frac{1}{2}([0]+[2])_{S_1} \frac{1}{2}([1]+[3])_E \\ 
 P_{S_1 E}^{(3,1)}=&\frac{1}{2}([0]+[3])_{S_1} \frac{1}{2}([1]+[2])_E.
\end{align}

It follows that after the swap operation on the system and environment, the marginal state of $S_2$, for each value $j$ of $J$ and for $K=1$, denoted $P_{S_2}^{(j,1)}$, is as follows:
%The marginal states of the system (for each value of $J$) after the system-environment interaction are given by
\begin{align} 
 P_{S_2}^{(1,1)}=&\frac{1}{2}([2]+[3])_{S_2}  \\ 
 P_{S_2}^{(2,1)}=&\frac{1}{2}([1]+[3])_{S_2}  \\ 
 P_{S_2}^{(3,1)}=&\frac{1}{2}([1]+[2])_{S_2}.
\end{align}

The input-output relation, therefore, is 
\begin{align} \label{inoutex}
\mathfrak{R}=\{ &(\frac{1}{2}([0]+[1]),\frac{1}{2}([2]+[3])), \nonumber \\
& (\frac{1}{2}([0]+[2]),\frac{1}{2}([1]+[3])),\nonumber \\
& (\frac{1}{2}([0]+[3]),\frac{1}{2}([1]+[2])) \}.
\end{align}

This relation is consistent with a linear map, 
\beq
\Gamma_{S_2|S_1}[\vec{p}_{S_1}] = 
\left[\begin{array}{rrrr}
0 & 0 & 0 & 0 \\
0 & 0 & 1 & 1 \\
0 & 1 & 0 & 1	\\
0 & 1 & 1 & 0
\end{array}\right] \cdot \vec{p}_{S_1},
\eeq
where the probability distribution over the four physical states of $S_1$ is expressed as a vector, $\vec{p}_{S_1}$.
However, this map is not stochastic (e.g., the columns do not each sum to one), and in fact there is {\em no} stochastic map consistent with the relation $\mathfrak{R}$ above. To see this, note that the first ordered pair in $\mathfrak{R}$ guarantees that physical state 0 has no probability of mapping to 0 or 1, the second ordered pair guarantees that it has no probability of mapping to 0 or 2, and the third ordered pair guarantees that it has no probability of mapping to 0 or 3. But every stochastic map is certain to map physical state 0 to {\em some} other physical state, in contradiction with these three constraints.

\subsection{A classical example where the standard proposal implies a map that is not linear} \label{c:NC}

Next, we provide a classical example in which the standard proposal yields a map $\Gamma_{S_2|S_1}$ that fails to be linear. This is analogous to the failure of linearity in the quantum example of Section~\ref{NC}, and it is realized in a classical circuit of exactly the same form (shown in Fig.~\ref{NonlinearExample}).

The system-environment composite is prepared in one of two possible states, depending on the value of $J$:
\begin{align}\label{tnl0}
 P_{S_1 E}^{(0)} &= [0]_{S_1} \otimes [0]_E \\
 P_{S_1 E}^{(1)} &= \frac{1}{2}([0]_{S_1} +[1]_{S_1} )\otimes [1]_E,
\end{align}

The system-environment composite then undergoes the following joint evolution: if $E=0$, then $S_2=S_1$, while if $E=1$, then $S_2=S_1\oplus_3 1$, where $\oplus_d$ denotes summation modulo $d$. (Note that the principal system has three distinct states in this example.)
% then interact in such a way that the system's physical states are left unchanged if $E=0$, but undergo the permutation $0 \rightarrow 1 \rightarrow 2 \rightarrow 0$ if $E=1$.
%; namely, the interaction described by the conditional probability distribution $P_{S_2|S_1 E}=\delta_{E,0} \delta_{S_2,S_1} + \delta_{E,1} \delta_{S_2,S_1 \oplus 1}$ (with arithmetic mod 3). 

The marginal states of the system (for each value of $J$) after the system-environment interaction are given by 
\begin{align} \label{tnl}
 P^{(0)}_{S_2} &= [0]_{S_2} \\
 P^{(1)}_{S_2} &= \frac{1}{2}([1] +[2])_{S_2}
\end{align}

The input-output relation defined by this transformation is 
\begin{align}
\mathfrak{R}=\big\{ &([0],[0]),\nonumber\\
&\big(\frac{1}{2}([0] +[1]),
\frac{1}{2}([1] +[2])\big)\big\}.
\end{align}
Any map consistent with this relation must take overlapping distributions to non-overlapping distributions, and every such map is nonlinear.
% satisfy $[0] \to [0]$, and $\frac{1}{2}([0] +[1]) \to \frac{1}{2}([1] +[2])$, so it must take overlapping distributions to non-overlapping distributions, and every such map is nonlinear.

\subsection{A classical example where the standard proposal does not define a map} \label{c:CNOT}

Finally, we provide a classical example in which the standard proposal fails to yield any map at all. This is an {\em exact} analogue of the quantum example of Section~\ref{CNOT},
%(which involved states and transformations that were all diagonal in the same basis),
 so it is realized in a classical circuit of exactly the same form 
 %as that of the quantum example 
 (shown in Fig.~\ref{OneToMany}).

As in the quantum example, $K$ is trivial, and for each value of $J \in \{0,1\}$, one prepares a distinct initial joint state by performing a controlled-NOT gate on the system. In this way, the system-environment composite is prepared in one of the distributions $\{ P_{S_1 E}^{(j)}\}_j$, namely,
\begin{align} 
 P_{S_1 E}^{(0)}=&\frac{1}{2} [0]_{S_1} \otimes [0]_E + \frac{1}{2} [1]_{S_1} \otimes [1]_E
\\
 P_{S_1 E}^{(1)}=&\frac{1}{2} [1]_{S_1} \otimes [0]_E + \frac{1}{2} [0]_{S_1} \otimes [1]_E
\end{align}
As before, the system-environment interaction is a controlled-NOT gate with the environment qubit as control and the system qubit as target, so that if $E=0$, then $S_2=S_1$, while if $E=1$, then $S_2=S_1\oplus_2 1$.
% (with arithmetic mod 2).
%the system's physical states are left unchanged if $E=0$, but undergo the permutation $0 \leftrightarrow 1$ if $E=1$.
%, described by the conditional probability distribution $P_{S_2|S_1 E}=\delta_{E,0} \delta_{S_2,S_1} + \delta_{E,1} \delta_{S_2,S_1 \oplus 1}$.

The marginal states of the system (for each value of $J$) after the system-environment interaction are given by 
\begin{align}
 &P^{(0)}_{S_2} = [0]_{S_2}\\
 &P^{(1)}_{S_2} = [1]_{S_2}.
\end{align}

The corresponding input-output relation is 
\begin{align} \label{crel3}
\mathfrak{R}=\big\{&(\frac{1}{2} ([0] +[1]),[0]),\nonumber\\
&(\frac{1}{2} ([0] +[1]),[1])\big\}.
\end{align}
Since it is one-to-many, there is no map consistent with this relation.

\subsection{Every classical pathology can be obtained from every classical circuit type} \label{cgeneric}

We have designed these specific classical examples to make the point as simply as possible (while maintaining a close analogy with the three quantum examples presented previously). As we show in Appendix~\ref{cex}, however, there are infinite families of examples which exhibit each of the three pathologies, no matter which type of operational scenario ($(a),(b)$, and $(c)$ in Section~\ref{genericrep} above) one considers. These generic classical examples are analogous to the generic quantum examples in Appendix~\ref{qex}.

\section{The correct definition of the evolution map in the classical sphere} \label{causalpersp}

In the quantum literature, some of the problematic examples we described earlier led researchers to question the notion that the evolution of quantum states is always described by a completely positive map. By contrast, no one has previously seen fit to cite the kinds of problematic classical examples that we have just described as a reason to question the claim that evolution of classical probability distributions is always described by a stochastic map. %In other words, there has never been a proposal for non-Markovian exceptionalism in the classical sphere. 
Why not? Could it simply be a failure to recognize the existence of problematic classical examples? 

No. The reason no one has questioned the adequacy of stochastic maps is because in the classical sphere no one was inclined to endorse the classical analogue of the standard proposal for how to define
%technique of the standard argument for inferring
 the evolution map. Rather, the framework of causal modeling and the so-called {\em do calculus} was developed~\cite{Pearl}, and this provided a scheme for inferring the evolution map on the system even in the presence of initial correlations between system and environment. 
%This is the scheme that disentangles correlation from causation.
 It is found to differ significantly from that of the standard proposal, and in particular it avoids all of the problematic implications. In this section, we will present this scheme, explain why it clearly yields the correct notion of an evolution map in the classical realm, and identify the mistaken assumptions in the standard proposal. 

The correct definition of an evolution map in the classical sphere generalizes easily to the quantum sphere, as we will show in Section~\ref{sec:quantumdomaps}.

%the argument in favor of the classical definition of an evolution map holds also in quantum theory. In Section~\ref{sec:quantumdomaps}, we will generalize the argument in order to define the correct quantum evolution map.
 
\subsection{The causal perspective on defining an evolution map} \label{sec:causalp}

We begin with some reflections on the notion of an evolution map in a classical statistical theory. 
These are the ideas that underlie the calculus for causal reasoning that 
%Based on these ideas, a calculus for causal reasoning in classical scenarios 
has been developed by Pearl~\cite{Pearl}\footnote{Classical causal models are typically defined in terms of directed acyclic graphs, rather than circuit diagrams. We use the latter representation because it facilitates the comparison with the quantum case.
%, while representing the essential causal facts that we will require for our analysis.
}. 

An evolution map is not a mere description of the {\em actual} statistical states of the input and the output of a process, but a prescription for determining the statistical state of the output for {\em any} statistical state of the input, in analogy to what distinguishes a law of motion from a historical account (as we already noted in Section~\ref{problemimp}).  Just as a law of motion is autonomous from the initial conditions, an evolution map is autonomous from the state at its input: for any variation of the state of its input, the map is unchanged.  

To define the evolution map from $A$ to $B$ in an arbitrary circuit, therefore, one considers a counterfactual scenario involving the minimal modification  to the circuit which allows one to freely vary the statistical state of $A$, while keeping the rest of the circuit unchanged.  This minimal modification consists of altering the causal mechanism that determines $A$, while keeping every other causal mechanism in the circuit unchanged. (We have here made explicit use of the fact that the different causal mechanisms in the circuit are autonomous.)  Because the way that $B$ depends on $A$ is only a function of these other causal mechanisms (and not a function of the mechanism that determines $A$), it is the same in the counterfactual circuit as it is in the actual circuit. 
Therefore, the evolution map from $A$ to $B$ in the actual circuit can be identified with the map from $A$ to $B$ in the counterfactual circuit. We denote the evolution map from $A$ to $B$ by $\Gamma_{B|{\rm do}A}(\framebox(5,5){}_A)$, in deference to the notion of a {\em `do-conditional'} from classical causal modeling. 

\begin{figure}[htb!]
\centering
\includegraphics[width=0.45\textwidth]{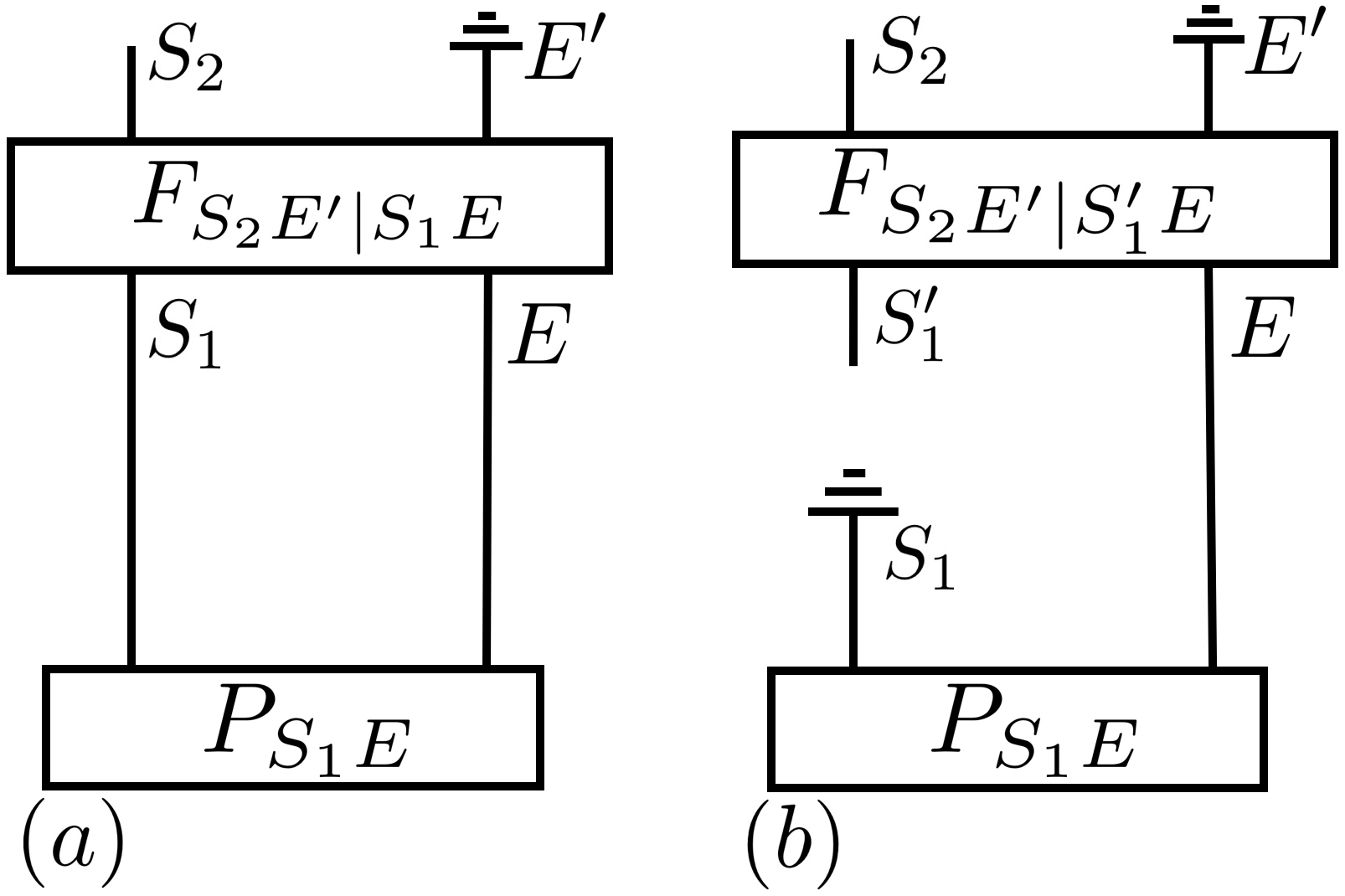}
\caption{(a) The classical analog of the circuit in Fig.~\ref{onestate}. (b) The hypothetical circuit which aids in defining the evolution map for the situation in (a). 
%This circuit is identical to that in (a), except that the system has been subjected to an intervention: the natural mechanism determining its state has been replaced by a new mechanism that ensures that it is statistically independent of the systems which, in the natural course of events, would be its causal parents.
}
\label{conestatemap}
\end{figure}

Recall the very first scenario we introduced, shown in Fig.~\ref{onestate}, and consider its classical analogue, shown in Fig.~\ref{conestatemap}(a). 
%We seek to define the evolution map in the circumstance wherein one has a complete description of the causal circuit, both its form and the identity of the circuit elements appearing within it (namely $P_{S_1 E}$ and $F_{S_2 E'|S_1 E}$). 
For such a circuit, one computes the evolution map from $S_1$ to $S_2$ by applying the prescription just given, as follows. 
One imagines that the naturally occurring state of $S_1$ is ignored, and instead a new state for $S_1$ is prepared, in a manner uncorrelated with the systems that would under natural circumstances be its causal parents. The counterfactual scenario being imagined is depicted in Fig.~\ref{conestatemap}(b). 
The wire labeled $S_1$ in the original circuit is 
replaced, in the counterfactual circuit, by 
%mapped onto two wire fragments 
a pair of wires that are disconnected. 
%in the counterfactual circuit. 
The first, which inherits the label $S_1$, maintains the causal ancestry of the original, but is marginalized over
%traced out
 (denoted by the ground symbol from electronics). The second, which is labeled by $S_1'$, maintains the causal descendants of the original, and is an input to the counterfactual circuit.
%in the counterfactual scenario.

The evolution map from $S_1$ to $S_2$ in the original circuit of Fig.~\ref{conestatemap}(a), $\Gamma_{S_2|{\rm do}(S_1)}:\mathfrak{P}_{S_1} \to \mathfrak{P}_{S_2}$, is defined as the map 
which is isomorphic (under the identification of $S_1$ and $S_1'$) to the map $\Gamma_{S_2|S_1'}: \mathfrak{P}_{S_1'} \to \mathfrak{P}_{S_2}$ 
in the counterfactual circuit of Fig.~\ref{conestatemap}(b), 
that is, 
\beq
\Gamma_{S_2|{\rm do} S_1}(\framebox(5,5){}_{S_1}\!) :=\Gamma_{S_2|S_1'}(\framebox(5,5){}_{S_1'}\!).
\eeq

Given that 
\begin{equation} 
\Gamma_{S_2|S_1'}(\framebox(5,5){}_{S_1'}\!) = \sum_{E'}F_{S_2 E'| S_1' E} (\framebox(5,5){}_{S_1'}\! \otimes P_E)
\end{equation}
where 
$P_E := \sum_{S_1} P_{S_1 E} $, %and where $P_{S_1 E}$ is the probability distribution on the system-environment composite.
%Given this isomorphism, \red{that $\Gamma_{S_2|{\rm do}S_1}:=\Gamma_{S_2|S_1'}$}, 
it follows that the evolution map for the scenario of Fig.~\ref{conestatemap}(a) is
\begin{equation} \label{onetruemap}
\Gamma_{S_2|{\rm do}S_1}(\framebox(5,5){}_{S_1}\!) := \sum_{E'}F_{S_2 E'| S_1 E} (\framebox(5,5){}_{S_1}\! \otimes P_E).
\end{equation}

The classical causal models framework typically focuses on the {do-conditional} $P_{S_2|{\rm do}S_1}$, rather than the evolution map $\Gamma_{S_2|{\rm do}S_1}$, but the latter is simply related to 
%obtained from
 the former via
% which it defines via 
\begin{align}\label{doconditional}
\Gamma_{S_2|{\rm do}S_1} (\framebox(5,5){}_{S_1}\!) := \sum_{S_1} P_{S_2|{\rm do}S_1} \framebox(5,5){}_{S_1}\!.
\end{align}

Eq.~\eqref{onetruemap} implies that the evolution map can be computed directly from the identity of the circuit elements in the original circuit of Fig.~\eqref{onestate}.  This is because it depends only on $F_{S_2 E'| S_1 E}$ and $P_E$ (which is obtained from $P_{S_1 E}$ by marginalization).
%Note that the correct definition of the evolution map, Eq.~\eqref{onetruemap}, can be inferred from a complete specification of the elements in the circuit of Fig.~\eqref{onestate}, even though there is no variation in the state of $S_1$ in the scenario. The evolution map depends {\em only} on the function $F_{S_2 E'|S_1E}$ and the marginal $P_E$
%~\footnote{Some previous works (e.g. Refs.~\cite{stelm,beyond}) sought to define the evolution map as a function of the marginal state of the environment {\em and also} as a function of the initial system-environment correlations present in the initial state of the composite. This prescription is distinct from that of the standard argument, but it does not correspond to our prescription. In particular, it does not reproduce the correct evolution map (of Eq.~\eqref{onetruemap}) because the latter depends {\em only} on the marginal state of the environment.}.

It follows, in particular, that there is no need to consider an input-ouput relation.
%Therefore, 
 %it makes perfect sense for the circuit of Fig.~\eqref{onestate}. One can compute the evolution map directly from the given circuit, with no need for an input-output relation.
%Therefore, it can be inferred for the circuit of Fig.~\eqref{onestate}. One can compute the evolution map directly from the given circuit, with no need for an input-output relation.
This is in contrast to the standard proposal, which (because of its insistence on starting with the input-output relation) {\em cannot} define a map uniquely unless the scenario explicitly involves a {\em set} of initial states. This is why, as we noted in Sec.~\ref{sec:standard}, the standard proposal cannot define a map for the circuit of Fig.~\eqref{onestate} and why it is forced to consider circuits of the form of Fig.~\eqref{fig:standard}, where there are variables $J$ and/or $K$ that allow one to introduce variation in the input state.

Note that we are here discussing how to {\em define} the evolution map when one has {\em a complete description of the form of the circuit and the identity of each of its elements}. The question of {\em how one obtains such a description} is not relevant to the definitional question, and it is important not to confuse the two questions.
%the {\em definition} of the evolution. 
We will discuss the question of how to infer the quantum evolution map from experimental data in Section~\ref{practical}.
%characterize an unknown circuit in Section~\ref{practical}, where we consider how a quantum evolution map can be inferred from a tomographic scheme.) 
 
\subsection{Contrasting the evolution map with the inference map}\label{contrastingevolutionandinference}

In the causal modeling framework~\cite{Pearl}, the primary motivation for introducing do-conditionals was to distinguish them from standard conditionals. We pause here to describe the distinction because it is critical to our analysis of the mistakes of the standard proposal. 

Suppose that one considers the map defined by the standard conditional $P_{S_2|S_1}$, namely,
\begin{align}\label{standardconditional}
\Gamma_{S_2|S_1} (\framebox(5,5){}_{S_1}\!) := \sum_{S_1} P_{S_2|S_1} \framebox(5,5){}_{S_1}\!.
\end{align}
%(Note the notational difference between this and the expression for the evolution map, in Eq.~\eqref{doconditional}.)
The correct way of interpreting the map defined in Eq.~\eqref{standardconditional} is as an {\em inference} map. It answers the question: Given a particular state of knowledge of $S_1$, what is the appropriate state of knowledge to assign to $S_2$? In particular, if one updates one's description of $S_1$ (based on passively observing it, for instance), this map specifies how one should update one's description of $S_2$. If $S_1$ and $S_2$ are related causally by both a cause-effect connection and a common cause of the two, then the inference map describes what $S_1$ can teach you about $S_2$ through {\em either} causal pathway. By contrast, the evolution map specifies {\em only} what $S_1$ can teach you about $S_2$ through the cause-effect pathway. 

This distinction is often illustrated by Simpson's paradox.   This is a scenario in which there is a positive correlation between a treatment variable and a recovery variable, even though the causal influence of the treatment is to {\em reduce} the probability of recovery (see Sec. 6 of \cite{Pearl}). The positive correlation is the result of there also being a common cause acting on the two variables. For instance, if men are more likely than women to seek the treatment and are also more likely than women to recover regardless of treatment, then gender acts as a common cause. In this circumstance, learning that an individual in the sample population got the treatment warrants assigning a higher likelihood to the proposition that that individual recovered, simply because learning that someone got the treatment is positive evidence for them being male, which in turn is positive evidence for them recovering. 

To see this mathematically, let $R$ denote the recovery variable and let $T$ denote the treatment variable.  If one na\"{i}vely computes $P_{R|T}=P_{RT}/P_R$, or equivalently, the map $\Gamma_{R|T}$ (via Eq.~\eqref{standardconditional}), it is clear that this does {\em not} represent the cause-effect relation that exists between $T$ and $R$ and hence cannot be used to make assessments of the effectiveness of the treatment on recovery. Only the map $\Gamma_{R|{\rm do}T}$, computed from $P_{R|{\rm do}T}$ (via Eq.~\eqref{doconditional}), wherein the intervention on $T$ rules out the possibility of inference via a common cause, represents state updating based purely on the cause-effect relation. 

%It is precisely for this reason that correlation does not imply causation. 
In short, Simpson's paradox reminds us that correlation does not imply causation. To draw conclusions about whether a given treatment contributes causally to recovery, one must consider what would occur in a trial wherein the value of the treatment variable is assigned at random (drug or placebo, for instance), independently of any preferences of the individual. The do-conditional describes what would occur in such randomized trials.

Hence, we see that the evolution map is generally distinct from the inference map.
%, and that the two allow for different sorts of inferences.

Finally, note that---just as we saw for the evolution map---the inference map in classical scenarios can be computed directly from the original circuit elements, so that one need not consider any sort of input-output relation. To see this, first 
%If one knows the circuit to be of the form of Fig.~\ref{conestatemap}(a), one can compute the inference map as follows. First, 
note that the conditional $P_{S_2|S_1}$ can be expressed as
\begin{align}\label{compstandardcond}
P_{S_2|S_1} &= \sum_E P_{S_2|S_1E} P_{E|S_1}\\
&= \sum_{E' E} P_{S_2 E'|S_1E} P_{E|S_1}
\end{align}
The assumption that the circuit elements in Fig.~\ref{conestatemap}(a) are known implies that $F_{S_2E'|S_1E}$ and $P_{S_1E}$ are known.   $P_{S_2 E'|S_1E}$ in the above equation is just the conditional associated to the stochastic map $F_{S_2E'|S_1E}$,
%via Eq.~\eqref{standardconditional}, 
while $P_{E|S_1}$ can be computed from $P_{S_1E}$ by $P_{E|S_1} = P_{S_1E} /P_{S_1}$ where $P_{S_1} = \sum_E P_{S_1E}$.
The inference map $\Gamma_{S_2|S_1}$ associated to the conditional $P_{S_2|S_1}$ is then found, via Eq.~\eqref{standardconditional}, to be:
\begin{align} \label{itogivens}
\Gamma_{S_2|S_1} (\framebox(5,5){}_{S_1}\!) &:= \sum_{S_1 E E'} P_{S_2 E'|S_1E}P_{E|S_1}\framebox(5,5){}_{S_1}\! \nonumber\\
&=\sum_{E'} F_{S_2E'|S_1E} (P_{E|S_1}\framebox(5,5){}_{S_1}\!).
\end{align}

%\begin{equation} \label{itogivens}
%\Gamma_{S_2|S_1} (\framebox(5,5){}_{S_1}\!) := \sum_{E'} F_{S_2E'|S_1E} (P_{E|S_1}\framebox(5,5){}_{S_1}\!).
%\end{equation}

\begin{comment}
\begin{align}\label{compstandardcond}
P_{S_2|S_1} = \sum_E P_{S_2|S_1E} P_{E|S_1},
\end{align}
The assumption that the circuit elements in Fig.~\ref{conestatemap}(a) are known implies that $F_{S_2E'|S_1E}$ and $P_{S_1E}$ are known, and hence that one can compute $P_{E|S_1}$ via $P_{E|S_1} = P_{S_1E} /P_{S_1}$, and $P_{S_2|S_1E}$ via 
{$P_{S_2|S_1E}=
%\sum_{E'}  \delta_{S_2,f(S_1,E)}\delta_{E',g(S_1,E)} =
 \delta_{S_2,f(S_1,E)}$, where $f$ is the function describing the deterministic dynamics that defines the map $F_{S_2E'|S_1E}$ (see Eq.~\eqref{defnstoch}).}
%$P_{S_2|S_1E}=\sum_{E'} P_{S_2 E'|S_1E}$, computed for the particular conditional $P_{S_2 E'|S_1E}$ defined by $F_{S_2E'|S_1E}$ (using Eq.~\eqref{defnstoch}). 
One can thereby compute the standard conditional $P_{S_2|S_1}$ using Eq~\eqref{compstandardcond}, and the inference map $\Gamma_{S_2|S_1}$ can be computed from it by Eq.~\eqref{standardconditional}: 
\begin{equation}
\Gamma_{S_2|S_1} (\framebox(5,5){}_{S_1}\!) := \sum_{S_1 E} P_{S_2|S_1E} P_{E|S_1}\framebox(5,5){}_{S_1}\!.
\end{equation}
In terms of the circuit elements given in our problem,% (e.g. in Fig.~\ref{}), 
\begin{equation} \label{itogivens}
\Gamma_{S_2|S_1} (\framebox(5,5){}_{S_1}\!) := \sum_{E'} F_{S_2E'|S_1E} (P_{E|S_1}\framebox(5,5){}_{S_1}\!).
\end{equation}
\end{comment}

\section{The mistaken assumptions of the standard proposal in the classical sphere} \label{sec:prob}

With the causal point of view in mind and the correct definition of the evolution map in hand, it is very instructive to revisit the standard proposal in the classical sphere and to isolate the mistaken assumptions therein. 

\subsection{The mistake of confusing the evolution map with the inference map}

%Let us return to the very first circuit form we considered, that of Fig.~\ref{onestate}, but where we imagine the systems to be classical. 
%Consider what the standard argument entails in this case. 
%for this circuit in the classical sphere. 
Consider what the standard proposal entails for the simple circuit of Fig.~\ref{conestatemap}(a). There is only a single distribution $P_{S_1}$ and a single distribution $P_{S_2}$, and so the input-output relation consists of a {\em single} input-output pair, namely, $\{ (P_{S_1},P_{S_2})\}$. 
The standard proposal is committed to the notion that this input-output pair can be interpreted as a constraint on the evolution map, namely, that the evolution map acting on $P_{S_1}$ must yield $P_{S_2}$. 

Classical probability theory dictates that the relationship which holds between $P_{S_1}$ and $P_{S_2}$ is
\begin{align}
P_{S_2} = \sum_{S_1} P_{S_2|S_1} P_{S_1},
\end{align}
and consequently, if $\Gamma_{S_2|S_1}$ is the stochastic map associated to the conditional $P_{S_2|S_1}$ by Eq.~\eqref{standardconditional}, then
\begin{align}
P_{S_2} = \Gamma_{S_2|S_1} (P_{S_1}).
\end{align}
But $\Gamma_{S_2|S_1}$ is the {\em inference } map from $S_1$ to $S_2$, which can differ from the evolution map $\Gamma_{S_2|{\rm do} S_1}$, as discussed in Sec.~\ref{contrastingevolutionandinference}. Consequently, the input-output pair $\{ (P_{S_1},P_{S_2})\}$ is only guaranteed to be consistent with the inference map and is {\em not} guaranteed to be consistent with the evolution map. 
%from $S_1$ to $S_2$. It does {\em not} necessarily provide a constraint on the evolution map $\Gamma_{S_2|{\rm do} S_1}$. 
 Indeed, if there is a common cause acting on $S_1$ and $S_2$, as is the case in the circuit of Fig.~\ref{conestatemap}(a), then generically the evolution map and the inference map {\em do} differ, and the input-output pair $\{ (P_{S_1},P_{S_2})\}$ is {\em only} consistent with the inference map and not the evolution map.

To assume, as the standard proposal does, that the input-output pair $\{ (P_{S_1},P_{S_2})\}$ constitutes a constraint on the evolution map is to make a mistake akin to inferring causation from correlation alone. For instance, it is akin to inferring, from a positive correlation between treatment and recovery, that treatment has a positive causal influence on recovery even though  
%only that there is a positive correlation between them and one knows 
there is a common cause 
%based only on a positive correlation between them and where there is a common cause 
(such as gender) that could account for this positive correlation. 

This is the first mistake of the standard proposal. 

As we discuss in Appendix~\ref{coincide}, the inference map and the evolution map coincide if and only if $E$ and $S_1$ are marginally independent, $P_{S_1E}=P_{S_1}\otimes P_E$. This is ironic, because proponents of the standard proposal were interested in characterizing precisely those scenarios which violated this condition.
%: those with initial system-environment correlations.

\subsection{The mistake of taking input-output pairs from different inference maps as constraints on a single map}

%In the discussion surrounding Fig.~\ref{onestate}, where the quantum version of Fig.~\ref{conestatemap}(a) was being considered, it was noted that a single input-output pair is not a constraint on an evolution map. It {\em is} a constraint on the inference map, but it is an extremely weak constraint, and certainly doesn't define the inference map uniquely. 
%In an attempt to obtain constraints on the evolution map, those who 
Proponents of the standard proposal sought to define a map through an input-output relation $\mathfrak{R}$
% focused attention only on circuits of the form of Fig.~\ref{cgeneral}, 
where variability in the input state of the system $S_1$ (and therefore also the output state of $S_2$) was introduced via variation in the values of $J$ and $K$ in a circuit of the form of Fig.~\ref{cgeneral}(b). 
 %In doing so, they obtained a set of input-output pairs which defined the input-output relation $\mathfrak{R}$. 
Since, as noted in the previous section, an individual input-output pair in $\mathfrak{R}$ is not a constraint on the evolution map, the {\em set} of such pairs obviously does not constrain the evolution map either. 
This would seem to leave open the possibility that the set of input-output pairs might still constrain the inference map. However, this is not the case either. 
%But could it be that the set of input-output pairs defines the correct inference map? No, they do not. 
If confusing inference with evolution had been the {\em only} mistake of the standard proposal, 
%But each input-output pair in $\mathfrak{R}$ {\em is} a constraint on the inference map, so might it be that all of the input-output pairs together form a set of constraints which defines an inference map? The answer is no. 
%If confusing inference with evolution had been the only mistake of the standard argument, one would have found that the input-output relation was always consistent with a stochastic map, because every inference map is a stochastic map. 
then because every inference map is a stochastic map, one would never have found any failure of the input-output relation to define a stochastic map,
%have found that the input-output relation was always consistent with a stochastic map, 
contrary to what is found in the pathological examples.
%However, we saw that this is generally not the case (e.g. in our examples of relations that were consistent with linear but non-stochastic maps, with nonlinear maps, or with no map at all). 

The standard proposal not only fails to yield the evolution map, it {\em also} fails to yield the inference map, due to a second mistaken assumption, which we now discuss.

As we argued in Section~\ref{contrastingevolutionandinference}, the inference map in classical scenarios can be computed directly from the circuit, without introducing a set of initial states indexed by $J$ and $K$. Therefore, introducing explicit variation in the initial state was unnecessary. Moreover, it is the root of all the pathological implications of the standard proposal. 

The means by which variability was introduced in the input state {\em inadvertently} led to variability in the inference map as well. That is, for generic examples of circuits to which the standard proposal has been applied (those of the form of Fig.~\ref{cgeneral}(b)), the variation in the values of $J$ and $K$ leads {\em not only} to variation in the marginal state of the system $S_1$, but also to variation in $P_{E|S_1}$, and hence, given 
Eq.~\eqref{itogivens},
%Eq.~\eqref{compstandardcond}, 
to variation in the inference map.

To show this, we determine the formula that relates the conditional probability $P_{S_1|JK}$ to $P_{S_2|JK}$ for circuits of the form of Fig.~\ref{cgeneral}(b).

By definition, 
\begin{align} \label{cccmap1}
P_{S_2|JK} &=\sum_{S_1 E} P_{S_2 S_1 E|JK}
\end{align}
To express $P_{S_2|JK}$ as a function of $P_{S_1|JK}$, we simply make a repeated application of an identity from Bayesian probability theory, namely, $P_{AB|C}=P_{A|BC}P_{B|C}$.  We infer that $P_{S_2 S_1 E| JK} = P_{S_2 |S_1 E JK}P_{E |S_1 JK} P_{S_1|JK}$ and consequently that
\begin{align} \label{cccmap2}
P_{S_2|JK} &= \sum_{S_1 E} P_{S_2 | S_1 E JK}P_{E |S_1 JK}P_{S_1|JK}.
\end{align}
For a circuit of the form of Fig.~\ref{cgeneral}(b), the causal structure ensures that $S_2$ is conditionally independent of $JK$ given $S_1 E$, so that $P_{S_2| S_1 E JK} = P_{S_2| S_1 E}$, and we conclude that
\begin{align} \label{cccmap3}
P_{S_2|JK} &= \sum_{S_1 E} P_{S_2 | S_1 E} P_{E |S_1 JK}P_{S_1|JK}.
\end{align}
This defines a $JK$-dependent map from distributions on $S_1$ to distributions on $S_2$.  Specifically, if we define $P_{E|S_2}^{(j,k)}$ by
\begin{align}
&\forall e,s,j,k: P_{E|S_2}^{(j,k)}(e|s) := P_{E|S_2 JK}(e|sjk),
%&P_{S_1}^{(j,k)}(s) := P_{S_1|JK}(s|jk)\nonumber\\
%&P_{S_2}^{(j,k)}(s) := P_{S_2|JK}(s|jk)\nonumber
\end{align}
the map is 
\begin{align} \label{cccmap4}
\Gamma_{S_2|S_1}^{(j,k)}(\framebox(5,5){}_{S_1}\!) &= \sum_{S_1 E} P_{S_2 |S_1 E} P^{(j,k)}_{E|S_1}\; \framebox(5,5){}_{S_1}\!.
\end{align}
Rewriting in terms of the circuit elements given in our scenario (as was done in Eq.~\eqref{itogivens}), this map has the form:
\begin{align}  \label{cccmap5}
\Gamma_{S_2|S_1}^{(j,k)}(\framebox(5,5){}_{S_1}\!) &= \sum_{E'} F_{S_2 E'|S_1 E} (P^{(j,k)}_{E|S_1}\; \framebox(5,5){}_{S_1}\!).
\end{align}

For one who knows that $J=j$ and $K=k$,
%$JK$ take values $(j,k)$, 
the latter map is the correct way of making inferences from $S_1$ to $S_2$. The inference map 
clearly depends on the values of $J$ and $K$.
%depends on one's knowledge of $J$ and $K$.

It follows from Eqs.~\eqref{cccmap3} and \eqref{cccmap4}
 that if $P_{S_1}^{(j,k)}$ and $P_{S_2}^{(j,k)}$ are defined by
\begin{align}
&\forall s,j,k: P_{S_1}^{(j,k)}(s) := P_{S_1|JK}(s|jk)\label{defnjk1}\\
&\forall s,j,k: P_{S_2}^{(j,k)}(s) := P_{S_2|JK}(s|jk),\label{defnjk2}
\end{align}
then the input-output pair $\{ (P^{(j,k)}_{S_1}, P^{(j,k)}_{S_2})\}$ 
%$\{ (P_{S_1|J=j,K=k}, P_{S_2|J=j,K=k})\}$ 
is consistent with the map $\Gamma_{S_2|S_1}^{(j,k)}$ in the sense that 
\begin{align}\label{manymaps}
P^{(j,k)}_{S_2} = \Gamma_{S_2|S_1}^{(j,k)}(P^{(j,k)}_{S_1}).
%P_{S_2|J=j,K=k} = \Gamma_{S_2|S_1}^{(j,k)}(P_{S_1|J=j,K=k}).
\end{align}
However, the prescription of the standard proposal was to find a {\em single} map $\Gamma_{S_2|S_1}$ such that
\begin{align}
\forall j,k : P^{(j,k)}_{S_2} = \Gamma_{S_2|S_1}(P^{(j,k)}_{S_1}).
%\forall j,k : P_{S_2|J=j,K=k} = \Gamma_{S_2|S_1}(P_{S_1|J=j,K=k}).
\end{align}

%This is where the real trouble with the standard argument lies. 
This last step is the origin of the 
%most severe
 pathologies of the standard proposal. 
As Eq.~\eqref{manymaps} shows, in general the input-output pairs for different values of $JK$ describe the input and output states of {\em different} maps. But the prescription of the standard proposal asks us to collect all of these input-output pairs into a single set, the input-output relation, and to try and find a {\em single} map that is consistent with all of them. Given the origin of these pairs, there is no guarantee that there is {\em any} such map, and even if there does happen to be one, there is no guarantee that it is linear or stochastic. 

To summarize, the second mistake of the standard proposal is to have inadvertently introduced variability in the inference map and then to have tried to define a {\em unique} map from the input-ouput pairs that are associated to these different inference maps. 

%In general, the standard argument makes both the first and the second mistake.

\subsection{The correct evolution map(s) for the scenarios wherein the standard proposal led to pathologies}
%considered by the standard argument} 
\label{sec:sac}

In the previous section, we noted that the inference map can be $JK$-dependent in the scenario of Fig.~\ref{cgeneral}(b). It turns out that the evolution map can also be $JK$-dependent in this scenario.
%Proponents of the standard argument sought to find a single map to describe the scenario of Fig.~\ref{cgeneral}, where one imagines conditionning on the values of $J$ and $K$. 
%However, it follows immediately from the correct definition of evolution that 
Generally, therefore, {\em there is no single evolution map} to be characterized in this scenario. 
Rather, following the prescription of Section~\ref{sec:causalp}, one finds that for each valuation $(j,k)$ of $J$ and $K$, there is a (generally) distinct marginal state $P_E^{(j,k)}$ of the environment, which leads to distinct evolution maps, namely
%there is a distinct map for each valuation $(j,k)$ of $J$ and $K$, namely,
\begin{align} \label{cdomapik}
\Gamma_{S_2|{\rm do}S_1}^{(j,k)}(\framebox(5,5){}_{S_1}\!) &= \sum_{E'}F_{S_2E'|S_1 E} (\framebox(5,5){}_{S_1}\!\otimes P_E^{(j,k)}).
\end{align}
%where $P_E^{(j,k)}$ is the marginal on $E$ conditional on $J=j$ and $K=k$.

At first glance, this may seem problematic, but in fact the knowledge-dependence of evolution maps is ubiquitous in both classical and quantum physics. We give several examples in Appendix~\ref{knowldepend}.
%Explicitly, applying the prescription of Section~\ref{sec:causalp} to the classical analog of the circuit in Fig.~\ref{cgeneral} gives the true description of the evolution, which is a {\em set} of evolution maps, one for each valuation of $J$ and $K$:

\subsection{Illustrating the mistakes with the classical examples considered previously}
\label{illust}%the three classical examples of Sec.~\ref{sec:classical}}

In this section, we illustrate the general discussion just given by explicitly analyzing classical examples introduced previously, highlighting interesting features along the way.
%This analysis of the second classical example also illustrates the subtle implications of causal relations in a careful treatment of inference and evolution.

Consider again the third classical example, discussed in Section~\ref{c:CNOT}, where the input-output relation failed to identify any map at all.

Recall that the example has no $K$ variable, so only $J$ is relevant. Consider the inference map from $S_1$ to $S_2$ for a particular value $j$ of $J$. 
Specializing Eq.~\eqref{cccmap5}, this has the form
\begin{align} \label{ex3Gamma}
\Gamma_{S_2|S_1}^{(j)}(\framebox(5,5){}_{S_1}\!) &= \sum_{E'} F_{S_2 E' |S_1 E} ( P^{(j)}_{E|S_1}\; \framebox(5,5){}_{S_1}\!).
\end{align}

The coupling of system and environment implies that $S_2= S_1 \oplus_2 E$; that is, it implies that $S_2$ tracks the {\em parity} of $S_1$ and $E$. The controlled-NOT operation from $J$ to $S_1$ toggles this parity in a $J$-dependent way, but without changing the marginal state of $S_1$. Thus, as one varies $J$, one has variability in the state on $S_2$, but no variability in the state on $S_1$, leading to the one-to-many relation that fails to correspond to any map. 

One way to understand the fact that different values of $J$ lead to different states on $S_2$ is that $E$ is correlated with $J$ given $S_1$, 
so that $\exists e,s: P_{E| S_1J}(e|s,0) \ne P_{E| S_1J}(e|s,1)$, or equivalently (given Eq.~\eqref{defnjk1}), $P^{(0)}_{E| S_1} \ne P^{(1)}_{E| S_1}$,
%so $P_{E| S_1,J=0} \ne P_{E| S_1,J=1}$ in Eq.~\eqref{ex3Gamma}, 
and consequently the inference map in Eq.~\eqref{ex3Gamma} becomes $J$-dependent.

Indeed, as we show in Appendix~\ref{app:classical}, the inference maps for the two values of $J$ are
\begin{align} \label{infmap3}
\Gamma^{(j)}_{S_2|S_2} (\framebox(5,5){}_{S_1}\!) = \delta_{S_2,j},
%&= \sum_E \delta_{S_2, S_1 \oplus E} \delta_{E,S_1 \oplus j}\\
%&= \delta_{S_2,j}\\
%&= [j]_{S_2},
\end{align} 
%as we work out explicitly in Appendix~\ref{app:classical}. 
so that for  $J=0$ ($J=1$), one updates one's description of $S_2$ to the state $[0]_{S_2}$ ($[1]_{S_2}$) {\em regardless} of one's state of knowledge of $S_1$.
%This establishes that when $J=0$ ($J=1$), the inference map is such that, regardless of one's state of knowledge of $S_1$, one updates one's description of $S_2$ to the state $[0]_{S_2}$ ($[1]_{S_2}$).
 
Recall the input-output relation for this example, Eq.~\eqref{crel3}.
% \begin{align}
%\mathfrak{R} =  &\{( \frac{1}{2}[0]_{S_1} +\frac{1}{2}[1]_{S_1}, [0]_{S_2})\nonumber\\
%&( \frac{1}{2}[0]_{S_1} +\frac{1}{2}[1]_{S_1}, [1]_{S_2})\}.
%\end{align}
%pairs for different values of $J$ for this example were:
 %\begin{align}
%J=0: \quad &( \frac{1}{2}[0]_{S_1} +\frac{1}{2}[1]_{S_1}, [0]_{S_2})\nonumber\\
%J=1: \quad &( \frac{1}{2}[0]_{S_1} +\frac{1}{2}[1]_{S_1}, [1]_{S_2}).
%\end{align}
Clearly, the first ($J=0$) input-output pair is consistent with the inference map for $J=0$, $\Gamma^{(0)}_{S_2|S_1}$, while the second ($J=1$) input-output pair is consistent with the inference map for $J=1$, $\Gamma^{(1)}_{S_2|S_2}$. However, if one mistakenly considers {\em both} input-output pairs to be associated to a {\em single} map, one finds a contradiction.
%, because the two pairs have the same input state and different output states. 
This illustrates the second mistake of the standard proposal. 
%The fact that the standard argument then mistakenly takes the input-output pairs to be consistent with a single $J$-independent map is what causes the input-output relation to be inconsistent with any stochastic map.

The first mistake of the standard proposal is also illustrated in this example.%, since each individual input-output pair is not a constraint on the evolution map.
% (as opposed to the corresponding inference map).
%, as was mistakenly assumed in the standard argument. 

The correct evolution map is straightforward to identify using the prescription of Section~\ref{sec:causalp}. For both values of $J$, the marginal state of the environment is uniformly random, so the evolution map is $J$-independent and equal to the 
 %in this unique evolution map is the 
 randomizing map
\begin{equation} \label{randchan}
\Gamma_{S_2|{\rm do}S_1}(\framebox(5,5){}_{S_1}\!) = \frac{1}{2}([0]_{S_2}+[1]_{S_2}),%\sum_{S_1} (P_{S_1}) 
\end{equation}
 which takes any input probability distribution on $S_1$ to a uniformly random distribution on $S_2$. This is distinct from either of the inference maps $\Gamma^{(0)}_{S_2|S_1}$ or $\Gamma^{(1)}_{S_2|S_1}$, and so confirms that the evolution map is not constrained {\em at all} by the input-output pairs considered in the standard proposal. 

We now turn our attention to the second classical example, discussed in Section~\ref{c:CNOT}.%, where the input-output relation failed to identify any map at all. 

Again, only the $J$ variable is nontrivial in this example. However, whereas $J$ was a cause of $S_1$ alone in the third classical example, here it is a cause of $E$ as well. In fact, $J$ is the complete common cause of $S_1$ and $E$ and consequently $E$ is conditionally independent of $S_1$ given $J$, $P_{E|S_1 J} = P_{E|J}$. Thus, 
Eq.~\eqref{cccmap5} specializes to the expression 
\begin{align} \label{ex2Gamma}
\Gamma_{S_2|S_1}^{(j)}(\framebox(5,5){}_{S_1}\!) &= \sum_{E'} F_{S_2 E' |S_1 E} (P^{(j)}_{E} \otimes \framebox(5,5){}_{S_1})\!
\end{align}
for the inference map from $S_1$ to $S_2$ in this example. Different values of $J$ lead to different 
marginals on $E$,
%states of knowledge of $E$, 
%and given that 
and therefore different inference maps.
%inferences about $S_2$ depend on one's state of knowledge of $E$, the inference map from $S_1$ to $S_2$ is $J$-dependent. 

The inference map for $J=j$ is
\begin{align} \label{infmap2}
\Gamma^{(j)}_{S_2|S_1} (\framebox(5,5){}_{S_1}\!) %&= \sum_E \delta_{S_2, (S_1 \oplus E){\rm mod} 3} \delta_{E,J}\\
%= \delta_{S_2, (S_1 \oplus j){\rm mod}3}.
= \delta_{S_2, (S_1 \oplus_3 j)},
\end{align} 
as shown in Appendix~\ref{app:classical}.
%The inference map is such that, 
For any state of knowledge of $S_1$, one should assign the same state of knowledge to $S_2$ if $J=0$, and the same state of knowledge modulo an increase of the value by 1 (in arithmetic modulo 3) if $J=1$.

Again, one can check explicitly that in this example an individual input-output pair for $J=j$, (as in Eqs.~\eqref{tnl0} and \eqref{tnl}), is a constraint on the inference map for $J=j$. However, if one mistakenly considers both input-output pairs to be associated to a single map, then the only maps consistent with the constraint are nonlinear.

%One can therefore verify that the first input-output pair from this example, $( [0]_{S_1}, [0]_{S_2})$, is consistent with the inference map for $J=0$, $\Gamma^{(0)}_{S_2|S_1}$, while the second input-output pair, $( \frac{1}{2}[0]_{S_1} +\frac{1}{2}[1]_{S_1}, [1]_{S_2}+[2]_{S_2})$, is consistent with the inference map for $J=1$, $\Gamma^{(1)}_{S_2|S_2}$. However, if one mistakenly considers both input-output pairs to be associated to a single map, then the only maps consistent with the constraint are nonlinear.%, because no linear map can increase the distinguishability of two distributions. 

In this example, it happens that the evolution map for a particular value $j$ of $J$ coincides with the inference map for that same value,
\begin{align}
\Gamma^{(j)}_{S_2|{\rm do}S_1} = \Gamma^{(j)}_{S_2|S_1}.
\end{align}
The reason is that conditioning on $J$ makes $S_1$ and $E$ independent, so that in the presence of this conditioning, one need not intervene on the system to achieve this independence.

The analysis of the first classical example is similar, and is also provided in Appendix~\ref{app:classical}.

%{\bf Summary}

%To summarize, the situation in the classical sphere is as follows. 
%Insofar as the standard argument is proposed as a means for determining the evolution of a system, the object of interest should be the classical evolution map, and 
%We have argued that the only sensible notion of an evolution map classically is the one defined by a do-conditional. We have pointed out that the information used by the standard argument to try to infer the system's evolution---input-output pairs---do not constrain the evolution map but instead only constrain the inference map. Furthermore, {\em even if} one were content to try to identify the inference map rather than the evolution map, the standard argument fails to do so because it takes a number of input-output pairs for {\em different} inference maps and mistakenly takes them all to be constraints on a single map.

The situation in the classical sphere can be summarized as follows. The only sensible notion of an evolution map classically is the one defined by a do-conditional, but the information used in the standard proposal to try to infer the system's evolution---input-output pairs of statistical states---does not constrain the evolution map but instead only constrains the inference map. Furthermore, {\em even if} one were content to try to identify the inference map rather than the evolution map, the prescription of the standard proposal does not provide a means of doing so because it takes a number of input-output pairs for {\em different} inference maps and mistakenly takes them all to be constraints on a single map.

\section{The correct definition of the evolution map in the quantum sphere}\label{sec:quantumdomaps}

The definition of a classical evolution map, presented in Section~\ref{causalpersp}, generalizes naturally to the definition of a quantum evolution map.  It suffices to substitute the quantum analogues of the relevant classical notions (quantum states for statistical states and unitary dependences for functional dependences) in the definition provided there. 

The ideas underlying the definition are the same as those outlined in Section~\ref{sec:causalp}:
%conceptual motivation 
%for the quantum evolution map
 %is the same as in Section~\ref{sec:causalp}: 
 an evolution map for quantum states of a system is a prescription for determining the quantum state of the output for {\em any} quantum state of the input, and it is autonomous from the state at its input.

Just as in the classical case, one imagines the counterfactual scenario with the minimal modification to the circuit which allows one to freely vary the quantum state of $A$, while keeping the rest of the circuit unchanged; the evolution map from $A$ to $B$ in the actual circuit can then be identified with the map from $A$ to $B$ in the modified circuit. 
We denote an evolution map from quantum system $A$ to quantum system $B$ by $\mathcal{E}_{B|{\rm do}A}(\framebox(5,5){}_A)$, where the ``do$A$'' on the right of the conditional parallels the classical notation and is a reminder that the definition requires contemplating the counterfactual scenario just described.
%there is an intervention on $A$. 

\begin{figure}[htb!]
\centering
\includegraphics[width=0.45\textwidth]{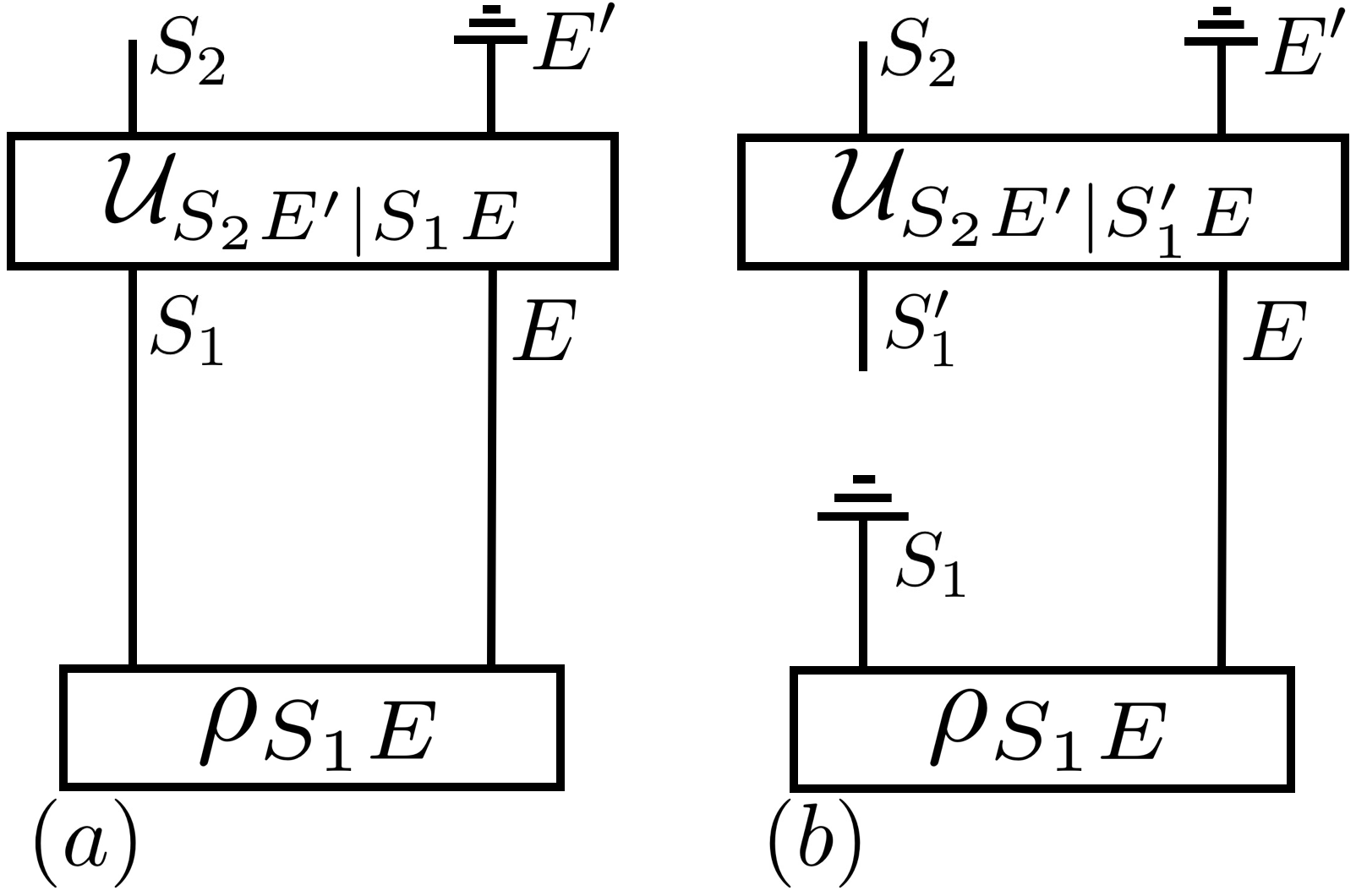}
\caption{(a) Repeat of Fig.~\ref{onestate}, for ease of reference. (b) The hypothetical circuit which aids in defining the evolution map for the situation in (a).}
%This circuit is identical to that in (a), except that the input system has been intervened upon: namely, it is ignored and reprepared in an arbitrary quantum state, independently of its causal past in the original circuit.}
\label{qonestatemap}
\end{figure}

Let us return to the most basic quantum circuit wherein the system and environment are initially correlated, that of Fig.~\ref{onestate}, reproduced here as Fig.~\ref{qonestatemap}(a). For this case, the relevant counterfactual scenario is depicted in Fig.~\ref{qonestatemap}(b). We again denote the version of the system which is varied counterfactually by $S_1'$, and here the ground symbol from electronics represents the trace over a subsystem.

%The evolution map from $S_1$ to $S_2$ in the original circuit of Fig.~\ref{conestatemap}(a), $\Gamma_{S_2|{\rm do}(S_1)}:\mathfrak{P}_{S_1} \to \mathfrak{P}_{S_2}$, is defined as the map 
%which is isomorphic (under the identification of $S_1$ and $S_1'$) to the map $\Gamma_{S_2|S_1'}: \mathfrak{P}_{S_1'} \to \mathfrak{P}_{S_2}$  in the counterfactual circuit of Fig.~\ref{conestatemap}(b), 
%that is, 

The evolution map from $S_1$ to $S_2$ in the circuit of Fig.~\ref{qonestatemap}(a), denoted $\mathcal{E}_{S_2| {\rm do}S_1}:  \mathcal{L}(\mathcal{H}_{S_1'}) \to \mathcal{L}(\mathcal{H}_{S_2})$, is defined as the map which is isomorphic (under the identification of $S_1$ and $S_1'$) to the map $\mathcal{E}_{S_2|S_1'}: \mathcal{L}(\mathcal{H}_{S_1'}) \to \mathcal{L}(\mathcal{H}_{S_2})$ in the counterfactual circuit of Fig.~\ref{qonestatemap}(b); that is,
%The evolution map from $S_1$ to $S_2$ in the circuit of Fig.~\ref{qonestatemap}(a), which we will denote by $\mathcal{E}_{S_2| {\rm do}S_1}$, is defined by the input-output functionality for input $S_1'$ and output $S_2$ of the circuit of Fig.~\ref{qonestatemap}(b),
\beq\label{defnquantumdomap}
%\mathcal{E}_{S_2|{\rm do}(S_1)}(\framebox(5,5){}_{S_1}\!) := \mathcal{E}_{S_2|S_1'}(\framebox(5,5){}_{S_1'}\!),
\mathcal{E}_{S_2|{\rm do}(S_1)} := \mathcal{E}_{S_2|S_1'}.
\eeq
This prescription unambiguously leads one to identify the evolution map to be
\beq \label{qdomapik}
\mathcal{E}_{S_2|{\rm do}(S_1)}(\framebox(5,5){}_{S_1}\!)={\rm Tr}_{E'}(\mathcal{U}_{S_2E'|S_1 E} (\framebox(5,5){}_{S_1}\!\otimes \rho_E)),
\eeq
where $ \rho_E :={\rm Tr}_{S_1}(\rho_{S_1E})$. This is the quantum analogue of Eq.~\eqref{onetruemap}. It is always completely positive.
 
The quantum evolution map can be deduced from the identity of the circuit elements in the original circuit of Fig.~\ref{qonestatemap}(a) because it depends only on $\mathcal{U}_{S_2 E'| S_1 E}$ and on $\rho_E$ (which is obtained from $\rho_{S_1 E}$ by taking a partial trace). No input-output relation is needed.

Furthermore, like its classical counterpart, the quantum evolution map is knowledge-dependent insofar as it depends on one's information about the environment, encoded in the quantum state $\rho_E$.\footnote{Some previous works (e.g. Refs.~\cite{stelm,beyond}) sought to define the evolution map as a function of the marginal state of the environment {\em and also} as a function of the initial system-environment correlations present in the initial state of the composite. This prescription is distinct from that of the standard proposal, but it does not correspond to our prescription. In particular, it does not reproduce the correct evolution map (of Eq.~\eqref{qdomapik}) because the latter depends {\em only} on the marginal state of the environment.} 
As we discuss in Appendix~\ref{knowldepend}, this knowledge-dependence of one's description of evolution is ubiquitous in physics, including textbook quantum mechanics.

For completeness,  in Appendix~\ref{mapsinqSA} we determine the correct evolution map for each of the three quantum examples from Section~\ref{sec:standard}. 
%To do so, one simply applies the procedure of hypothetically intervening on system $S_1$ in the relevant circuit. As we show in the Appendix~\ref{mapsinqSA}, 
We also show that the most general circuit considered in the standard proposal, Fig.~\ref{fig:standard}, is described by a distinct evolution map for each valuation $(j,k)$ of $J$ and $K$:
\begin{equation}
\mathcal{E}^{(j,k)}_{S_2|{\rm do}(S_1)}(\framebox(5,5){}_{S_1}\!)={\rm Tr}_{E'}(\mathcal{U}_{S_2E'|S_1 E}(\framebox(5,5){}_{S_1}\!\otimes \rho_E^{(j,k)})).
\end{equation}

\subsection{What is an inference map quantumly?} \label{qinfmap}

Once one takes a causal point of view, the description of quantum evolution is 
scarcely more complicated than the description of classical evolution. 
Devising a complete framework for describing {\em inference} in quantum theory, however, is a much more complicated venture, and remains an open problem~\cite{bayes,Horsman20170395,Leifer2014}. 
(Note that, in keeping with the framework laid out in section \ref{causalpersp}, both evolution and inference are formalized as maps from states of $S_1$ to states of $S_2$, and it is this type of object that we are interested in here. We discuss maps with a more general input type in Section~\ref{causalmaps}.) 
%[COMMENT: I'm not sure what is the best place for this disclaimer, but I do think we should make this clear quite early on in the discussion, so as to minimise misunderstandings on the readers' part.] [Maybe point out already that there are solutions for maps from two copies of $S_1$ to $S_2$?]}

The root of the problem lies in how one acquires the information about $S_1$ that is input into the map. Classically, passive observation of a variable does not change the dependence of that variable on its causal parents. So, there is a way to update one's knowledge of $S_1$, and therefore the distribution $P_{S_1}$, without changing any of the causal mechanisms that relate $S_1$ to the other variables of interest.
In the quantum realm, however, it is unclear whether there is an appropriate analogue of passive observation given that every attempt to gain information about a system changes its state.
%how \red{the state of the system (after the observation)} depends on its causal parents.
 Consequently, it would seem to be impossible to update one's knowledge of $S_1$, and therefore the state $\rho_{S_1}$, without changing any of the causal mechanisms that relate $S_1$ to the other systems of interest. (For example, measuring a quantum system generally leads to a different post-measurement state, altering the way in which the system affects its causal children, while preparing the system in a known state nullifies the influence of its causal parents.)

It is worth noting that, even classically, it is only for a limited set of probing schemes on $S_1$ (which includes `passive' measurements, which are non-disturbing and maximally informative) that the question of what one can infer about $S_2$ (given the outcome of the probing scheme) can be answered by an inference map that takes statistical states on $S_1$ as input. 
%(For general probing schemes, one can consider a different kind of map, as we show in Section~\ref{causalmaps}.)
%can always consider an inference map which acts on statistical states for {\em two copies} of $S_1$, the pre-probing version and the post-probing version.)
Since this is mathematically a limiting case of the full quantum treatment, one can see already that for quantum systems, too, only a {\em limited set} of probing schemes on $S_1$ could possibly admit inferences about $S_2$ that can be expressed by a map with quantum states on $S_1$ as input. Moreover, since quantum states generally contain richer information about a system than their classical limit, the set of scenarios that admit such an inference map is likely to be even more restricted.
%It follows that in the quantum case, it is also only for a {\em limited set} of probing schemes on $S_1$ that the question of what one can infer about $S_2$ could possibly be modeled by an inference map that takes quantum states on $S_1$ as input. 
Nonetheless, it is conceivable that there are {\em intrinsically quantum} scenarios---those that cannot be reduced to an effectively classical description---wherein such an inference map can be defined.  
%{\color} [COMMENT: Katja is wondering whether we actually make two contradictory points here. At the beginning of this paragraph we point out that only a limited set of probing schemes admit inference maps (with states on a single copy of $S_1$ as input) even in the classical case, suggesting that one shouldn't be surprised by such  restrictions in the quantum case. But the main point of this section is that ]}

For {\em generic} probing schemes (both classical and quantum), there is no such map which takes a state of $S_1$ as input. 

\subsection{The mistakes of the standard proposal in the quantum sphere}

The set of quantum states and evolution maps includes the set of classical states and evolution maps as special cases (wherein all operators are diagonal in some fixed basis). The fact that the standard proposal fails to identify the correct evolution map in various classical examples, therefore, implies that it is not a valid prescription for identifying the correct evolution map in the quantum sphere. Furthermore, the mistaken assumptions of the standard proposal that we have identified by considering classical examples remain mistaken assumptions in the quantum sphere. % because of the inclusion of these classical examples. 

%What about intrinsically quantum examples {\color} \textendash{} that is, examples wherein not all operators are diagonal in some fixed basis? Although 
Outside the classical subtheory, the story is more subtle, but analogues of the two mistaken assumptions can be identified. We begin with the quantum analogue of the first mistake. Although we have left open the question of whether one can make sense of the notion of an inference map from $S_1$ to $S_2$ in such examples, it is still clear that, whenever one has initial system-environment correlations, the quantum systems $S_1$ and $S_2$ are causally related not only as cause and effect but by a common cause as well.  As such, an individual input-output pair of states in such a scenario will generally not reflect the cause-effect relation alone.  But given that the correct definition of the evolution map depends {\em only} on the cause-effect relation, there is no reason to think that an individual input-output pair of states is a constraint on the evolution map when there are initial system-environment correlations. 

Indeed, as we show in Appendix~\ref{coincide2}, the necessary and sufficient condition for an individual input-output pair to be a constraint on the evolution map is that the joint state on $S_1 E$ factorizes, $\rho_{S_1 E} =\rho_{S_1}\otimes \rho_{E}$.  The ironic conclusion (reached also in our discussion of the classical case)  is that  the only circumstance in which the first mistake of the standard proposal would be innocuous is the case of {\em no} initial system-environment correlations.

If there is (as speculated in the previous section) a subset of intrinsically quantum scenarios where an inference map can be defined, and the scenario in question is presumed to be within this subset, then {\em perhaps} an input-output pair of states for a given value of $JK$ could be taken as a constraint on the inference map.  But it would still be the case that the inference map would vary with $JK$, and consequently it would be inappropriate to consider all of these input-output pairs as constraints on a single map. %, contrary to what is assumed in the standard argument.
This is the second mistake of the standard proposal.
%
% \color{} [I disagree with the following because the question about an inference map is a question about whether there exists a probing scheme relative to which it makes sense.  
% %Could this be said in the section on causal maps?
% ] \st{Generically, the input-output pairs in quantum scenarios are {\em not} constraints on an inference map from (a single copy of) $S_1$ to $S_2$, since there is generally no such inference map (as argued in the previous section). }%Unless such a subset of intrinsically quantum scenarios is discovered, however, the individual input-output pairs in scenarios outside the scope of classical circuits and passive interventions cannot be taken as constraints even on an inference map.}
%\color{}

\section{Discussion and Future Work} \label{disc}
 
We here summarize some of the lessons of our analysis. First, one cannot define an evolution map simply by listing a set of marginal states at its input and at its output, since if there is a common cause acting on the input and the output, then such pairs of states do not constitute constraints on the evolution map. Second, one should not restrict the input domain of the evolution map, as this would violate the counterfactual conception of evolution, embodied for instance in the idea that a law of motion should be autonomous from the initial conditions. Further, such restrictions are not needed to preserve complete positivity. Third, the marginal state of the environment is relevant to the evolution map, while the system-environment correlations are {\em not}. This latter fact is not assumed, but rather is derived from our conception of evolution, and holds true in all possible operational scenarios and for all possible initial states. Fourth, because the marginal state one assigns to the environment describes one's information about the environment, the evolution map one assigns will depend on one's state of knowledge. If a scenario includes a variable that contains information about the environment, then that scenario is associated to a {\em set} of evolution maps rather than a single evolution map, one for each value of the variable.
%  One consequence of this fact is that one must always be careful to note whether a given scenario is described by a single evolution map, or by a set of evolution maps.

We hope that these lessons might be valuable outside the scope of this work.

%{\color{red} [Rob, did you want to add some lessons here? Or, perhaps to the conclusion or elsewhere?]}

\subsection{Composition of evolution maps}

\begin{figure}[htb!]
\centering
\includegraphics[width=0.16\textwidth]{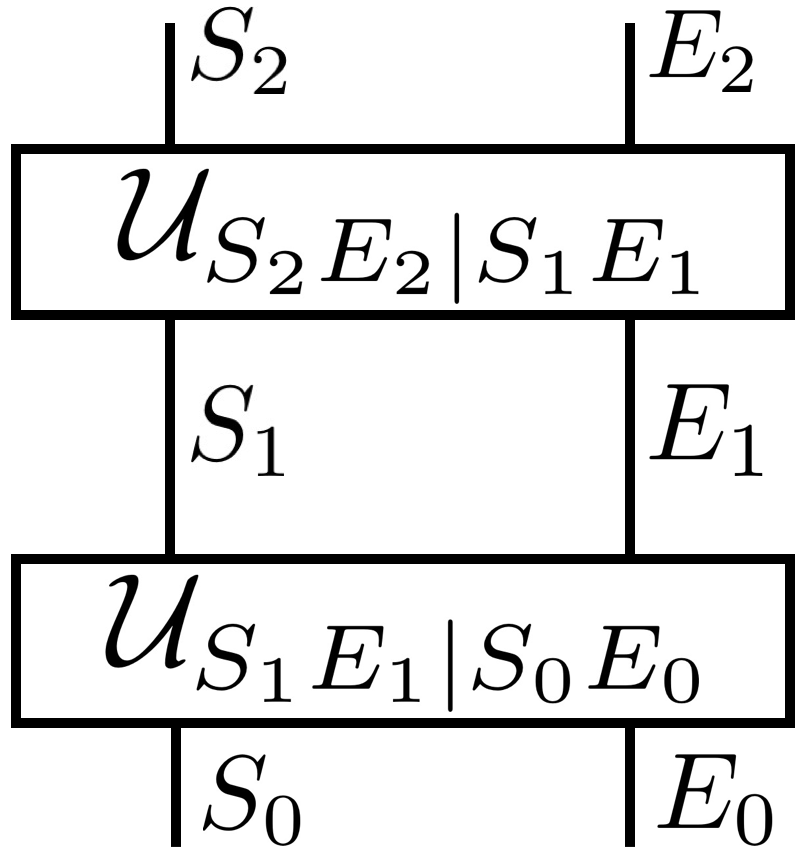}
\caption{Composition of system-environment interactions, leading to non-Markovian evolution on the system.}
\label{composition}
\end{figure}

The definition of the quantum evolution map, 
 %we have provided in 
 Eq.~\eqref{qdomapik}, has the following feature: for a circuit of the form of Fig.~\ref{composition}, the composition of the evolution map from $S_0$ to $S_1$ with the evolution map from $S_1$ to $S_2$ does not generally yield the evolution map from $S_0$ to $S_2$; that is, the compositional property $\mathcal{E}_{S_2|{\rm do}S_0} = \mathcal{E}_{S_2|{\rm do}S_1}\circ \mathcal{E}_{S_1|{\rm do}S_0}$ fails to hold. 

This compositional property only holds if $S_1$ constitutes the {\em complete} causal mediary between $S_0$ and $S_2$. For the circuit in Fig.~\ref{composition}, however, $E_1$ is also such a causal mediary, and so the only sequential decomposition of the evolution map from $S_0$ to $S_2$ that holds is the decomposition into an evolution map from $S_0$ to $S_1 E_1$ and an evolution map from $S_1 E_1$ to $S_2$, that is, $\mathcal{E}_{S_2|{\rm do}S_0} = \mathcal{E}_{S_2|{\rm do}S_1 doE_1}\circ \mathcal{E}_{S_1 E_1|{\rm do}S_0}$. 

It is only with the correct definition of the evolution map that one can properly pose the question of how it can be decomposed into a temporal sequence of evolution maps. Therefore, we expect that our results will have relevance to questions about the divisibility of quantum channels and the detection of nonMarkovianity~\cite{wolf2008dividing,wolf2008assessing}. 

\subsection{
%Quantum combs
Comparison with prior work
} \label{causalmaps}

As mentioned in the introduction, some prior work has also criticized the standard argument for the inadequacy of completely positive maps~\cite{Modi1,Modi4}.
%, and proposed a formalism which avoids the problematic implications of non-completely positive maps. 
%Unlike our work, these works have {\em not} directly disputed the validity of the standard argument that one cannot represent the dynamical evolution of a system in terms of a completely positive map with the usual input space of $\mathcal{H}_{S_1}$ and output space $\mathcal{H}_{S_2}$ ~\footnote{For instance, Ref.~ concurs with the standard argument, stating that: ``The dynamical map is the mapping from the initial states of S to the final states of S, resulting from unitary dynamics of the SE state, [Eqs. 2-4 ...].  ...This means that B may not be a completely positive map when $\chi^SE \ne 0$, nevertheless it fully describes the dynamics of $S$.'' Additionally, PRA 97, 012127 (2018) states that
%"The presence of initial S-E correlations indicates one of the simplest non-Markovian processes: the initial correlations are a record of the past interactions between S and E. In such cases, the CPTP description of the dynamics breaks down. Pechukas has shown that, in order to describe the dynamics in the presence of initial S-E correlations, we must give up something [7,8], e.g., complete positivity or linearity [9].''}
%However, these works {\em have} 
These works have argued against the standard proposal on the grounds that it lacks operational meaning, for instance, by pointing out that one cannot vary the marginal state of the system while keeping the system-environment correlations fixed. 
%They have also pointed out that the precise sort of laboratory intervention that one implements is relevant to the final state of the system. 
In contrast, our criticism of the standard proposal is that (i) it fails to satisfy the criterion of universality, e.g., it only seeks to answer question $Q'$ (concerning Fig.~\ref{fig:standard}) rather than question $Q$ of Section~\ref{standardargument4inadequacyCP} (concerning Fig.~\ref{onestate}), and even for question $Q'$ it sometimes fails to identify {\em any} map as the description of the evolution of the system; and (ii) in the classical limit, it fails to reproduce the evolution map that is implied by the framework of classical causal modeling.
%gives the wrong answer in the classical limit, 
%is not universally applicable (e.g., it fails to answer question Q of Section~\ref{}; in the classical limit, it fails to provide the answer given by the framework of classical causal modeling; it fails to satisfy the criterion of universality insofar as it 
 %and that even for cases to which it is applicable, the standard proposal 
 %sometimes fails to identify {\em any} map as the description of the evolution of the system.}

%Based on these operational criticisms of the meaningfulness of the dynamical map in the standard proposal,
This prior work, e.g. Refs.~\cite{ModiChar,Modi2,Modi1,Modi4,Modi3,ModiChir}, has also proposed a new framework for analyzing dynamics in the presence of system-environment correlations,
% was proposed, 
where one focuses on an altogether different type of map that {\em does} have a clear operational meaning and which is always completely positive.
%, but which has a different input space.
% than a usual dynamical map. 
 Specifically, whereas maps purporting to describe the dynamics of system $S$ typically take states on $S_1$ as input and have states on $S_2$ as output, 
  %have input space $\mathcal{H}_{S_1}$ and output space $\mathcal{H}_{S_2}$, 
  these new maps take instruments from $S_1$ to $S'_1$ as input and have states on $S_2$ as output.  In other words, the domain of the map is the space of operators on {\em two} rather than one copy of the system's Hilbert space, that is, on $\mathcal{H}_{S_1}\otimes \mathcal{H}_{S'_1}$ rather than on $\mathcal{H}_{S_1}$.
  % whereas the domain 
  %(equivalently, they take an operator on {\em two} copies of the system's Hilbert space).
% that is, it has input space H_S1 \otimes H?_S1 and output space H_S2.

\begin{figure}[htb!]
\centering
\includegraphics[width=0.5\textwidth]{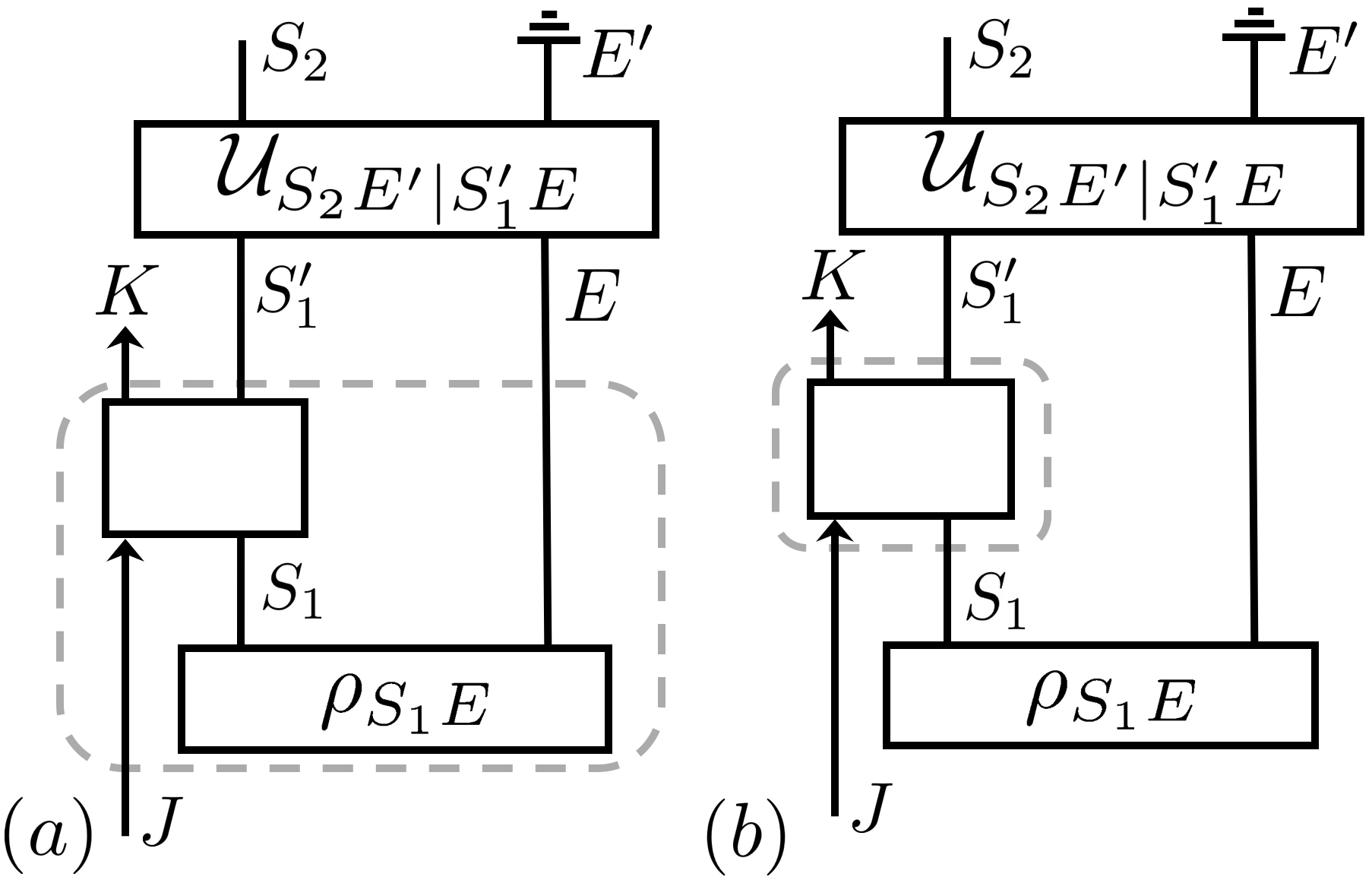}
\caption{(a) The subset of circuits considered in the standard proposal wherein the setting variable $J$ determines the measurement on the system and one post-selects on the measurement outcome $K$. (b) The same operational scenario as in (a), but reconceptualized as an intervention with setting $J$ and outcome $K$ on the system, rather than a preparation of a set of states labeled by $J$ and $K$.}
\label{Modi}
\end{figure}

One can motivate this new perspective by considering the circuit in Fig.~\ref{Modi}(a), which is clearly a special case of Fig.~\ref{fig:standard}, and reconceptualizing the experiment not as a preparation of a set of initial joint states on $S'_1 E$, but rather as an interventional probing scheme on the system. 
This idea is represented in Fig.~\ref{Modi}(b), where the intervention on the system is denoted inside the dashed box, and is associated for each value $k$ of $K$ and $j$ of $J$ with a map $\mathcal{E}^{(j,k)}_{S'_1|S_1}: \mathcal{L}(\mathcal{H}_{S_1})\to\mathcal{L}(\mathcal{H}_{S'_1})$, and where the circuit fragment outside the dashed box is associated with a map $\mathcal{E}_{S_2S_1|S_1'}: \mathcal{L}(\mathcal{H}_{S_1'})\to\mathcal{L}(\mathcal{H}_{S_1}\otimes\mathcal{H}_{S_2})$. The circuit fragment defines a map from each element of the instrument describing the intervention, $\mathcal{E}^{(j,k)}_{S'_1|S_1}$, to a state $\rho^{(j,k)}_{S_2}$ using the link product of Ref.~\cite{splitnode3}, $\mathcal{E}_{S_2S_1|S_1'} \star \mathcal{E}^{(j,k)}_{S'_1|S_1}=\rho^{(j,k)}_{S_2}$, where the $S_1$ output of the circuit fragment is fed into the $S_1$ input of the instrument, and the $S'_1$ output of the instrument is fed into the $S'_1$ input of the circuit fragment.
%This idea is represented in Fig.~\ref{Modi}(b), where the intervention on the system is denoted inside the dashed box, and the circuit fragment outside the dashed box is associated with a map $\mathcal{E}_{S_2S_1|S_1'}: \mathcal{L}(\mathcal{H}_{S_1'})\to\mathcal{L}(\mathcal{H}_{S_1}\otimes\mathcal{H}_{S_2})$. 
%{\color{red}[Note that I cut the explicit form of this map in terms of circuit elements. It was problematic to introduce, and was never used anyway.]}
%, namely
%\beq
%\mathcal{E}_{S_2 S_1|S'_1}(\framebox(5,5){}_{S'_1|S_1}) := {\rm Tr}_{E'} \circ \mathcal{E}_{S_2 E'| S_1 E} (\framebox(5,5){}_{S'_1|S_1} \rho_{S_1 E} ).
%\eeq
%\beq
%\mathcal{E}_{S_2 S_1|S'_1}(\framebox(5,5){}_{S'_1|S_1}) := {\rm Tr}_{E'} \circ \mathcal{U}_{S_2 E'|S_1E} \circ (\framebox(5,5){}_{S'_1|S_1} \otimes {\rm id}_E)(\rho_{S_1 E} ).
%\eeq
%{\color{red} What is going on in this equation?}
This sort of map has been studied in many formalisms~\cite{Katja1,splitnode1,splitnode2,splitnode3,splitnode4,splitnode5,splitnode6,splitnode7,Modi1} and is often termed a {\em quantum comb}~\cite{splitnode3} or a {\em process matrix}~\cite{splitnode4}. When it was introduced for the study of open quantum systems in Ref.~\cite{Modi1}, this object was termed the $M$-map.

As we understand it, the formalism of $M$-maps differs from our own framework in two key respects.

Firstly, the $M$-map and our evolution map seem to have been proposed as answers to different questions: the $M$-map was introduced to describe aspects of the system's dynamics in an experiment of the form of Fig.~\ref{Modi}(b), where an intervention is actually made on the system, whereas our proposal is intended to describe the evolution of the system in the natural experiment of Fig.~\ref{onestate}.  In other words, whereas we aim to answer question $Q$ from Sec.~\ref{standardargument4inadequacyCP} (concerning Fig.~\ref{onestate}), the $M$-map is proposed as an answer to a question of the form of $Q'$, but with the experiment given by Fig.~\ref{Modi}(b) (rather than Fig.~\ref{fig:standard}). 
%  As discussed above, such an experimental scenario is a special case of the circuit of Fig.~\ref{}, and hence the $M$-map is meant to address question $Q'$ from Section~\ref{}, rather than question $Q$. In contrast, we have argued that question $Q$ is the more interesting question, and is the question addressed by the evolution map we have introduced. 

Secondly, we have defined the evolution map to be a complete description of the inferences one can make from $S_1$ to $S_2$ based purely on the cause-effect relation between the two, while the $M$-map does not capture only those types of inferences, but, rather includes (as noted in, e.g., Ref.~\cite{Modi1}) information about the initial correlations between the system and environment, and hence describes one's inferences based on both the cause-effect and on the common-cause relation between $S_1$ and $S_2$.
%the parameters in the $M$-map do not describe all and only those aspects of inference which are based on the cause-effect relation between $S_1$ and $S_2$, as we have argued that an evolution map should do. 
%As noted in Refs.~\cite{}, the $M$-map also contains information about the initial correlations between the system and environment, and hence it also encodes information about inferences through a common cause. 
One consequence of this distinction is that our notion of the quantum evolution map reduces in the classical limit (where all operators are diagonal in some fixed basis) to the classical evolution map, which we identify with Pearl's do-conditional, whereas the $M$-map does not reduce to the do-conditional in the classical limit.
%In particular, if one considers the $M$-map in the classical limit, it does not reduce to the do-conditional, which we have argued (as Pearl has also extensively argued) is the correct description of classical evolution.
Indeed, from our perspective, the very fact that the $M$-map takes an instrument from $S_1$ to $S_1'$ as input rather than a state on $S_1$ precludes it from being a candidate for the evolution map.
 %direct description of evolution, 
% since we have conceived of evolution as a map taking states of the system at an early time to states of a system at a late time. In contrast, the $M$-map is operationally motivated, and instead describes how laboratory interventions at an early time lead to states of the system at a later time. (say more? Put this elsewhere?)

%In the same spirit as the M-map, 
Ref.~\cite{Mprep2} 
%(what about Mprep1 and Mrole)... these are basically repeat articles, published later, so no point including them
also asks a question of type $Q'$; namely, it asks what map one should use to describe scenarios in which certain special types of interventions are performed on the system. 
%{\color{blue}[or, system-environment composite? Why do they say that the state of E depends on the pin map? Confused...]}. 
One such set of interventions considered in Ref.~\cite{Mprep2} and called `stochastic preparations' consist of a trace operation on $S_1$ and a repreparation of $S'_1$ in one of an informationally complete set of states. 
%In the particular case wherein a trace operation is performed on $S_1$ and where $S'_1$ is reprepared in one of an informationally complete set of states, 
For such a set of interventions, the map that is tomographically reconstructed from data is
\beq \label{extractE}
\mathcal{E}_{S_2|S'_1}(\framebox(5,5){}_{S'_1}) = {\rm Tr}_{S_1} \circ \mathcal{E}_{S_2 S_1|S'_1}(\framebox(5,5){}_{S'_1}).
\eeq
Based on the notation used for this object in Ref.~\cite{Modi1}, we call it the $L$-map.  Unlike the $M$-map, it is identical in form to the evolution map we have introduced in Eq.~\eqref{qdomapik}. However, just as we noted in our discussion of the M-map, we read this prior work as asserting that the L-map describes the dynamics only for a very limited scope of experiments, namely, those wherein a `stochastic preparation' was actually implemented and not that it does so for the the general case involving no intervention, depicted in Fig.~\ref{onestate}.

This prior work appears to be motivated by a type of operationalism wherein an entity is only meaningful insofar as one can specify the experimental procedure that would allow one to measure it.  In particular, it defines both the $M$-map and the $L$-map in terms of the tomographic procedure by which one could identify it. 
%come to deduce it when the circuit components are unknown.  
By contrast, the causal modelling perspective that we adopt in this work is not wedded to this positivist notion of meaningfulness and therefore aims to answer question $Q$ directly, without recourse to interventions, but rather through the use of counterfactuals. 

Note, however, that if one adopts the positivist attitude merely as a means for discovering the form of the $L$-map, but one then kicks away this empiricist ladder and simply treats the $L$-map as the correct description of dynamics even in the case of Fig.~\ref{onestate} (where there are no interventions on which any operational definitions could be based), then one comes to the same conclusion  regarding the correct definition of the evolution map as we have reached here by adopting the causal modelling perspective.
 
%imagines the interventions that define the $L$-map as mere counterfactuals, 

%In Refs.~\cite{}, a special case of the $M$-map was considered which {\em does} describe all and only those aspects of inference from $S_1$ to $S_2$ that is  based on the cause-effect relation between them. This map, termed the $L$-map, is identical in form to the evolution map we have introduced in Eq.~\eqref{}.  However, like the M-map, we read this prior work as asserting that the L-map describes the dynamics only for a limited scope of experiments, namely, those wherein a `stochastic preparation' was implemented.~\footnote{E.g., Ref.~\cite{} states that... INSERT QUOTE} 
%The key innovation of our work is to realize that 

%Note, however, that if one were to reconceptualize the aforementioned stochastic preparation not as an actual laboratory intervention, but rather as a hypothetical introduced merely as a handle on the scenario of Fig.~\ref{onestate}, then (in our terminology) the $L$-map would be the correct description of evolution in general scenarios. 

%Similarly, one might hope to 
Similarly, one might appeal to such a ladder-kicking procedure to justify considering the $M$-map as pertinent not just to the class of experiments shown in Fig.~\ref{Modi}(b), but also to the general case of Fig.~\ref{onestate}. 
%To do so, one would need to argue that the interventions in the dashed box of Fig.~\ref{}(b) need not be physically realized, but rather that they were simply counterfactual possibilities  introduced (again) as a handle on the scenario of Fig.~\ref{onestate}. 
%Strictly speaking, the $M$-map, 
Even reconceptualized in this way, however, the $M$-map is not an evolution map in our sense, for the reasons described above.  
So what does the map $\mathcal{E}_{S_2 S_1|S'_1}$ describe from the causal modelling perspective? We address this question in the next section.

\subsection{The status of the quantum comb 
%$\mathcal{E}_{S_2 S_1|S'_1}$ 
from the causal point of view}

Viewed as a description of the general case of Fig.~\ref{onestate}, Eq.~\eqref{extractE} shows how the map $\mathcal{E}_{S_2 S_1|S'_1}$ can be used to recover the evolution map, which describes inferences which can be made solely on the basis of the cause-effect relation between $S_1$ and $S_2$. Additionally, $\mathcal{E}_{S_2 S_1|S'_1}$ can be used to recover an inference map which describes inferences which can be made solely on the basis of the common cause relation between $S_1$ and $S_2$.
%The map $\mathcal{E}_{S_2 S_1|S_1}$ can also be used to define an object that specifies what inferences can be made from $S_1$ to $S_2$ based only on the {\em common cause} relation that holds between them (whereas the evolution map specifies what inferences can be made based only on the {\em cause-effect} relation that holds between them).  
In order to be assured that one's inferences about $S_2$ are based purely on knowledge of $S_1$, and not on knowledge of $S_1'$, one presumes a state of complete ignorance regarding $S_1'$, represented by the completely mixed state $\rho_{S_1'}= \tfrac{1}{d_{S_1'}}\mathbb{1}_{S_1'}$, where $d_{S_1'}$ is the dimension of $\mathcal{H}_{S_1'}$.  The object that allows one to make inferences based solely on the common cause relation, therefore, is the joint state
 \beq
 \rho_{S_1 S_2}  := \mathcal{E}_{S_2S_1|S_1'}(\tfrac{1}{d_{S_1'}}\mathbb{1}_{S_1'}),
 \eeq
In particular, if one implements a measurement on $S_1$ and one obtains an outcome $X=x$ associated to a positive operator $E^{(x)}_{S_1}$, then one should update one's description of $S_2$ to 
\beq
\rho^{(x)}_{S_2} = \mathcal{N} {\rm Tr}_{S_1} (\rho_{S_1 S_2}(E^{(x)}_{S_1}\otimes \mathbb{1}_{S_2})),
\eeq 
where $\mathcal{N}$ is a normalization constant.  This defines a map from effects on $S_1$ to states on $S_2$ which has been termed the {\em steering map} and discussed in \cite{Katja2}.  It is linear and co-CP (that is, the composition of the map with the tranpose operation is completely positive).
 
 Based on these facts, one might hope that the map $\mathcal{E}_{S_2 S_1|S'_1}$ simply represents the elusive quantum inference map that we discussed in Section~\ref{qinfmap}. However, it does not.
 
Here, one must again distinguish between two sorts of maps that might describe such inferences---one whose inputs are operators on a single copy of $S_1$, and one whose inputs are operators on two copies of $S_1$.  It is the possibility of the former that we discussed in Section~\ref{qinfmap}, whereas the map $\mathcal{E}_{S_2 S_1|S'_1}$, if it is to be interpreted as an inference map, is of the latter variety. 
 Even if it turns out that the only sort of inference map which can sensibly be defined in generic quantum scenarios has as input a state on $\mathcal{H}_{S_1} \otimes \mathcal{H}_{S'_1}$~\footnote{Which is a reasonable conjecture: even in the classical sphere, disturbing interventions are most naturally described by an inference map from states of knowledge on {\em two} copies of random variable $S_1$ to $S_2$.}, the natural inference map would still be of the form $\mathcal{E}_{S_2| S_1S'_1}$, rather than $\mathcal{E}_{S_2 S_1|S'_1}$. Defining such a map likely requires a theory of quantum Bayesian inversion (e.g. to determine the correct input state on $\mathcal{H}_{S_1} \otimes \mathcal{H}_{S'_1}$ from a specification of an instrument from $S_1$ to $S'_1$).

\subsection{Experimentally determining quantum combs and evolution maps} \label{practical}
 
Our focus herein has been on elucidating the correct {\em definition} of the evolution map when the circuit is completely specified.
%of the system when the circuit in which the system is embedded is specified. 
A secondary problem concerns deducing the evolution map from experimental data. (It is secondary insofar as it can only be addressed %answered 
once one has the correct definition of the evolution map).
%A secondary question---which one could only hope to answer once one already knows how to correctly define the evolution map---concerns the problem of {\em deducing} the evolution map from experimental data. 

It is clear that if one can experimentally characterize the different components of the circuit (the state $\rho_{S_1 E}$ and the unitary map $\mathcal{U}_{S_2 E'|S_1E}$), then one can deduce the evolution map from Eq.~\eqref{qdomapik}. Although this is sufficient, it is not necessary for characterizing the evolution map. The question arises, therefore, of what is the {\em minimal} experimental effort that suffices.
%procedure would suffice.

To begin with, consider the idealization wherein one can perfectly implement any desired laboratory operation. 
%
%\noindent{\bf Characterizing the causal map and the evolution map via tomography}
%{\bf Experimental determination of the causal map}
%
The causal map $\mathcal{E}_{S_2S_1|S_1'}$ has input space $\mathcal{L}(\mathcal{H}_{S_1'})$ and output space $\mathcal{L}(\mathcal{H}_{S_1}\otimes\mathcal{H}_{S_2})$ 
%\red{[Or do we want $S_1$ as an output...]}. 
Hence, if one intervenes on the system by performing an informationally complete measurement on $S_1$ and then repreparing one of an informationally complete set of states on $S_1'$, and 
%for each such intervention one
one also implements
 (for each state on $S_1'$)
  an informationally complete measurement on $S_2$, then one can tomographically reconstruct the causal map. 
This was termed {\em causal tomography} in Ref.~\cite{Katja2}.
%Ried et al.
%[...] Need informational completeness for operator space of $S_1 S_1'$. 

%\noindent{\bf Experimental determination of the evolution map}

If one has experimentally determined the causal map, then the evolution map can be computed directly from it via Eq.~\eqref{extractE}.
%, as mentioned above. 
A less demanding experimental method, however, is to simply ignore 
%to trace over 
$S_1$ (rather than measuring it) and prepare $S_1'$ in one of an informationally complete set of states. For each such state, one then implements an informationally complete measurement on $S_2$. This achieves process tomography on the evolution map. Note that this corresponds to experimentally implementing the intervention that is contemplated in the counterfactual circuit that defines the evolution map. It is analogous to measuring a classical do-conditional by implementing a randomized trial.
% ,  defined in Eq.~\eqref{qdomapik}.

%The following experiment is one which yields sufficient information to achieve the reconstruction. Intervene on $S_1$ by tracing over it and repreparing one of an informationally complete set of states. For each such state, implement an informationally complete measurement on $S_2$. This achieves process tomography on the map defined in Eq.~\eqref{qdomapik}. An intervention on $S_1$ necessitates that we split $S_1$ into a pair of systems, the one that is the input to the intervention and the one that is its output. Each intervention has a predictive aspect, which concerns the version of $S_1$ that is the output of the intervention, and a retrodictive aspect, which concerns the version of $S_1$ at its input. Critically, to characterize the evolution map from $S_1$ to $S_2$, only the predictive aspect of the intervention on $S_1$ is relevant. The retrodictive aspect can only teach one about the version of $S_1$ at the input of the intervention, and although such information can support inferences about $S_2$, they are inferences that are mediated by the common cause of $S_1$ and $E$. The evolution map, on the other hand, concerns {\em only} the inferences from $S_1$ to $S_2$ that are based on the cause-effect relation that holds between them. 

%\noindent{\bf Experimentally determining causal maps and evolution maps when laboratory operations are imperfect}

%A final important question is that of 
What about experimentally determining the causal map or evolution map when the experimenter does {\em not} have ideal laboratory operations?
%---which one could only hope to answer once one already knows how to experimentally determine  the evolution map in ideal scenarios---
For example, what if they are not able to localize their operations to particular systems, or to  characterize their laboratory operations accurately?
%they may not be able to localize their operations to particular systems, or they may not be able to  characterize their laboratory operations accurately. 
Because such limitations can result in inadvertently preparing initial system-environment correlations, they have previously served as a motivation for this field.
%the study of the evolution of systems initially correlated with their environments.    
Specifically, it has been suggested that if no completely positive map fits the relation of the standard proposal, this should be taken as evidence that one has inadvertently introduced system-environment correlations~\cite{Cory,Mrole}.%~\footnote{\color{red}Note that our analysis challenges the logic of any such diagnostic which is based on the assumptions of the standard argument.}.

% when an experimenter is targetting 
%(Both of these have been motivations for work in this field.)

Given our demonstration that the prescription of the standard proposal does not yield the evolution map, we advocate against analyzing experimental data in the manner it proposes, even if only as a diagnostic for imperfections in one's laboratory operations.   The solution to the problem of imperfect laboratory operations is to use a form of tomography in which various features of the probing scheme (such as the identities of the laboratory operations and the dimensionality and nature of the systems being probed) are not {\em presupposed}, but are rather taken to be hypotheses whose plausibility is assessed on the basis of the data. 
% which are evaluated on the 
%but rather extracts all these from the analysis of the data. 
In identifying the evolution map that provides the best fit to the experimental data, the maps over which one varies must be {\em constrained} to be completely positive  (by virtue of the definition of an evolution map),
%.  Because every such fit makes assumptions about the probing scheme (e.g., the nature and dimension of the system being probed and possibly also the identity of the quantum measurements and preparations appearing therein), 
and if the quality of the fit is poor (as is indicated, for instance, by a bad $p$-value), then the correct reaction is not to entertain the possibility that completely positive maps are inadequate for describing evolution, as has previously been suggested~\cite{Cory,Mrole}, but to reject one or more of the assumptions about features of the probing scheme. 
A demonstration of how to implement tomography in this way is provided in Ref.~\cite{bootstrap}.

%\noindent{\bf Deducing an evolution map by passive observation}

Finally, we note that in classical scenarios, it is sometimes possible to identify the evolution map from purely passive observations if these are made on the right systems. (The more general problem of identifying the do-conditional for some pair of variables embedded in a given causal structure is known as 
%This well-understood problem is known as 
the ``identifiability problem''~\cite{Pearl}.)  The significance of this fact is that {\em although evolution maps are defined in terms of hypothetical nonpassive interventions, these nonpassive interventions need not necessarily be performed in order to identify the map.} 
%This fact is critical to classical causal modeling. 
Whether there is a sensible quantum analogue of the identifiability problem and whether it is possible to sometimes deduce the quantum evolution map without nonpassive interventions remains unclear.  In particular, to even pose the question, one must explore whether there is anything in quantum theory that ought to be considered an analogue of passive observation.
%it is not clear whether there are  {\em any} interventions in quantum theory that should be considered a kind of passive observation. 
%anything in quantum theo
%However, it is unclear what role it plays in quantum scenarios, where there is no perfect analogue of passive observation.

\section{Conclusion}

We have argued that, for a general circuit wherein a system interacts with its environment such as the one in Fig.~\ref{onestate}, the evolution map on the system should be defined as in Eq.~\eqref{qdomapik}. This map is always linear and completely positive, {\em even} in the presence of arbitrary initial system-environment correlations. %Hence, we have shown that the description of quantum evolution in terms of such maps 
Thus, we have shown that the common wisdom that such initial correlations constitute an exception to the rule of complete positivity is mistaken. 

Our results provide good reasons to abandon a host of questions that have previously been the primary focus of research in this field. There is no reason to find necessary or sufficient conditions 
%on the circuit elements
%(e.g., on a fixed system-environment interaction together with a family of initial states) 
under which one can find a completely positive map (or a linear map, a positive map, etc.) that is consistent with the input-output relation of the standard proposal, because this relation has no bearing on the evolution map. There is also no reason to worry about the physical meaning of evolution maps that are not completely positive or not linear, because these never arise.

Instead, we advocate for pursuing a new set of questions, inspired by the drive to generalize the do-calculus of Ref.~\cite{Pearl} to the quantum realm~\cite{QCM2,QCM1}. Are there circumstances in which one can define an intrinsically quantum inference map? 
%{\color{red} [COMMENT: What do we mean by that?]}
%How are inference maps defined in quantum theory? 
%Is there an analogue of the identifiability problem in quantum theory? When can one sometimes identify the true evolution map even without implementing the hypothetical intervention invoked in its definition?
Can one sometimes identify the evolution map even without implementing the hypothetical intervention invoked in its definition?
 What are the implications of our results for continuous-time dynamics of open quantum systems, and in particular, nonMarkovian dynamics?

%\st{Just as our analysis was general enough to apply to both classical and quantum theory, so too does it apply to generalized probabilistic theories}~\cite{Lucien}, \st{and therefore one can ask the same sorts of questions about how to generalize concepts from the framework of classical causal modelling to such theories. }
%. Can one answer these questions in post-quantum theories?

%-Some prior work in the literature gives the impression that one could hope to (and should) define an evolution map directly from the input-output relation $\mathfrak{R}$, without specifying any further details~\cite{assign1,sud,afraid,lidar,standard1,stelm,beyond,Kraus} (e.g., the quantum circuit under consideration).

\textit{Acknowledgment}---We acknowledge useful comments by K. Modi, F. Costa and F. Buscemi on a draft of this article, in particular, concerning how our results relate to previous work.
% (though they may not agree with our characterization).
 Of course, any inaccuracies in our account of this relation are entirely our fault.  
  D.S. is supported by a Mike and Ophelia Lazaridis Fellowship. K.R. is partially supported by the Austrian Science Fund (FWF) through the SFB FoQuS F4012. This research was also supported by a Discovery grant of the Natural Sciences and Engineering Research Council of Canada and by Perimeter Institute for Theoretical Physics. Research at Perimeter Institute is supported by the Government of Canada through the Department of Innovation, Science and Economic Development Canada and by the Province of Ontario through the Ministry of Research, Innovation and Science. 

\appendix
\addcontentsline{toc}{section}{ \quad Appendices}

\section{Generic families of quantum examples in which the standard proposal fails} \label{qex}

We here illustrate the breadth of the pathologies of the standard proposal, and show that there is nothing special about the examples we have given. 

In Section~\ref{genericrep}, we argued that there exist operationally realizable relations $\mathfrak{R} = \{ (\sigma_j, \tau_j)\}_j$ for which %where for quantum states $\sigma_j \in \mathcal{L}(\mathcal{H}_{S_1})$ and $\tau_j \in \mathcal{L}(\mathcal{H}_{S_2})$, 
\begin{enumerate}
\item there is no completely positive map $\mathcal{E}$ such that $\forall j:\mathcal{E}(\sigma_j) = \tau_j. $ 
\item there is no linear map $\mathcal{E}$ such that $\forall j:\mathcal{E}(\sigma_j) = \tau_j. $ 
\item there is no map $\mathcal{E}$ such that $\forall j:\mathcal{E}(\sigma_j) = \tau_j. $ 
\end{enumerate}
For notational simplicity, we have dropped the system labels on the quantum states in the relation; throughout, note that $\sigma_j \in \mathcal{L}(\mathcal{H}_{S_1})$ and $\tau_j \in \mathcal{L}(\mathcal{H}_{S_2})$.

We further asserted that all three of these failures can arise in each of the three operational circuit types described in Section~\ref{genericrep}; namely, circuits where the initial state of the system is varied by 
{\renewcommand{\labelenumi}{\alph{enumi}.}
\begin{enumerate}
\item the choice of transformation on the system-environment composite.\\
\item the choice of measurement on the system (and the choice of post-selection on its outcome).\\
\item the choice of transformation on the system alone.
\end{enumerate}}
All of these are, of course, special cases of the general circuit in Fig.~\ref{fig:standard}. We have highlighted these three specific circuit types because they cover the range of operational scenarios considered in the literature. Roughly, scenarios of type (a) are considered in Refs.~\cite{Modi2,ModiChar,Modi1,Modi3,Mprep}; of type (b) in Refs.~\cite{Mprep,Alicki,assign1,stelm,brodutch}, and of type (c) in Refs.~\cite{Modi2,ModiChar,Modi1,Modi3,Mprep}. Furthermore, {\em all} examples in the literature, to our knowledge, fit within one of these three categories.

We now provide these nine families of examples (one for each pairing of $1,2,3$ and $a,b,c$), in order to cover the full diversity of examples in the literature and to demonstrate explicitly that one cannot salvage the standard proposal simply by restricting to one of the families of operational circuits ($a,b,c$).% the more refined {\em causal} conditions we found in Section~\ref{SAworks} are really required.

For simplicity of presentation, we represent the system-environment interaction in these examples by a general quantum channel $\mathcal{F}_{S_2|S_1E}$ rather than a unitary channel $\mathcal{U}_{S_2E'|S_1E}$. (Of course, any such quantum channel could be dilated to recover a unitary description of the same example, although this dilation might require an increase in the dimensionality of the environment, to accommodate the ancilla required to achieve the dilation.)

Examples of types $1,2,3$ will be given, respectively, by providing an operationally realizable relation such that 
\begin{enumerate}
\item the set $\{ \sigma_j \}_j$ is informationally complete, and $\forall j: \tau_j := T(\sigma_j)$, where $T$ denotes the transpose map (relative to some basis), so that $\mathcal{E}$ is this transpose map and consequently is not completely positive. 
\item the $\{ \tau_j \}_j$ are more distinguishable\footnote{That is, if there exists a pair $(\tau_j,\tau_k)$ which has a smaller fidelity than the corresponding pair $(\sigma_j,\sigma_k)$.} than the $\{ \sigma_j\}_j$, so that $\mathcal{E}$ is necessarily a nonlinear map. 
\item the $\sigma_j$ are independent of $j$ (so $\forall j: \sigma_j :=\bar{\sigma}$ for some fixed $\bar{\sigma}$) but $\tau_j$ varies nontrivially with $j$, so that the relation is one-to-many, and there can be no map $\mathcal{E}$ consistent with it.
\end{enumerate}

%\subsubsection{Setting variable affects the system and the environment} \label{sec:SEq}
\subsubsection{Operational scenarios in which the setting variable $J$ affects the system and the environment and $K$ is trivial} \label{sec:SEq}

Here we provide quantum examples of types $(a,1)$, $(a,2)$, and $(a,3)$.

If the setting variable $J$ influences both the principal system and the environment, one can prepare any initial joint state on the system and environment. Then, one can easily generate any relation $\mathfrak{R}$ {\em at all}, including all three types from above. The following arguments can be seen as a generalization of arguments originating with Pechukas~\cite{Pechukas}.

For example, taking the environment system to have dimension equal to the cardinality of the set $\{ \sigma_j \}_j$, then for any set $\{ \sigma_j \}_j$ one can prepare the system-environment joint state $\sigma_j \otimes |j\rangle \langle j|$ when $J=j$. The system and environment may then interact via the controlled channel $\mathcal{F}_{S_2|S_1E}(\framebox(5,5){}_{S_1 E}\!)= \sum_j \langle j| ({\rm Tr}_{S_1}(\framebox(5,5){}_{S_1 E}\!)) |j\rangle_E (\tau_j)_{S_2}$, for any set of states $\{ \tau_j \}_j$ on $S_2$. Hence, for $J=j$ the final marginal state of the system is $\tau_j$. Since both sets $\{ \sigma_j \}_j$ and $\{ \tau_j \}_j$ are completely unconstrained, one can certainly satisfy any of the conditions articulated in 1, 2, and 3. 
%, then the relation achieved by this physical scenario is $\mathfrak{R} = \{ (\sigma_j, \tau_j) \}_j$ where $\tau_j = \mathcal{U}_{S_2|S_1}^{(j)}(\sigma_j)$.
%Because each state in the relation's domain is transformed via a distinct channel $\mathcal{U}_{S_2|S_1}^{(j)}$, one can trivially tailor the set of these channels such that each state in the domain is mapped to an arbitrary state, including those satisfying the conditions of 1, 2, and 3. 

%\subsubsection{Setting variable determines the measurement on the system, and the measurement outcome is post-selected upon} 
\subsubsection{Operational scenarios in which the setting variable $J$ determines the measurement on the system and one post-selects on the measurement outcome $K$} \label{sec:PSq}

Here we provide quantum examples of types $(b,1)$, $(b,2)$, and $(b,3)$.

Suppose that the initial state of the system-environment composite is a maximally entangled state. By the Hughston-Josza-Wootters theorem~\cite{HJW}, it is possible, by implementing a $j$-dependent measurement on the system and post-selecting on outcome $k$, to steer the environment to any arbitrary state $\tau_{j,k}$\footnote{ \label{footsteer} Note that for a given choice $j$ of measurement, the ensemble to which one steers is not entirely arbitrary: the weighted average of the states in the ensemble is fixed.
However, we can avoid this restriction by post-selecting on a particular outcome for each choice of measurement. }. 
Further, the update rule of the measurement (which affects only the system) can be arbitrary. Given an outcome $k$ of the $j$th measurement, one can find an update rule which ignores the state of the system and simply reprepares it in the state $\sigma_{j,k}$. If the subsequent system-environment interaction is a swap gate, one can generate the input-output relation $\mathfrak{R} = \{ (\sigma_{j,k}, \tau_{j,k})\}_{j,k}$ for completely unconstrained sets $\{ \sigma_{j,k} \}_{j,k}$ and $\{ \tau_{j,k} \}_{j,k}$. Hence, one can certainly satisfy the conditions of 1, 2, and 3 (where $j$ is replaced with $j,k$). 
%Hence, we can again generate examples of each of the three types, even when the setting variable has a causal influence on the system alone.

%\subsubsection{Setting variable affects only the system}
\subsubsection{Operational scenarios in which the setting variable $J$ affects only the system and $K$ is trivial} \label{QSysAloneNopost-selection}

Here we provide quantum examples of types $(c,1)$, $(c,2)$, and $(c,3)$.

Suppose that one of four possible states are prepared by implementing a transformation on the system, as follows. One applies a `preparation' channel $\mathcal{G}_j \otimes \rm{id}_E$, for $j \in \{1,2,3,4\}$ and for $\mathcal{G}_j: \mathcal{L}(\mathcal{H}_{S_0})\to \mathcal{L}(\mathcal{H}_{S_1})$ , to the fiducial state $|\Phi^+\rangle_{S_0E} := 2^{-1/2} ( |0\rangle_{S_0} |0\rangle_E + |1\rangle_{S_0} |1\rangle_E)$, where 
\beq
\mathcal{G}_j(\framebox(5,5){}_{S_0}\!)= X_j (\framebox(5,5){}_{S_0}\!) X_j^{\dagger} 
\eeq
and where $\{ X_j \}_j $ is the standard set of four Pauli matrices. In this case, there are four possible system-environment states, corresponding to the elements of the Bell basis, 
\begin{align}
(\rho_j)_{S_1E} = (X_j \otimes \mathbb{1}) |\Phi^+ \rangle \langle \Phi^+ |_{S_0 E} (X^{\dag}_j \otimes \mathbb{1}).
\end{align}
Each of these has the same marginal on $S_1$---the completely mixed state---so
\begin{align}
\forall j: \quad \sigma_j = \frac{1}{2} \mathbb{1}.
\end{align}

Now suppose that the system-environment interaction consists of a measurement of the Bell basis on the system-environment composite, and then a repreparation of the system in a state depending on the outcome that was obtained,
\beq \label{inter}
\mathcal{F}_{S_2|S_1\! E}(\framebox(5,5){}_{S_1\!E}\!)\!=\!\!\sum_j\! \bra{\phi^+}\! (X_j \otimes \mathbf{1}) (\framebox(5,5){}_{S_1 \!E}\!) (X_j^{\dagger}\otimes \mathbf{1})\!\ket{\phi^+}_{S_1 \!E}\! (\tau_j)_{S_2},
\eeq
where $\{ \tau_j\}_j$ denotes an arbitrary set of four states on $S_2$. 

The input-output relation in this case, therefore, is easily verified to be
\beq
\mathfrak{R} = \{ (\frac{1}{2} \mathbb{1}, \tau_j) \}_{j}.
\eeq
If the $\tau_j$ depend nontrivially on $j$, then the relation is one-to-many, so we have an example of type 3.

We can modify this example slightly to get one of type 2.
Simply let the initial preparation channel on the system, $\mathcal{G}_j$, be a Pauli unitary with probability $(1-\epsilon)$ and a repreparation of an arbitrary state $\tilde{\sigma}_j$ with probability $\epsilon$:
\beq
\mathcal{G}_j(\framebox(5,5){}_{S_1}\!)= (1-\epsilon) X_j (\framebox(5,5){}_{S_1}\!) X_j^{\dagger} + \epsilon\tilde{\sigma}_j,
\eeq
where the $\tilde{\sigma}_j$ can be drawn from an arbitrary set of four states $\{\tilde{\sigma}_j\}_j$.
 In this case, the four initial system-environment states one generates are
\begin{align} \label{jointstate1}
(\rho_j)_{S_1E} = (1-\epsilon) (X_j \otimes \mathbb{1}) |\Phi^+ \rangle \langle \Phi^+ |_{S_0 E} (X^{\dag}_j \otimes \mathbb{1}) 
+ \epsilon \tilde{\sigma}_j \otimes \frac{1}{2} \mathbb{1}.
\end{align}
The four corresponding marginals on $S_1$ are
\begin{align}
\sigma_j = (1-\epsilon) \frac{1}{2} \mathbb{1} + \epsilon \tilde{\sigma}_j.
\end{align}

The measurement of the Bell basis on the system-environment composite gives a uniform distribution over its outcomes if the joint state is of the form $\tilde{\sigma}_j \otimes \frac{1}{2} \mathbb{1}$, so that applying Eq.~\eqref{inter} with a set $\{ \tilde{\tau}_j\}_j$ rather than $\{\tau_j\}_j$ to the state in Eq.~\eqref{jointstate1} gives the four final marginal states of the system 
$\tau_j = (1-\epsilon) \tilde{\tau}_j + \epsilon \frac{1}{4} \sum_j \tilde{\tau}_j.$ Taking $\frac{1}{4} \sum_j \tilde{\tau}_j = \frac{1}{2} \mathbb{1}$ as a simple special case, we have
\begin{align}
\tau_j = (1-\epsilon) \tilde{\tau}_j + \epsilon \frac{1}{2} \mathbb{1}.
\end{align}

The input-output relation in this case is therefore
\begin{align}
\mathfrak{R} = \{ \big((1-\epsilon) \frac{1}{2} \mathbb{1} + \epsilon \tilde{\sigma}_j, (1-\epsilon) \tilde{\tau}_j + \epsilon \frac{1}{2} \mathbb{1}\big) \}_{j}.
\end{align}
Since both sets $\{ \tilde{\sigma}_j \}_j$ and $\{ \tilde{\tau}_j \}_j$ are completely unconstrained sets of four states, one can certainly choose them (for any $\epsilon$) to satisfy condition $2$. One can also choose them to satisfy condition $1$, since a set of four states can be informationally complete for a two-dimensional Hilbert space. The simplest such construction would take the set $\{\tilde{\sigma}_j\}_j$ to be an informationally complete set of pure states, $\epsilon= \frac{1}{2}$, and $\tilde{\tau}_j = T(\tilde{\sigma}_j)$ for all $j$, where $T$ is the transpose operation. 

\section{Generic families of classical examples in which the standard proposal fails}
\label{cex}

%As stated in Section~\ref{cgeneric}, 
All nine families of quantum examples from Appendix~\ref{qex} have close classical analogues. 
%Even in generic classical examples, then, the standard argument

For simplicity of presentation, we represent the system-environment interaction in these examples by a conditional probability distribution $P_{S_2|S_1E}$ rather than the deterministic map $F_{S_2E'|S_1E}$. (Of course, the stochastic map induced by $P_{S_2|S_1E}$ can always be dilated to a deterministic map $F_{S_2E'|S_1E}$.)%(Note that this dilation might require an increase in the dimensionality of the environment, to accommodate any ancilla used in the dilation.)}

We show that (for probability distributions $\omega_j \in \mathfrak{P}(S_1)$ and $\nu_j \in \mathfrak{P}(S_2)$) there exist operationally realizable input-output relations $\mathfrak{R} = \{(\omega_j,\nu_j)\}_j$ for which
\begin{enumerate}
\item there is no stochastic map $\Gamma$ such that $\forall j:\Gamma(\omega_j) = \nu_j. $ 
\item there is no linear map $\Gamma$ such that $\forall j:\Gamma(\omega_j) = \nu_j. $ 
\item there is no map $\Gamma$ such that $\forall j:\Gamma(\omega_j) = \nu_j. $ 
\end{enumerate}
%Here are simple, concrete, and operationally realizable examples of each type:
Each point is demonstrated, respectively, by providing an operationally realizable input-output relation such that 
\begin{enumerate}
%\item If the set $\{ \omega_j \}_j$ is informationally complete, and $\forall j: \nu_j =$, then $\Gamma$ is not a stochastic map. [All we need to do to generate examples is consider a set of epistemic states whose convex hull has symmetries that are not shared by the symmetries of the simplex of ontic states. Then any map corresponding to such a symmetry does not supervene on the ontic dynamics and consequently is not a stochastic map. 
\item the set $\{ \omega_j \}_j$ is informationally complete (that is, forms a basis for the space of functions on $S_1$), and the convex hull of the $\{ \omega_j \}_j$ has symmetries that are not shared by the full simplex of probability distributions on $S_1$, so that any map implementing such a symmetry transformation cannot be a mixture of permutations of the physical states, and consequently cannot be a stochastic map.\footnote{This holds because transformations generated by symmetries of the full space of probability distributions $\mathfrak{P}(S_1)$ are reversible, and the only reversible transformations on the physical states correspond to permutations---that is, symmetries of the simplex; further, all stochastic maps can be expressed as a mixture of permutations. We give a specific example of this form in Appendix~\ref{SonlyClassical}. } %[do we need to specify more about the regular simplex? e.g. that it contains the convex hull of the $\omega_j$?]
\item the $\{ \nu_j \}_j$ are more distinguishable than the $\{ \omega_j\}_j$ \footnote{That is, there exists a pair $(\nu_j,\nu_k)$ which has a smaller classical fidelity than the corresponding pair $(\mu_j,\mu_k)$.}, so that $\Gamma$ is necessarily a nonlinear map. 
\item the $\omega_j$ are independent of $j$ (so $\forall j: \omega_j =\bar{\omega}$ for some fixed $\bar{\omega}$) but $\nu_j$ varies nontrivially with $j$, so that the relation is one-to-many, and there can be no map $\Gamma$ consistent with it.
\end{enumerate}

We again demonstrate that all three of these failures occur in each of the three operational circuit types (listed as $a,b,c$ in Appendix~\ref{qex}).

\subsubsection{Classical operational scenarios in which the setting variable $J$ affects the system and the environment and $K$ is trivial} \label{SEClassical}

Here we provide classical examples of types $(a,1)$, $(a,2)$, and $(a,3)$.

If the setting variable $J$ influences both the principal system and the environment, one can prepare any initial joint probability distribution on the system and environment. Then, as in the quantum case, one can easily generate any relation $\mathfrak{R}$ {\em at all}, including all three types from above.

For example, taking the environment system to have dimension equal to the cardinality of the set $\{ \omega_j \}_j$, then for any set $\{ \omega_j \}_j$ one can prepare the system-environment joint state $\omega_j \otimes [j]$ when $J=j$. The system and environment may then interact via the channel $\Gamma_{S_2|S_1E}(\framebox(5,5){}_{S_1E}\!)= \sum_j\sum_{S_1E} \delta_{j,E} \ \framebox(5,5){}_{S_1E} (\nu_j)_{S_2}$, for any set of states $\{ \nu_j \}_j$. Hence, for $J=j$ the final marginal state of the system is $\nu_j$. Since both sets $\{ \omega_j \}_j$ and $\{ \nu_j \}_j$ are completely unconstrained, one can certainly satisfy the conditions of 1, 2, and 3. 

\subsubsection{Classical scenarios in which the setting variable $J$ determines the measurement on the system and one post-selects on the measurement outcome $K$} \label{PSClassical}

Here we provide classical examples of types $(b,1)$, $(b,2)$, and $(b,3)$.

Suppose that the initial state of the system-environment composite is a maximally correlated probability distribution. By the natural classical analogue of the Hughston-Josza-Wootters theorem, it is possible, by implementing a $j$-dependent measurement on the system and post-selecting on outcome $k$, to steer the environment to any arbitrary probability distribution $\nu_{j,k}$ (see footnote \ref{footsteer}). Further, the update rule of the measurement (which affects only the system) can be arbitrary. Given an outcome $k$ of the $j$th measurement, one can find an update rule which ignores the physical state of the system and simply reprepares the system in the state $\omega_{j,k}$. If the subsequent system-environment interaction is a swap gate, one can again generate the input-output relation $\mathfrak{R} = \{ (\omega_{j,k}, \nu_{j,k})\}_{j,k}$ for completely unconstrained sets $\{ \omega_{j,k} \}_{j,k}$ and $\{ \nu_{j,k} \}_{j,k}$. Hence, one can certainly satisfy the conditions of 1, 2, and 3 (where $j$ is replaced with $j,k$). 

\subsubsection{Classical operational scenarios in which the setting variable $J$ affects only the system and $K$ is trivial} \label{SonlyClassical}

Here we provide classical examples of types $(c,1)$, $(c,2)$, and $(c,3)$.

Suppose that one of four possible states are prepared by implementing a transformation on the system, as follows. For setting $j \in \{1,2,3,4\}$, one applies a transformation $\Gamma_{\Pi_j}\otimes \mathcal{I}$ to the fiducial state $P_{S_1E} := \frac{1}{4} ([11]+[22]+[33]+[44] )$, where $\mathcal{I}$ denotes the identity map and each $\Gamma_{\Pi_j}$ is the map on probability distributions over system $S_1$ induced by 
the corresponding permutation of ontic states (in cycle notation):
\begin{align} \label{perms}
\Pi_1 &:= (1)(2)(3)(4)\\ \nonumber
\Pi_2 &:= (12)(34)\\ \nonumber
\Pi_3 &:= (13)(24)\\ \nonumber
\Pi_4 &:= (14)(23).
\end{align}
In this case, there are four possible system-environment distributions, corresponding to four maximally correlated probability distributions,
\begin{align} \label{cbasis1}
(P_j)_{S_1E} = (\Gamma_{\Pi_j}\otimes \mathcal{I})(P_{S_1E}), \quad \ \ j = \{1,2,3,4\}.
\end{align}
Each of these has the same marginal on $S_1$, namely, the uniform probability distribution 
\begin{align}
\omega_j = \frac{1}{4}([1]+[2]+[3]+[4]) := \bar{\omega}
\end{align}

Now suppose that the system-environment interaction consists of a measurement distinguishing the supports of the maximally correlated states defined by Eq.~\eqref{cbasis1} (that is, a measurement whose outcome $j$ can be associated to the response function $\delta_{\Pi_j(S_1),E}$), followed by a repreparation of the system in the state $\nu_j$, thereby enacting the stochastic map
\beq \label{inter2}
\Gamma_{S_2|S_1 E}(\framebox(5,5){}_{S_1E}\!)=\sum_j \sum_{S_1E} \delta_{\Pi_j(S_1),E}\; \framebox(5,5){}_{S_1E} (\nu_j)_{S_2},
\eeq
where $\{ \nu_j \}_j$ denotes an arbitrary set of four probability distributions. 

The input-output relation in this case is
\beq
\mathfrak{R} = \{ (\bar{\omega},\nu_j) \}_{j}.
\eeq
If the $\nu_j$ depend nontrivially on $j$, then the relation is one-to-many, so we have an example of type 3.

We can modify this example slightly to get one of type 2.
Simply let the initial transformation $\Gamma_{\Pi_j}$ on the system be generated by the permutation $\Pi_j$ of Eq.~\eqref{perms} with probability $(1-\epsilon)$ and a repreparation of an arbitrary state $\tilde{\omega}_j$ with probability $\epsilon$, so that 
\beq
\Gamma_j(\framebox(5,5){}_{S_1}\!)= (1-\epsilon) \Gamma_{\Pi_j}(\framebox(5,5){}_{S_1}\!) + \epsilon\tilde{\omega}.
\eeq
In this case, the four initial system-environment states are
\begin{align} \label{cjointst}
(P_{j})_{S_1E} = (1-\epsilon) (\Gamma_{\Pi_j}\otimes \mathcal{I})(P_{S_1E})+ \epsilon \tilde{\omega} \otimes \bar{\omega}
\end{align}
The marginals on $S_1$ are
\begin{align}
\omega_j = (1-\epsilon) \bar{\omega} + \epsilon \tilde{\omega}.
\end{align}

The measurement defined above gives a uniform distribution over its outcomes if the joint probability distribution is of the form $\omega_j \otimes \bar{\omega}$, so that applying Eq.~\eqref{inter2} with $\nu_j$ replaced by $\tilde{\nu}_j$ to the state in Eq.~\eqref{cjointst} gives the four final marginal states of the system $\nu_j = (1-\epsilon) \tilde{\nu}_j + \epsilon \frac{1}{4} \sum_j \tilde{\nu}_j.$ Taking $\frac{1}{4} \sum_j \tilde{\nu}_j = \bar{\omega}$ as a simple special case, we have
\begin{align}
\nu_j = (1-\epsilon) \tilde{\nu}_j + \epsilon \bar{\omega}.
\end{align}

The input-output relation in this case, therefore, is 
\begin{align} 
\mathfrak{R}= \{ \big((1-\epsilon) \bar{\omega} + \epsilon \tilde{\omega}_j, (1-\epsilon) \tilde{\nu}_j + \epsilon \bar{\omega} \big) \}_{j}.
\end{align}
Since both sets $\{ \tilde{\omega}_j \}_j$ and $\{ \tilde{\nu}_j \}_j$ are completely unconstrained sets of four states, one can certainly choose them (for any $\epsilon$) to satisfy condition $2$.

One can also choose them to satisfy condition 1. As one example, let $\epsilon = \frac{1}{2}$, and let the $\tilde{\omega}_j$ be 
\begin{align} \label{domaino}
\tilde{\omega}_1 &:= \frac{1}{2}[1] + \frac{1}{2}[2],\\ \nonumber
\tilde{\omega}_2 &:= \frac{1}{2}[1] + \frac{1}{2}[3],\\ \nonumber
\tilde{\omega}_3 &:= \frac{1}{2}[1] + \frac{1}{2}[4],\\ \nonumber
\tilde{\omega}_4 &:= \frac{1}{2}[3] + \frac{1}{2}[4].%,\\ \nonumber
%\omega_5 &:= \frac{1}{2}[2] + \frac{1}{2}[4],\\ \nonumber
%\omega_6 &:= \frac{1}{2}[2] + \frac{1}{2}[3],
\end{align}
The set $\{ \tilde{\omega}_j\} $ is a basis for the space of distributions over the 4 physical states. Let the $\nu_j$ be $\tilde{\nu}_j = 2\bar{\omega} - \tilde{\omega}_j$, so that
\begin{align} \label{cbasis}
\tilde{\nu}_1 &:= \frac{1}{2}[3] + \frac{1}{2}[4],\\ \nonumber
\tilde{\nu}_2 &:= \frac{1}{2}[2] + \frac{1}{2}[4],\\ \nonumber
\tilde{\nu}_3 &:= \frac{1}{2}[2] + \frac{1}{2}[3],\\ \nonumber
\tilde{\nu}_4 &:= \frac{1}{2}[1] + \frac{1}{2}[2].%,\\ \nonumber
%\nu_5 &:= \mu^{(2)},\\ \nonumber
%\nu_6 &:= \mu^{(3)}.
\end{align}

%Note that because the set $\{ \omega_j\} $ is a basis for the space of distributions over the 4 physical states, so too is the set of first components of the elements of $\mathfrak{R}$, that is, $\{ \frac{1}{2}(\bar{\omega} + \omega_j ) \}_j$. 

The relation in such circumstances is
\begin{align} \label{crels}
&\mathfrak{R} =\\ \nonumber
&\{\Big(\frac{1}{2}(\bar{\omega} + \frac{1}{2}[1] + \frac{1}{2}[2] ), \frac{1}{2}(\bar{\omega} + \frac{1}{2}[3] + \frac{1}{2}[4] )\Big),
\nonumber\\
&\Big(\frac{1}{2}(\bar{\omega} + \frac{1}{2}[1] + \frac{1}{2}[3] ),\frac{1}{2}(\bar{\omega} + \frac{1}{2}[2] + \frac{1}{2}[4] )\Big),
\nonumber\\
&\Big(\frac{1}{2}(\bar{\omega} + \frac{1}{2}[1] + \frac{1}{2}[4] ), \frac{1}{2}(\bar{\omega} + \frac{1}{2}[2] + \frac{1}{2}[3] )\Big), 
\nonumber\\
&\Big(\frac{1}{2}(\bar{\omega} + \frac{1}{2}[3] + \frac{1}{2}[4] ), \frac{1}{2}(\bar{\omega} + \frac{1}{2}[1] + \frac{1}{2}[2] )\Big)%,
\nonumber
%&\Big(\frac{1}{2}(\bar{\omega} + \frac{1}{2}[2] + \frac{1}{2}[4] ),\frac{1}{2}(\bar{\omega} + \frac{1}{2}[1] + \frac{1}{2}[3] )\Big),
%\nonumber\\ \nonumber
%&\Big(\frac{1}{2}(\bar{\omega} + \frac{1}{2}[2] + \frac{1}{2}[3]),\frac{1}{2}(\bar{\omega} + \frac{1}{2}[1] + \frac{1}{2}[4] )\Big) 
\}.
\end{align}

There is a linear map which fits the input-output relation: it is given by the matrix 
\beq
%\Gamma(\vec{p}_{S_1}) = 
\left[\begin{array}{rrrr}
\sfrac{-1}{2} & \sfrac{1}{2} & \sfrac{1}{2} & \sfrac{1}{2} \\
\sfrac{1}{2} & \sfrac{-1}{2} & \sfrac{1}{2} & \sfrac{1}{2} \\
\sfrac{1}{2} & \sfrac{1}{2} & \sfrac{-1}{2} & \sfrac{1}{2}	\\
\sfrac{1}{2} & \sfrac{1}{2} & \sfrac{1}{2} & \sfrac{-1}{2}
\end{array}\right] %\cdot \vec{p}_{S_1}.
\eeq
(acting on probability distributions over the four physical states of $S_1$, expressed as four-component vectors).
This linear map is not stochastic, since the elements in the matrix are not all positive. Because the set of input distributions in the relation form a basis for the space of all distributions over physical states, this is the {\em unique} linear map satisfying the relation. Hence, in such an example, the only linear map consistent with the standard proposal is non-stochastic. 
%Could give an intuitive argument, like that of the older example: However, this $\Gamma$ is not a stochastic matrix, and in fact there is {\em no} stochastic map consistent with the relation $\mathfrak{R}$ above. To see this, note that Eq.~\eqref{trans1} guarantees that physical state 0 has no probability of transforming to 2 or 3, while Eq.~\eqref{trans3} guarantees that physical state 0 has no probability of transforming to 1 or 2. Hence, any single map on the physical states which respects these two facts cannot affect physical state 0 in any way. Every such map, however, is inconsistent with Eq.~\eqref{trans2}, which requires that physical state 0 be nontrivially transformed with probability at least $\frac{1}{2}$. 

Geometrically, the convex hull of the distributions in Eq.~\eqref{domaino} and Eq.~\eqref{cbasis} forms an octahedron inside the 4-simplex of all probability distributions. The evolution map one finds by the standard proposal (defined by Eq.~\eqref{crels}) corresponds to a reflection through the origin, which is a reversible injective map. However, this map (extended uniquely by linearity to the whole space) does not correspond to a symmetry of the 4-simplex. Hence the linear map in this example does not correspond to any permutation of the physical states.

\section{When does the evolution map coincide with the inference map classically?} \label{coincide}

%\red{[Is this section in the right place?]}
Consider the classical circuit of Fig.~\ref{conestatemap}(b). Generally, it is possible to have inferences along a common cause pathway and therefore possible that the inference map might differ from the evolution map. What are the necessary and sufficient conditions on $P_{S_1 E}$ for them to be the same for all $F_{S_2 E'|S_1E}$?
The answer is simply that $E$ and $S_1$ must be marginally independent, 
\begin{align}
P_{S_1 E}= P_{S_1} \otimes P_{E}.
\end{align}

A sufficient condition for this independence to hold is that $S_1$ and $E$ have no common ancestors in the causal structure. However, the independence relation can sometimes hold when this causal condition fails, for instance, using fine-tuned choices of circuit elements~\cite{Wood,Pearl}.

If one is considering a circuit that contains some variables which one is conditioning upon (like $J$ and $K$ in the examples we considered previously), then the sufficient condition for equality of the inference and evolution maps is simply that one must have $S_1$ independent of $E$ after conditioning on these variables. It is worth noting that this provides another example (besides the possibility of fine-tuning) of how a statistical independence can hold in spite of the variables failing to have disconnected causal ancestry---in classical example 2, for instance, $J$ is a common cause of $S_1$ and $E$, but if one conditions upon it, then $S_1$ and $E$ become independent.

Therefore, the precise circumstances under which confusing inference and evolution has no consequences are when there are no initial system-environment correlations. In other words, the first mistake of the standard proposal, conflating inference and evolution, is problematic precisely because the standard proposal sought to address scenarios with such correlations.

\section{Knowledge-dependence of evolution maps} ~\label{knowldepend}
We mentioned in Section~\ref{sec:sac} and Section~\ref{sec:quantumdomaps} that the classical and quantum evolution maps depend on one's knowledge of the state of the environment. We now provide several textbook examples of this phenomena, simply to show that it is not a cause for concern.

\begin{figure}[htb!]
\centering
\includegraphics[width=0.37\textwidth]{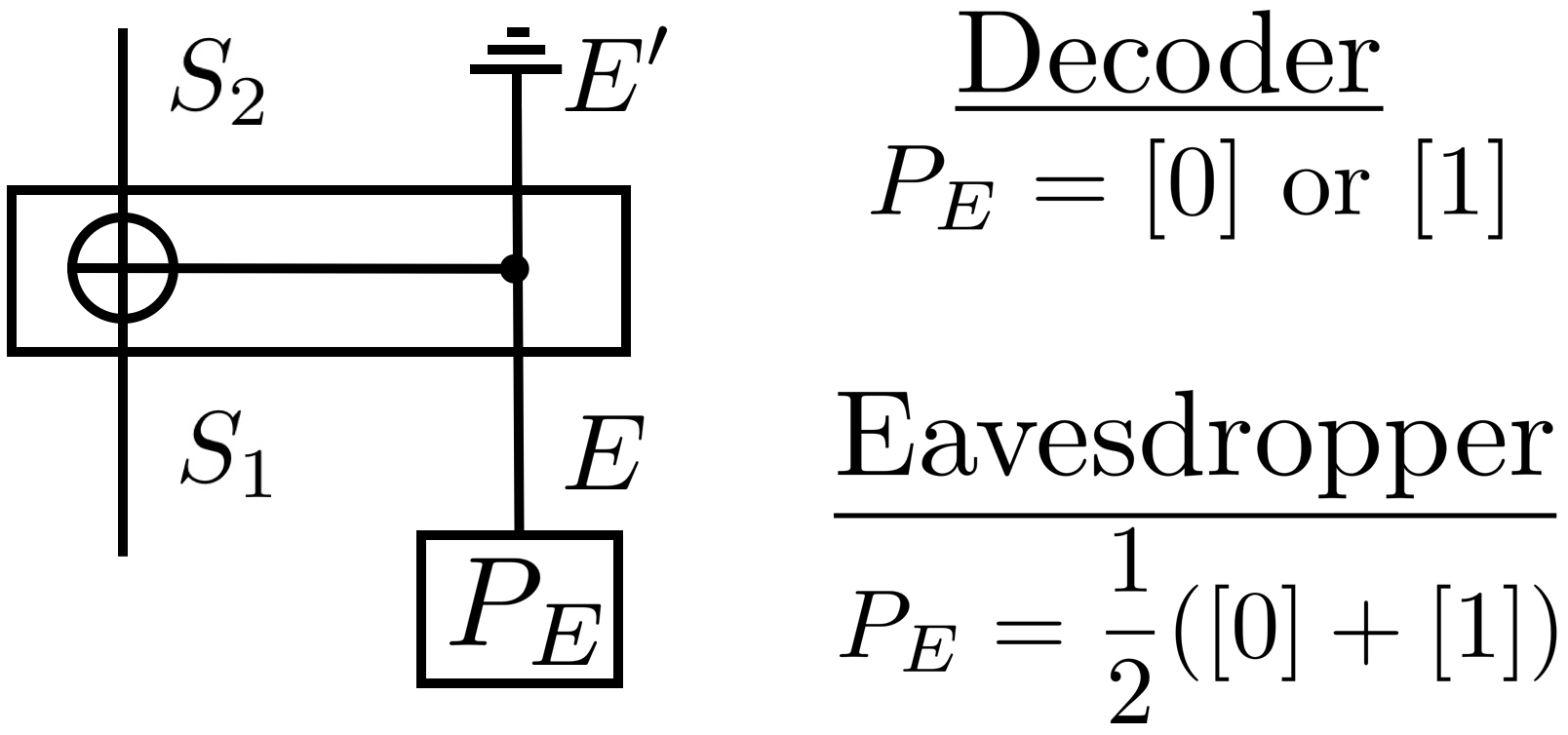}
\caption{(a) A one-time pad. The decoder and eavesdropper have distinct information about the key ($E$).}
%(a) In this realization of a one-time pad, the sender and recipient, who have complete information of the key $E$, describe the evolution from plaintext $S_1$ to cyphertext $S_2$ as reversible. The eavesdropper, who has no information about the key, describes the same evolution as totally irreversible. (b) The evolution map $\Gamma_{S_2|{\rm do} S_1}$ in this example depends on one's knowledge of the key, encoded in $P_E$. }
\label{pad}
\end{figure}

Consider encoding a single bit using a Vernam cypher (one-time pad). Let $S$ be the bit to be encoded (the plaintext) and $E$ the key, which is shared between the sender and the receiver. The value of $E$ is sampled uniformly at random. To encode the plaintext, a controlled-NOT is implemented on $S$ with the key $E$ as control, as shown in Fig.~\ref{pad}. Because the users of the cypher know the value of $E$, they describe the evolution of $S$, conditional on the value of $E$, as follows: if $E=0$ the evolution map is the identity, while if $E=1$ it is the bit-flip. An eavesdropper assigns to $E$ the uniformly random distribution $\frac{1}{2}([0]+[1])$, and therefore describes the evolution of $S$ as the randomization channel (the equal mixture of identity and bit-flip channels). 

The knowledge-dependence of one's description of evolution is also seen in many protocols for quantum cryptography. For example, a private quantum channel on a qubit~\cite{private} can be implemented as follows. The sender and receiver share a uniformly random two-bit key. The sender draws a Pauli matrix from the set $\{I,X,Y,Z\}$ based on the value of the key, and implements it on the system. To decode, the user applies the same Pauli. The users of the channel, who know the key's value, describe the evolution as $\rho \rightarrow \sigma_i \rho \sigma_i^T$. An eavesdropper who does {\em not} know the key's value, but knows only that the key was drawn uniformly at random, describes the evolution as $\rho \rightarrow \frac{1}{2}\mathbb{1}$.

As a final example, consider the transformation that a quantum state undergoes when it is measured. The map which one uses to update the state of the measured system depends on how much one knows about the outcome of the measurement. Consider for simplicity a rank-one projective measurement $\{ \Pi_k \}_k$ performed on some initial state $\rho$. If one has no information about which outcome occurred, then the correct update map is given by the non-selective update rule, $\rho \rightarrow \sum_k \Pi_k \rho \Pi_k$. On the other hand, if one knows that outcome $k$ occurred, then the correct update map is given by the selective update rule, $\rho \rightarrow \Pi_k \rho \Pi_k$. 

\section{Analysis of the three classical examples of Sec.~\ref{sec:classical}}\label{app:classical}

We now provide a derivation of the claims of Section~\ref{illust}, completing the analysis of the three classical examples from Sec.~\ref{sec:classical} and the mistakes of the standard proposal when applied to them. For convenience, we represent the system-environment interaction in each case by the probabilistic dependence $P_{S_2|S_1E}=\sum_{E'}P_{S_2E'|S_1E}$, where $P_{S_2E'|S_1E}$ is the conditional associated to the stochastic map $F_{S_2E'|S_1E}$; we also indicate conditioning on $J=j$ or $K=k$ in the subscripts (e.g., $P_{E|J=0}(e):=P_{E|J}(e|0)$).

{\bf Classical Example 3, from Section~\ref{c:CNOT}} 

Consider again the third classical example, introduced in Section~\ref{c:CNOT} and discussed in Section~\ref{illust}. %As we noted, this example yields a one-to-many input-output relation so that the standard argument fails to identify any map at all.

As indicated by Eq.~\eqref{ex3Gamma}, the inference map from $S_1$ to $S_2$ for a particular value $j$ of $J$ is
\begin{align} \label{ap:ex3Gamma}
\Gamma_{S_2|S_1}^{(j)}(\framebox(5,5){}_{S_1}\!) &= \sum_{S_1 E} P_{S_2 |S_1 E} P_{E|S_1 J=j}\ \framebox(5,5){}_{S_1}\!.
\end{align}

Given the nature of the coupling of system and environment, we have
 \begin{align}
P_{S_2|S_1 E} = \delta_{S_2, S_1 \oplus_2 E}.\label{PS2S1E}
 \end{align}
 Denote the early version of $S_1$ by $S_0$. Then, because the joint distribution over $S_0$ and $E$ is a state of perfect positive correlation with uniform marginals, 
 \begin{align}
 P_{S_0E} &= \frac{1}{2} [0]_{S_0} \otimes [0]_E + \frac{1}{2} [1]_{S_0} \otimes [1]_E\nonumber\\
&= (\frac{1}{2} [0]_{S_0}+\frac{1}{2} [1]_{S_0}) \delta_{E, S_0},
 \end{align}
and because the controlled-NOT from $J$ to $S_0$ is modeled by the conditional
 \begin{align}
 P_{S_1|S_0} = \delta_{S_1, S_0 \oplus_2 J},
 \end{align}
a simple application of Bayesian probability theory implies that
 \begin{align}
 P_{E|S_1} &= \sum_{S_0} P_{E|S_0} P_{S_0|S_1}\nonumber\\
 &= \sum_{S_0} \delta_{E,S_0} \delta_{S_0, S_1 \oplus_2 J}\nonumber\\
 &= \delta_{E, S_1 \oplus_2 J}.\label{PES1}
 \end{align}
Finally, substituting Eqs.~\eqref{PS2S1E} and \eqref{PES1} into Eq.~\eqref{ap:ex3Gamma}, we conclude that 
\begin{align}
\Gamma^{(j)}_{S_2|S_1} (\framebox(5,5){}_{S_1}\!) &= \sum_{S_1E} \delta_{S_2, S_1 \oplus_2 E} \delta_{E,S_1 \oplus_2 j} \framebox(5,5){}_{S_1}\!\\
&= \delta_{S_2,j}\\
&= [j]_{S_2}.
\end{align} 
This constitutes the proof of Eq.~\eqref{infmap3}.

The correct evolution map in this example is straightforward to identify using the prescription of Section~\ref{sec:causalp}. For both values of $J$, the marginal state of the environment is uniformly random, so there is a unique evolution map. It is easily seen to be the randomization channel, which takes any input probability distribution $P_{S_1}$ to a uniformly random distribution on $S_2$, as stated in Eq.~\eqref{randchan}.

{\bf Classical Example 2, from Section~\ref{c:NC}} 

We start with Eq.~\eqref{ex2Gamma}, in the form
\begin{align} \label{ap:ex2Gamma}
\Gamma_{S_2|S_1}^{(j)}(\framebox(5,5){}_{S_1}\!) &= \sum_{S_1 E} P_{S_2 |S_1 E} P_{E|J=j} \framebox(5,5){}_{S_1}\!.
\end{align}
Recalling the details of the example, one has
 \begin{align}
 P_{S_2|S_1 E} = \delta_{S_2, (S_1 \oplus_3 E)}.
 \end{align}
Furthermore, because $E$ simply tracks $J$, we have
 \begin{align}
 P_{E|J} &= \delta_{E,J}.
 \end{align}
Substituting these expressions into Eq.~\eqref{ap:ex2Gamma}, we conclude that 
\begin{align}
\Gamma^{(j)}_{S_2|S_1} (\framebox(5,5){}_{S_1}\!) &= \sum_{S_1E} \delta_{S_2, (S_1 \oplus_3 E)} \delta_{E,J}\framebox(5,5){}_{S_1}\!\\
&= \sum_{S_1} \delta_{S_2, (S_1 \oplus_3 j)}\framebox(5,5){}_{S_1}\!,
\end{align} 
confirming Eq.~\eqref{infmap2}

{\bf Classical Example 1, from Section~\ref{c:transpose}} 

Here, both $J$ and $K$ are nontrivial. The causal structure is the same as that of Fig.~\ref{Transpose}, and ensures that $E$ is conditionally independent of $S_1$ given $JK$, $P_{E|S_1 JK} = P_{E|JK}$. Thus, the inference map for $J=j$ and $K=k$ is
\begin{align} \label{ex1Gamma}
\Gamma_{S_2|S_1}^{(j,k)}(\framebox(5,5){}_{S_1}\!) &= \sum_{S_1 E} P_{S_2 |S_1 E} (P_{E|J=j, K=k} \otimes \framebox(5,5){}_{S_1}).
\end{align}

Because of the perfect correlation between $S$ and $E$ in the initial joint state, whatever one infers about $S$ from learning that $j=j$ and $K=k$, one learns the same thing about $E$ as well. Consequently, 
%for the case of $K=1$, if $J=1$, then
\begin{align}
P_{E|J=1,K=1} &= \frac{1}{2}[2]_{E}+\frac{1}{2}[3]_{E},\\
 P_{E|J=2,K=1} &= \frac{1}{2}[1]_{E}+\frac{1}{2}[3]_{E},\\
 P_{E|J=3,K=1} &= \frac{1}{2}[1]_{E}+\frac{1}{2}[2]_{E}.
 \end{align}
 The system and environment interact via a swap, so
 \begin{align}
 P_{S_2|S_1 E} = \delta_{S_2, E}.
 \end{align}
Substituting these expressions into Eq.~\eqref{ex1Gamma}, we have
\begin{align}
\Gamma^{(1,1)}_{S_2|S_1} (\framebox(5,5){}_{S_1}\!) &= \frac{1}{2}[2]_{S_2}+\frac{1}{2}[3]_{S_2},\label{ex1inferencemaps1}\\
\Gamma^{(2,1)}_{S_2|S_1} (\framebox(5,5){}_{S_1}\!) &= \frac{1}{2}[1]_{S_2}+\frac{1}{2}[3]_{S_2},\label{ex1inferencemaps2}\\
\Gamma^{(3,1)}_{S_2|S_1} (\framebox(5,5){}_{S_1}\!) &= \frac{1}{2}[1]_{S_2}+\frac{1}{2}[2]_{S_2}.\label{ex1inferencemaps3}
 \end{align}
The inference map for $K=1$ and for the different values of $J$ is simply the map that ignores the distribution on $S_1$ and outputs whatever distribution one infers for $E$ as the distribution on $S_2$. Clearly, therefore, the inference map is $JK$-dependent in this example.

Recalling Eq.~\eqref{inoutex}, one sees that each of the input-output pairs (for a given value of $J$ when $K=1$) is consistent with the corresponding inference map in Eqs.~\eqref{ex1inferencemaps1}-\eqref{ex1inferencemaps3}. But if one tries to find a {\em single} map that is consistent with {\em all three} input-output pairs, as the standard proposal mistakenly suggests to do, one finds that no stochastic map can do the job.

As with example 2, it happens that the evolution map for any given values of $J$ and $K$ coincides with the inference map for those values because the causal structure is such that $S_1$ and $E$ are conditionally independent given $JK$,
 %(as noted above) conditioning on $JK$ makes $S_1$ and $E$ independent, 
 so that an intervention on $S_1$ does not change the map after conditioning on $JK$,
\begin{align}
\Gamma^{(j,k)}_{S_2|{\rm do}S_1} = \Gamma^{(j,k)}_{S_2|S_1}.
\end{align}

\section{The correct quantum evolution maps for scenarios with nontrivial $J$ and $K$} \label{mapsinqSA}

%The map in the hypothetical scenario of Fig.~\ref{generalq}(b), 
Consider the scenario at play in the standard proposal, shown in Fig.~\ref{fig:standard}. Following the prescription of Section~\ref{sec:quantumdomaps}, one obtains (for $J=j$ and $K=k$) a map in the modified circuit given by
\beq 
\mathcal{E}^{(j,k)}_{S_2|S_1'}(\framebox(5,5){}_{S_1'}\!)={\rm Tr}_{E'}(\mathcal{U}_{S_2E'|S_1' E} (\framebox(5,5){}_{S_1'}\!\otimes \rho_E^{(j,k)}))
\eeq
where $ \rho_E^{(j,k)} ={\rm Tr}_{S_1}(\rho_{S_1E}^{(j,k)})$.
% and where $\rho_{S_1E}^{(j,k)}$ is the state of the system-environment composite.
Thus, by Eq.~\eqref{defnquantumdomap}, the evolution map from $S_1$ to $S_2$ in the circuit of Fig.~\ref{fig:standard}, given that the setting variable has value $j$ and the outcome was found to be $k$, is 
\beq \label{qdomapikSA}
\mathcal{E}^{(j,k)}_{S_2|{\rm do}(S_1)}(\framebox(5,5){}_{S_1}\!)={\rm Tr}_{E'}(\mathcal{U}_{S_2E'|S_1 E} (\framebox(5,5){}_{S_1}\!\otimes \rho_E^{(j,k)})).
\eeq
This is the quantum analogue of Eq.~\eqref{cdomapik}. 

As was the case classically, the evolution map generally depends on the values of the parameters $JK$, through the marginal state of the environment appearing in Eq.~\eqref{qdomapikSA}. 

\subsection{Resolution of the three quantum examples in Section~\ref{sec:standard}}

Using Eq.~\eqref{qdomapikSA}, we can compute the correct evolution maps for the three quantum examples of Section~\ref{sec:standard}.

In the first quantum example (Section~\ref{transpose}), the evolution map for $J=j$ and $K=1$ is
\begin{equation}
\mathcal{E}^{(j,1)}_{S_2|{\rm do}(S_1)}(\framebox(5,5){}_{S_1}\!) = \ket{\psi_{j,1}^T}\bra{\psi_{j,1}^T}_{S_2}.
\end{equation}
In other words, for each value $j$ of $J$ (which corresponded to preparing $\ket{\psi_{j,1}}_{S_1}$ as the marginal state of the system), one has a distinct map, which ignores the input on the system and reprepares the fixed state $\ket{\psi_{j,1}^T}_{S_2}$.

In the second quantum example (Section~\ref{NC}), the evolution maps for $J=0$ and $J=1$ are
\begin{align}
\mathcal{E}^{(0)}_{S_2|{\rm do}(S_1)}(\framebox(5,5){}_{S_1}\!) &= \framebox(5,5){}_{S_1}\! \\
\mathcal{E}^{(1)}_{S_2|{\rm do}(S_1)}(\framebox(5,5){}_{S_1}\!) &= (XH)\framebox(5,5){}_{S_1}\! (XH)^{\dagger}. 
\end{align}
That is, for $J=0$ the controlled operation is not activated, while for $J=1$ it is.
%As with the classical analog, this description is painfully obvious once one recognizes that there is no single map for all values of $J$.

For these two quantum examples, if one considers specific values for $J$ and $K$, the input-output pair associated to that valuation is consistent with the evolution map for that valuation. This is because in both examples, conditioning on $J$ and $K$ makes $S_1$ and $E$ marginally independent in the sense that 
\begin{align}
\rho_{S_1 E|J=j,K=k} =\rho_{S_1 |J=j,K=k}\otimes \rho_{E|J=j,K=k},
\end{align}
and therefore satisfies the condition of Eq.~\eqref{condequiv} for the equivalence of the evolution map and the inference map.

In the third quantum example (Section~\ref{CNOT}), there is a unique evolution map $\mathcal{E}^{(j,k)}=\mathcal{E}$, since $\rho^{(j,k)}_E = \rho_E$. It is
\begin{align}
\mathcal{E}_{S_2|{\rm do}(S_1)}(\framebox(5,5){}_{S_1}\!) = \frac{1}{2}\framebox(5,5){}_{S_1}\!+\frac{1}{2}Z \;\framebox(5,5){}_{S_1}\! Z^{\dagger},
\end{align}
the dephasing map in the Z basis. This is intuitively clear from the fact that we are performing a flip of the system in the Z basis, conditioned on an environment whose marginal state is completely mixed.

In this example, each individual input-output pair for a given value of $J$ and $K$ is {\em not} consistent with the evolution map. This is because, for a given value of $J$, if you came to know something about $S_1$, it would inform you about $S_2$ {\em both} by the cause-effect connection between $S_1$ and $S_2$ {\em and} by the common-cause connection between $S_1$ and $S_2$ that is mediated by $E$. The precise formalism for modeling this sort of inference in quantum theory, however, is not yet clear.

\section{When do individual input-output pairs of quantum states constrain the quantum evolution map?}
 \label{coincide2}

Here we show that the necessary and sufficient condition for an individual input-output pair appearing in the standard proposal to be a constraint on the evolution map is
%this to be the case is
%In quantum scenarios, the necessary and sufficient condition is analogous; it is
 that $S_1$ and $E$ are marginally independent,
 % i.e., their quantum state factorizes
\begin{align} \label{condequiv}
\rho_{S_1 E} =\rho_{S_1}\otimes \rho_{E}.
\end{align}
The proof is by contradiction.  If the joint state on $S_1 E$ did not factorize, then there would be a choice of  unitary in the circuit such that $S_2$ depended explicitly on the correlations between $S_1$ and $E$ for a given value of $JK$.  The input-output pair for that value of $JK$ would then reflect this dependence.  On the other hand, by the definition of the quantum evolution map, its output on $S_2$ {\em cannot} depend on the correlations between $S_1$ and $E$, and so cannot yield the same input-output pair.

\bibliographystyle{ieeetr}
\bibliography{bib}

\end{document}